\documentclass[12pt]{article} 

\pdfoutput=1
\usepackage{jheppub} % for details on the use of the package, please
\usepackage[utf8]{inputenc}
\usepackage{hyperref}
\usepackage{url}
\usepackage{setspace}
\usepackage{amsmath,amssymb,amscd,braket}
\usepackage{amsfonts}
\usepackage{color}
\usepackage{graphicx}
\usepackage{xcolor}
\usepackage{mathtools}
\usepackage{dsfont}

\numberwithin{equation}{section}

\newcommand{\vp}{\varphi}

\newcommand{\R}{\mathbb{R}}
\newcommand{\Z}{\mathbb{Z}}
\newcommand{\Pcr}{Poincar\'{e} $\space$}
\newcommand{\Riem}{\wt{\mbox{R}}\mbox{iem}}
\newcommand{\nA}{\wt{\nabla}}
\newcommand{\X}{\wt{X}}
\newcommand{\T}{\hat{T}}

\newcommand{\comment}[1]{ }

\newcommand{\wt}[1]{\widetilde{#1}}
\newcommand{\wh}[1]{\widehat{#1}}

 \def\p{\partial}
 \newcommand{\bea}{\begin{eqnarray}}
\newcommand{\eea}{\end{eqnarray}}
\newcommand{\be}{\begin{equation}}
\newcommand{\ee}{\end{equation}}
\newcommand{\ba}{\begin{align}}
\newcommand{\ea}{\end{align}}

\def\eg{{\it e.g.~}}

\newlength{\slength}
\DeclarePairedDelimiter\ceil{\lceil}{\rceil}

\renewcommand{\title}[1]{\vbox{\center\LARGE{#1}}\vspace{5mm}}
\renewcommand{\author}[1]{\vbox{\center#1}\vspace{5mm}}

\begin{document}

\pagestyle{plain}

	%----------------------------------------------------------------------%
	%  numbering equations with section number
	%----------------------------------------------------------------------%
	\makeatletter
	\@addtoreset{equation}{section}
	\makeatother
	\renewcommand{\theequation}{\thesection.\arabic{equation}}
	%----------------------------------------------------------------------%
	%  title page
	%----------------------------------------------------------------------%
	\pagestyle{empty}
	%\vspace*{1.0in}

\vspace{4cm}

\begin{center}
	\LARGE{\bf The Ambient Space Formalism\vspace{2mm}}\\
	\large{Enrico Parisini, Kostas Skenderis, Benjamin Withers\\[4mm]}
	\footnotesize{
	Mathematical Sciences and STAG Research Centre,\\ University of Southampton, Highfield, Southampton SO17 1BJ, UK}\\
	\footnotesize{\href{mailto:e.parisini@soton.ac.uk}{e.parisini@soton.ac.uk}, \href{mailto:k.skenderis@soton.ac.uk}{k.skenderis@soton.ac.uk}, \href{mailto:b.s.withers@soton.ac.uk}{b.s.withers@soton.ac.uk}}

\vspace*{20mm}

\small{\bf Abstract} 
\\[5mm]
\end{center}
\begin{center}
\begin{minipage}[h]{\textwidth}
We present a new formalism to solve the kinematical constraints due to Weyl invariance for CFTs in curved backgrounds and/or non-trivial states, and we apply it to thermal CFTs and to CFTs on squashed spheres. The ambient space formalism is based on constructing a class of geometric objects that are Weyl covariant and identifying them as natural building blocks of correlation functions. We construct (scalar) $n$-point functions and we illustrate the formalism with a detailed computation of 2-point functions. We compare our results for thermal 2-point functions with results that follow from thermal OPEs and holographic computations, finding exact agreement. In our holographic computation we also obtain the OPE coefficient of the leading double-twist contribution, and we discuss how the double-twist coefficients may be computed from the multi-energy-momentum contributions, given knowledge of the analytic structure of the correlator. The 2-point function for the CFT on squashed spheres is a new result.  We also discuss the relation of our work to flat holography. 

\end{minipage}
\end{center}

	\newpage
	%----------------------------------------------------------------------%
	%  Resetting of counters
	%----------------------------------------------------------------------%
	\setcounter{page}{1}
	\pagestyle{plain}
	\renewcommand{\thefootnote}{\arabic{footnote}}
	\setcounter{footnote}{0}
	%----------------------------------------------------------------------%
	%  Paper begins
	%----------------------------------------------------------------------%
	
	%\end{document}
	%\newpage
	
 \tableofcontents

\section{Introduction}
On flat spacetime, conformal invariance imposes strong restrictions on correlation functions for CFTs in vacuum states. 1-, 2- and 3-point functions are completely determined up to constants, while 4- and higher-point functions are given in terms of arbitrary functions of cross-ratios of the insertion locations \cite{Polyakov:1970xd,Ferrara:1973yt, DiFrancesco:1997nk}. These arise as the general solutions to Ward identities associated to conformal invariance.

But what happens to the structure of CFT correlation functions if the theory is on a curved background $g_{(0)}$ or in a non-trivial state? In this situation we no longer enjoy the constraints of conformal invariance\footnote{If the CFT is in a non-trivial state, it should still satisfy spontaneously broken conformal Ward identities, see \cite{Marchetto:2023fcw} for recent work, and CFT $n$-point functions in non-trivial states are equivalent to $(n+2)$-point functions in vacuum, with the additional operators representing the initial and final states. Moreover, if the background is weakly curved one can use conformal perturbation theory to connect the correlators to higher-point functions in flat space involving insertions of energy-momentum tensors. More generally, we expect $n$-point functions on non-trivial backgrounds to serve as generating functions of higher-point functions with additional energy-momentum tensor insertions. In any given example, such considerations may provide a way to go beyond the universal terms we determine in this paper. Here we aim to solve the generic kinematical constraints due to Weyl covariance. We expect the constraints we derive to be wholly compatible with those coming from known constraints on vacuum $n$-point functions on flat space.}, but the theory should still be invariant under Weyl transformations\footnote{This is true for unitary CFTs in $d\leq 10$\cite{Farnsworth:2017tbz}. There are counterexamples based on non-unitary CFTs, see \cite{graham}, \cite{Eastwood:2002su}, \cite{Gover2004ConformalDR} for mathematics literature and 
\cite{Karananas:2015ioa} for a physics discussion. The converse is always true, {\it i.e.} a Weyl invariant theory on a curved background is always a CFT when the metric is set to the flat metric. We would like to thank Kara Farnsworth for a discussion regarding this point.} and
we can obtain useful kinematic constraints from \emph{Weyl covariance}. In particular, we consider correlation functions of conformal primary operators $O_i$ which transform homogeneously under Weyl transformations\footnote{The Weyl transformation rule of CFT primary operators of sufficiently high dimension could acquire additional inhomogeneous terms that depend on the curvature tensor \cite{Farnsworth:2017tbz}. No such examples are currently known, but had such cases arise our discussion would need to be suitably amended.}, 
\begin{equation} \label{intro:weyl}
    \braket{O_1(x_1) \ldots O_n(x_n)}_{\Omega^2 g_{(0)}} =
    \Omega(x_1)^{-\Delta_1} \dots \Omega(x_n)^{-\Delta_n} 
    \braket{O_1(x_1) \ldots O_n(x_n)}_{g_{(0)}}.
\end{equation}
The aim of this work (a companion to \cite{Parisini:2022wkb}) is to present a universal solution to the constraints \eqref{intro:weyl} (we will explain shortly what we mean by ``universal''). 

The simplest case of \eqref{intro:weyl} is that of 1-point functions. On the vacuum and in flat space these are always zero (unless the operator is the unit operator).  On a curved background and/or non-trivial states 1-point functions are generically non-vanishing, and they transform as in \eqref{intro:weyl}. The classification of local Weyl \footnote{In the mathematics literature the terminology ``conformal'' is used instead of ``Weyl''. Here we reserve the terminology ``conformal'' to refer to the combination of a Weyl transformation and a diffeomorphisms that leave a given metric (typically the flat metric) invariant.} invariants constructed from the metric has a long history in mathematics. Fefferman and Graham \cite{Fefferman85,Fefferman:2007rka} mapped this problem to that of a classification of diffeomorphism invariants, using the so-called \emph{ambient space}, 
an associated Ricci-flat curved spacetime in $d+2$ dimensions. 

In this paper we will use the ambient space to provide solutions of \eqref{intro:weyl} for $n >1$ with prescribed leading singularity at coincident points, as dictated by physics.\footnote{Previous uses of the ambient space and related notions in physics include higher spin theories, holographic anomalies and Weyl-covariant theories \cite{Grigoriev:2011gp,Joung:2013doa,Bekaert:2017bpy,Grigoriev:2018mkp,Curry:2014yoa,Gover:2011rz,RodGover:2012ib,Ciambelli:2019bzz, Jia:2021hgy, Jia:2023gmk}.}  This will be done by building a set of Weyl covariant functions of the insertion points $x_i$, as geometric objects in $d+2$ dimensional ambient space. In a sense we generalise the Fefferman-Graham construction to the multi-local case. In doing so, our solutions build a bridge between results in conformal geometry and CFTs. We do not claim that solutions to \eqref{intro:weyl} generated in this way are the most general, however they are universal, applying to all CFTs in curved backgrounds and non-trivial states. 

In vacuum and on conformally flat spaces, a convenient way of obtaining solutions to \eqref{intro:weyl} (as well as to the conformal Ward identities) is by using the embedding space formalism \cite{Dirac:1936fq, Boulware:1970ty,Weinberg:2010fx,Costa2011,Costa2011a, Rychkov2016}. This exploits the realisation of the conformal group in $d$ dimensions $SO(1,d-1)$ as the action of Lorentz transformations in $d+2$ dimensions. For example, consider the square distance between insertions $O_i$ and $O_j$ in the embedding space,
\be
X_{ij} \equiv (X_i - X_j)^2. \label{introXij}
\ee
This object has definite weight $2$ under Weyl transformations and appears as the fundamental building block in all scalar correlation functions, introduced in appropriate combinations so as to solve \eqref{intro:weyl}. In fact, this procedure completely fixes the 2- and 3- point functions. The ambient space departs from the embedding space by having non-zero Riemann curvature. We propose that this curvature captures the effects of both the CFT metric $g_{(0)}$ and non-trivial state. On the curved ambient space, \eqref{introXij} ceases to be a Weyl covariant scalar and cannot be used to build solutions to \eqref{intro:weyl}. However, we can naturally improve it to $\X_{ij}$, the squared geodesic distance between the insertions, 
\be
\X_{ij} \equiv \ell(\X_i, \X_j)^2. \label{introwtXij}
\ee
Indeed, this has definite Weyl weight and is an appropriate building block for correlation functions, reducing back to \eqref{introXij} in the flat space limit. The trajectory of this geodesic explores the geometry of the ambient space, thus encoding the dependence of correlation functions on the CFT metric $g_{(0)}$ and state. 

The fact that the ambient formalism encodes the dependence of the CFT on the metric $g_{(0)}$ is natural, since by construction the formalism provides a systematic construction of local functions of $g_{(0)}$ that have well-defined Weyl transformation properties. The dependence on the state requires more explanation. In \cite{Fefferman85}, Fefferman and Graham presented two (equivalent) constructions: the ambient space in $d+2$ dimension, which involves a Ricci-flat metric, and a construction in $d+1$ dimensions that involves a hyperbolic metric. 
The latter construction has been instrumental in the setting up of the holographic dictionary in generality \cite{Haro2000}, and one of the outcomes is that a specific subleading term in the bulk metric encodes the state of the dual CFT. While this precise connection to the state of the CFT holds for holographic CFTs, here we aim to only encode kinematics and our considerations may be sufficient for that (this will be confirmed in an explicit example).
 
The scalars \eqref{introXij} are the only independent scalar building blocks in the embedding space. This is a consequence of the embedding space enjoying the full conformal group as isometries. The ambient space breaks all of these isometries in general, and so one should expect many more building blocks than just \eqref{introwtXij}. More building blocks means a richer set of allowed solutions to \eqref{intro:weyl}. 
The remaining invariants that we construct are sufficient to assess the curvature corrections at finite $\Delta$ to the geodesic approximation, and we explain how they contribute \emph{according to Weyl-covariance}.
To construct the new building blocks, we note (as we review later)  
that the ambient space always comes equipped with a homothetic vector, $T$, a conformal Killing vector with constant conformal factor \cite{stephani2009exact}. The transformation properties w.r.t. this scaling symmetry encodes the Weyl weights of local Weyl covariant functions.
We determine a new class of Weyl invariants through contractions of the parallel transport of $T$ and ambient covariant derivatives with powers of the ambient Riemann tensor. The ambient geodesic distance in \eqref{introwtXij} is also most naturally expressed in terms of $T$: it is the inner product of the parallel transported $T$ with itself.
As natural geometric objects in the ambient space, we argue that this class of invariants captures the universal contributions of multi-energy-momentum  tensors in correlation functions. 
Full detail of the generalisation of the embedding space construction to the ambient space and the proliferation of new invariants is given in the proceeding sections. 

In this work we also present explicit example applications, in order to show how to find the invariant building blocks and to construct ambient correlators in practice. One of them involves a CFT in a  non-trivial state, and the other a CFT in a non-trivial background.
One of the most widely-studied examples where conformal symmetry is broken are CFTs at finite temperature \cite{Witten:1998zw,El-Showk:2011yvt,Witczak-Krempa:2013nua,Katz:2014rla,Iliesiu2018}. We start by reviewing the general features of thermal CFTs. Then we will apply the ambient proposal. We check that the answer given is compatible with the thermal OPE and with a novel holographic computation of scalar correlators on the thermal state defined by an AdS black hole. We also see how the curvature invariants correct the geodesic approximation and account for finite $\Delta$ effects.
Another example we study are CFTs on squashed spheres \cite{Zoubos:2002cw,Zoubos2004,Hartnoll:2005yc,Bobev:2016sap,Bobev:2017asb,Bueno:2018yzo,Bueno:2020odt,Chester:2021gdw}, where there are only a few existing results for correlators.

There are many commonalities between the ambient space formalism presented here and efforts to construct holographic duals of asymptotically flat spacetimes. Here, we are employing a $d+2$ dimensional, Lorentzian, Ricci-flat spacetime in order to construct Weyl-covariant building blocks of CFT correlators in $d$ dimensions. In celestial holography \cite{Strominger:2017zoo,Raclariu:2021zjz,Pasterski:2021rjz,Pasterski:2021raf,McLoughlin:2022ljp}, CFT correlators in $d$ dimensions appear as the duals of scattering processes in $d+2$ dimensional Minkowski background. The common elements of these proposals are discussed in section \ref{sec_FlatHolography} with a view to build better links between them and generalise flat holography to Ricci-flat backgrounds.
\newline\newline
The layout of the paper is as follows. For orientation we provide a review of the embedding space formalism in section \ref{sec:embedding}, turning to an introduction of the construction and key properties of the ambient space in section \ref{sec:ambient}. We then use the ambient geometry to identify Weyl covariant and invariant ingredients for building correlation functions in section \ref{Sec:Acorrelators} where we make a concrete proposal for scalar 2-point functions organised by powers of the ambient Riemann tensor. We then apply these results in case studies of thermal CFTs in section \ref{sec:thermal} and to CFTs on squashed spheres in section \ref{Sec:SqSphere}. We discuss the connection between the ambient space approach and flat holography in section \ref{sec_FlatHolography}. Finally we discuss future prospects and open questions in section \ref{sec:outlook}.
\newline\newline
\emph{A note on conventions.} Lowercase Latin indices $i,j \dots$ are used for $d$ boundary/CFT directions, lowercase Greek indices $\mu,\nu\dots$ for $d+1$ bulk/AdS directions while uppercase Latin letters $M,N, A, B \dots$ for the $d+2$ embedding space or ambient directions. In $d+2$ dimensions, $X^A$ are Minkowski coordinates (used for the embedding space) while $\X^A = (t,\rho,x^i)$ refer to a Gaussian null coordinate system \eqref{AspGeneralFormTRHO} (which can be used for either the embedding space or the ambient space).

\section{The embedding space} \label{sec:embedding}

The key idea at the root of the embedding space construction is that conformal transformations on $\R^d$ form the group $SO(1,d+1)$, and hence can be realised as Lorentz transformations in the embedding space, $\R^{1,d+1}$. The advantage of this perspective is that conformally-covariant quantities on $\R^d$ can be easily represented as Lorentz tensors  \cite{Dirac:1936fq, Boulware:1970ty,Weinberg:2010fx,Costa2011,Costa2011a, Rychkov2016}. In this section we review how to embed  $\R^d$ into $\R^{1,d+1}$ and how this can be used to efficiently constrain CFT correlation functions.

To find an embedding, we parameterise $\R^{1,d+1}$ with a set of coordinates $X^M= \left(X^0, X^i, X^{d+1}\right)$ and the Minkowski metric, $ds^2 = \eta_{MN} dX^M dX^N$.  A Lorentz invariant locus of $\R^{1,d+1}$ is given by $X^2 = \text{const}$. This gives a $d+1$ dimensional space which we need to reduce further to $d$ dimensions. This is achieved by restricting to the lightcone, $X^2 = 0$ and picking a section $X^+ = {\cal F}(X^i)$, where $X^\pm = X^0 \pm X^{d+1}$.

\begin{figure}[ht!]
	\begin{center}
		\includegraphics[width=0.5\textwidth]{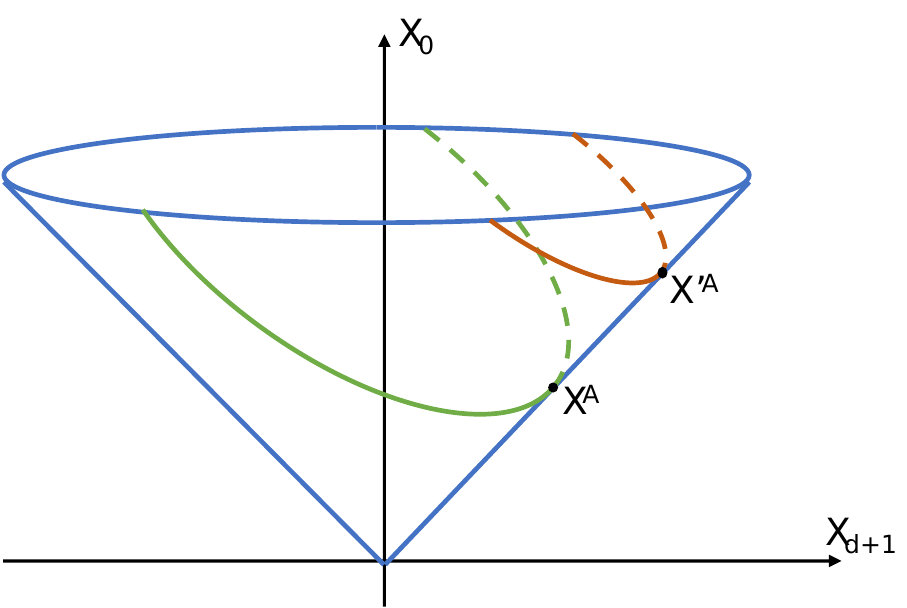}
		\caption{ The points $X^A$ and $X^{\prime A}$ lie on different lightcone sections but on the same light-ray. Hence they are represented by the same point on the projective slice.
			\label{FigProjectiveSlice}}
	\end{center}
\end{figure}

The only sectional choice which gives $\R^d$ and preserves conformal transformations $\tilde{X}^A = \Lambda^A_{\;\, B} X^B$ is given by a constant function, $X^+ = t$, with the embedding map
\begin{equation}\label{EspLCembedding}
	X^M
	= 	t \left(
		\frac{1+ x^2}{2}, x^i,\frac{1- x^2}{2}
	\right),
\end{equation}
where $x^i$ denote coordinates on $\R^d$ with the induced metric $g_{(0)ij}=t^2 \delta_{ij}$, and $x^2=\delta_{ij} x^i x^j$. Here, changing the choice of constant $t$ can be viewed as a gauge transformation. More precisely, one can define an equivalence of points in embedding space, based on whether they are connected by a light-ray,
\begin{equation}\label{}
	X^A \sim X^{\prime A} \qquad \iff \qquad X^{\prime A} = t X^A,
\end{equation}
for some non-vanishing real $t$ (see Figure \ref{FigProjectiveSlice}). This amounts to describing $\R^d$ with projective coordinates,
\begin{equation}\label{EspProjCoords}
	x^i = \frac{X^i}{X^+} \,,
\end{equation}
and one can simply work on this projective slice. This is a particularly useful perspective that will be adopted in the ambient space construction.

The more general choice of lightcone section $X^+={\cal F}(X^i) = \Omega(x^i)$ allows one to describe manifolds other than $\R^d$. In this case, the embedding map takes the form
\begin{equation}\label{EspWeylRescaledSection}
	X^M 
	= 
	\Omega(x)\left(
	\frac{1+ x^2}{2}, x^i, \frac{1- x^2}{2}
	\right),
\end{equation}
with a conformally flat induced metric, $ g_{(0)ij} = \Omega(x)^2 \delta_{ij}$. This is the most general class of $d-$dimensional spacetimes that can be embedded in the Minkowski lightcone preserving its structure. Thus, global rescalings of the embedding coordinates (generated by the dilation vector $X^M\partial_M$) end up describing the same projective slice, while local rescalings induce Weyl transformations on the CFT background.

The embedding space machinery outlined above allows one to write all kinematic constraints on conformal correlators in a simple and convenient fashion. In particular, conformal invariance is realised by Lorentz invariance in the embedding space, while Weyl covariance is realised by the freedom in the choice of the lightcone section, ${\cal F}(X^+)$. In what follows, we treat correlators on the embedding space as multi-local conformal densities depending on the insertion points on the lightcone $\{X_i\}$ and with dimensions $\{\Delta_i\}$, where $i = 1 \ldots n$ labels the insertion.

Invariance under Lorentz transformations, generated by $J_{MN} = X_M \partial_N - X_N\partial_M$, result in the following Ward Identities
\begin{equation}\label{EspWardIdConf}
\sum_{i=1}^n J_{MN}^{(i)} \braket{O_1(X_1)\dots O_n(X_n)} =0,
\end{equation}
where $J_{MN}^{(i)}$ acts on $X_i$. Thus, finding the form of correlators on the embedding space reduces to enumerating  the compatible Lorentz tensor structures. For  2- and 3-point functions of scalar primaries, the only available invariants consist in the pairwise products of the insertion points,
\begin{equation}\label{EspXij}
X_{ij} = -2	X_i\cdot X_j,
\end{equation} 
which are equal to the square distances $x_{ij}^2=|x_i-x_j|^2$ once reduced onto a $d$-dimensional section.

For Weyl transformations, correlators of a CFT on a background $g_{(0)}$ transform as
\begin{equation} \label{WeylCovAspace}
    \braket{O_1(x_1) \ldots O_n(x_n)}_{\Omega^2 g_{(0)}} =
    \Omega(x_1)^{-\Delta_1} \dots \Omega(x_n)^{-\Delta_n} 
    \braket{O_1(x_1) \ldots O_n(x_n)}_{g_{(0)}},
\end{equation}
regardless of the spin of the operators. 
In the embedding space, the correlator on the left-hand side is simply the embedding space correlator in a different lightcone section. Thus the transformation \eqref{WeylCovAspace} is realised by an adjustment to the function ${\cal F}(X^i)$, giving different embedding maps \eqref{EspWeylRescaledSection}. For instance the invariants $X_{ij}$ transform as
\begin{equation} \label{EspaceWeylTransfXij}
    X_{ij} \to \Omega(x_i) \Omega(x_j) X_{ij} \,
\end{equation}
and consequently constrain the form of the correlator.

Note that in the above discussion, we had to take into account the whole lightcone and not just the projective slice so as to make correlators well-defined on every $d-$dimensional conformally flat space. Being defined exclusively on the lightcone, correlators in the embedding space  are determined up to contributions $\sim X^2$. This gauge redundancy will play an interesting role when discussing the ambient space.

Let us consider some simple examples for illustration. For scalar 2-point functions with embedding insertions $X_1$ and $X_2$, Lorentz invariance implies that it must be a function of the invariant $X_{12}$. Furthermore, Weyl covariance fixes this function up to a multiplicative constant, and makes the 2-point function non-vanishing only for identical operators,
\begin{equation}\label{EspScalar2pt}
 \braket{O(X_1) O(X_2)} = \frac{C_\Delta}{\left(X_{12}\right)^\Delta},
\end{equation}
where $O$ is an operator of dimension $\Delta$. Following similar arguments for scalar 3-point functions, Lorentz invariance and Weyl covariance determine
\begin{equation}\label{EspScalar3pt}
	\braket{O_1 O_2 O_3} 
= \frac{C_{123}}{ 
	( X_{12})^{\alpha_{123}}
	( X_{13})^{\alpha_{132}}
	( X_{23})^{\alpha_{231}}
}, \qquad \alpha_{ijk} = \frac{\Delta_i+\Delta_j-\Delta_k}{2}\,.
\end{equation}
As is well known, scalar higher-point functions are only fixed by conformal symmetry up to functions of the cross-ratios. We can conveniently express them on the embedding space as
\begin{equation}\label{EspHigherptf}
    \braket{O_1(X_1) \ldots O_n(X_n)} =  \left(\prod_{i<j} (X_{ij})^{\alpha_{ij}}\right) f\left(u\right)\, ,
\end{equation}
where $\alpha_{ij}$ are defined through $\Delta_i = -\sum_{j=1}^n\alpha_{ij}$, and $u$ denotes the set of cross-ratios 
\begin{equation}
    u_{[pqrs]} = \frac{X_{pr}X_{qs}}{X_{pq}X_{rs}} \,.
\end{equation}
The function $f(u)$ is fixed by the dynamics of the CFT. The expression \eqref{EspHigherptf} automatically satisfies the requirement of Weyl-covariance \eqref{WeylCovAspace} as a consequence of the scaling property \eqref{EspaceWeylTransfXij}.

Without going into details, we point out that the embedding space formalism is particularly powerful for dealing with spinning correlators, since elaborate conformal tensor structures can be written as simple tensors on the embedding space, where useful differential operators can also be constructed \cite{Costa2011,Costa2011a,Karateev2017}. 

Finally we note that the embedding space is a useful tool for treating holographic duals of CFTs. This is because aside from the lightcone $X^2 = 0$ discussed above, another Lorentz-invariant locus in the embedding space is Euclidean AdS$_{d+1}$, given by the upper half-hyperboloid 
\begin{equation}\label{}
	X^2 = - R^2 \qquad \mbox{ with } X^0>0,
\end{equation}
and the \Pcr patch $ds^2 = \frac{R^2}{r^2}\left[dr^2 + \delta_{ij} dx^i dx^j\right]$ via the map
\begin{equation}\label{EspAdSSlicing}
	X^M = \left(X^0, X^i, X^{d+1}\right) = 
	R \left(
	\frac{1+ x^2+r^2}{2r}, \frac{x^i}{r},\frac{1- x^2-r^2}{2r}
	\right).
\end{equation}
A key observation is that the embedding space allows one to represent bulk and boundary point covariantly in the same language. Denoting by   $X$ and $P$ the bulk and boundary point respectively, the scalar bulk-to-boundary propagator reads
\begin{equation}\label{EspScalarB2b}
	K_\Delta(X,P) = \frac{C'_\Delta}{\left(-2 P \cdot X\right)^\Delta}.
\end{equation}
Note that modulo the normalization, its form matches that of  scalar 2-point functions \eqref{EspScalar2pt}.

\section{The ambient space} \label{sec:ambient}

Our aim is to extend the embedding formalism reviewed in section \ref{sec:embedding} to more general settings where conformal invariance may be broken, including non-conformally flat CFT backgrounds and generic states, so that we may usefully constrain the form of correlators in such settings. To achieve this aim we adopt the ambient space \cite{Fefferman85,Fefferman:2007rka} as our principal tool, a $(d+2)-$dimensional spacetime that replaces the role of the embedding space.

\subsection{Construction} \label{sec:ambient_construction}

There are two key defining features of the ambient space. The first is the existence of a null scaling isometry $T$ (called a homothetic vector in mathematics literature), obeying $\mathcal{L}_T \tilde{g} = 2 \tilde{g}$ where $\tilde{g}$ is the metric of the $(d+2)-$dimensional ambient space. Note that $T$ is a conformal Killing vector of the ambient space and is non-Killing. The existence of $T$ reflects the fact that CFT correlators on any background and state satisfy Weyl-covariance constraints, playing a role analogous to the embedding space's $X\cdot \p_X$ dilation vector. 

The second defining feature is Ricci-flatness. To depart from the embedding space we must depart from $\R^{1,d+1}$. Riemann-flatness is too restrictive, as this results in a formalism locally equivalent to the embedding space, leaving Ricci-flatness is the next most natural class of spacetimes. One may wish to consider further relaxing this by introducing matter with a energy-momentum-tensor, but we will not do so for the present discussion; we will comment on the role of such extensions in the concluding section \ref{sec:outlook}.

Given a $d$-dimensional conformal manifold with coordinates $x^i$ and a representative $g_{(0)ij}(x)$ of the conformal class of metrics $[g_{(0)ij}(x)]$, one is able to construct a new $d+2$-dimensional spacetime with the above two requirements, the ambient space \cite{Fefferman85,Fefferman:2007rka}. Parameterising the $d+2$ ambient space directions with the coordinates $\X^M = (t,x^i,\rho)$, the most general ambient space metric is given by
\be \label{AspGeneralFormTRHO}
\wt{g} = 2 \rho dt^2 + 2 t dt d\rho +t^2 g_{ij}(x,\rho) dx^i dx^j,
\ee
where $g_{ij}(x,0)  = g_{(0)ij}(x)$ and where $g_{ij}(x,\rho)$ is such that $\wt{R}_{MN}=0$. In these coordinates the homothetic vector is given by
\be
T = t \p_t,
\ee
and we note the useful property $\wt{\nabla}_A T_B = \tilde{g}_{AB}$. 

\begin{figure}
\begin{center}
   \includegraphics[width=0.5\textwidth]{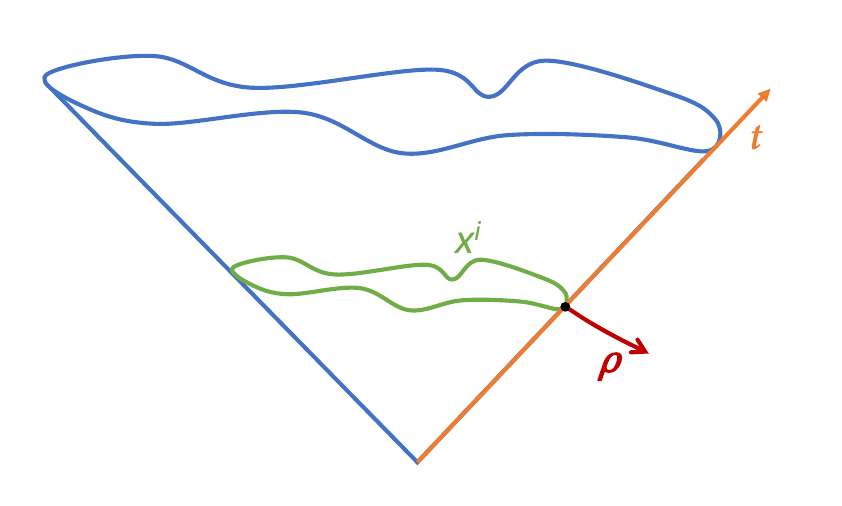}
\end{center}
\caption{ This cartoon  illustrates qualitatively the meaning of the ambient coordinates: $t$ and $x^i$ span the nullcone, while $\rho$ describes the distance from it.
			\label{FigAspaceLC}}
\end{figure}

We shall refer to the coordinates $\X^M$ as \emph{ambient coordinates}. The meaning $t$ and $\rho$ is depicted in Figure \ref{FigAspaceLC}. The coordinate $t$ is related to  ambient  scale transformations, generated by the homothety $T$. Intuitively, the coordinate $\rho$  describes the distance from the nullcone.\footnote{We call it nullcone to keep in mind the relation to the corresponding surface of the embedding space. As we will see shortly, in the ambient space it is a null submanifold with a degenerate induced metric, which can be represented as a cone space of the arbitrary $d$-dimensional manifold with a conformal class represented by $g_{(0)}$.} While $t$ is taken to be strictly positive, $\rho$ is real and we place the nullcone at $\rho=0$. Hence, projecting onto the $d$-dimensional spacetime of interest amounts to setting $t=1$ and $\rho=0$ where one recovers $g_{(0)ij}(x)$. As in the embedding space, the nullcone is obtained by rescaling $g_{(0)ij}$ and as such it is covered by the coordinates $t$ and $x^i$, in analogy with \eqref{EspLCembedding}. Choosing a specific $t$ corresponds to restricting to a specific section of the nullcone. 

Knowing the $d$-dimensional metric together with the presence of the dilations $T$ completely fixes the geometry on the nullcone.
The specific form of \eqref{AspGeneralFormTRHO} follows from a convenient gauge choice such that $\rho$ and $t$ are geodesic coordinates in a vicinity of the nullcone. By this we mean that starting at a fixed point $(t,x,0)$ on the nullcone, the curve $\gamma(\rho) = (t, x,\rho)$ is a geodesic for the ambient metric $\wt{g}$. Similarly, any curve $\gamma(t)=(t,x,\rho)$ starting at $(t_0,x,\rho)$ is a geodesic for the ambient metric $\wt{g}$. This fixes the ambient geometry to take the form of a Gaussian null foliation \cite{Parattu:2015gga,Chakraborty:2015aja} near the ambient lightcone, resulting in \eqref{AspGeneralFormTRHO}. Observe that the $t$-dependence is completely fixed by the choice of gauge and homogeneity. In particular, at $\rho=0$  a dilation generated by $T=t\p_t$ will produce a rescaling of the boundary metric as desired.

Solving the equations $\wt{R}_{MN}=0$ determines the components $g_{ij}(x,\rho)$, with boundary conditions given by the $d$-dimensional metric $g_{ij}(x,0)  = g_{(0)ij}(x)$. These are,
\begin{subequations} \label{EqRicciFlatness}
 \begin{align}
     \label{Eq317}
\widetilde{R}_{ij} &= \rho g^{\prime\prime}_{ij} 
- \rho g^{kl} g^{\prime}_{ik} g^{\prime}_{lj} 
+ \frac{1}{2} \rho g^{kl} g^{\prime}_{kl} g^{\prime}_{ij}  
- \left(\frac{d}{2} - 1\right) g^{\prime}_{ij} 
-\frac{1}{2} g^{kl} g^{\prime}_{kl} g_{ij}    + R_{ij} = 0\,,\\
 \label{Eq317b}
\widetilde{R}_{i\rho} &= \frac{1}{2} g^{kl} \left(
\nabla_k g'_{il} - \nabla_i g'_{kl}
\right)= 0\,,\\
 \label{Eq317c}
\widetilde{R}_{\rho\rho} &= -\frac{1}{2} g^{kl} g''_{kl} +
\frac{1}{4} g^{lk} g^{pq} g'_{kp} g'_{ql}= 0\,,
 \end{align}\end{subequations}
where the primes denote derivatives in $\rho$, while $R_{ij}$ and $\nabla_i$ indicate the Ricci tensor and the covariant derivative of $g_{ij}(x,\rho)$ evaluated at fixed $\rho$.

General properties of $g_{ij}(x,\rho)$ maybe studied through solutions of \eqref{Eq317}, \eqref{Eq317b}, \eqref{Eq317c} obtained in a perturbative expansion at small $\rho$, {\it i.e.} a near-nullcone expansion. In terms of the boundary metric $g_{(0)ij}(x)$, one has
\be \label{AspExpansionEvend}
g_{ij}(x,\rho) = g_{(0)ij}(x) + 2 P_{ij} \, \rho + \dots + \rho^{\frac{d}{2}}\left( g_{(d)ij} +  h_{(d)ij} \log\rho \right) + \dots
\ee
for even dimensions $d$, while for odd $d$ one has 
\be \label{AspExpansionOddd}
g_{ij}(x,\rho) = g_{(0)ij}(x) + 2 P_{ij} \, \rho + \dots + \rho^{\frac{d}{2}} g_{(d)ij}   + \dots,
\ee
where the coefficient of the expansion only depend on $x$, and $P_{ij}$ is the boundary Schouten tensor
\be\label{Schouten_def}
P_{ij}=\frac{1}{d-2}\left( R_{ij} - \frac{R}{2(d-1)}g_{(0)ij}\right).
\ee
Remarkably,  all the coefficients including $h_{ij}$ are completely fixed by the boundary metric up to order $O\left(\rho^{d/2}\right)$, while only the trace and divergence of $g_{(d)ij}$ are determined by $g_{(0)ij}$. The remaining transverse traceless piece of $g_{(d)ij}$ constitutes the second piece of boundary data required for general solutions of the set of second order differential equations \eqref{EqRicciFlatness}.
Note that $g_{ij}(x,\rho)$ is in general non-analytic at $\rho = 0$ since at order $O\left(\rho^{d/2}\right)$ logarithmic contributions are present for even $d$ and for non-conformally Einstein $g_{(0)}$, while half-odd powers of $\rho$ appears for odd $d$ starting at $O\left(\rho^{d/2}\right)$.

\begin{figure}
\begin{center}
\includegraphics[width=0.5\textwidth]{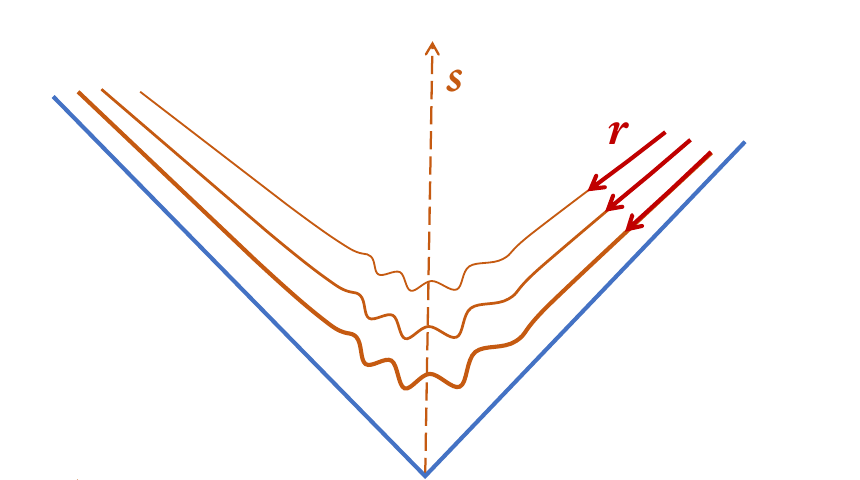} 
\end{center}
\caption{ This cartoon illustrates the meaning of the ambient coordinates in hyperbolic slicing: $r$ is the Fefferman-Graham radial coordinate along a hyperboloid, while $s$ defines the radius of different hyperboloids.
			\label{FigAspaceLCsr}}
\end{figure}

The similarity to the usual holographic expansion for asymptotically locally AdS spaces is striking, and we can make this relation more precise by performing the following coordinate transformation,
\be \label{AspCoordsAdSSlicing}
\rho = - \frac{r^2}{2}\,, \quad t = \frac{s}{r}\,,
\ee
with $r, s >0$, covering the region $\rho<0$. The ambient metric becomes
\be \label{AspMetricAdSSlicing}
\wt{g} = - ds^2 +s^2 \left( \frac{ dr^2 + g_{ij}(x,r) dx^i dx^j }{r^2} \right),
\ee
where the piece in parentheses must solve the vacuum Einstein equations with a negative cosmological constant in $d+1$ dimensions, as a consequence of Ricci-flatness in $d+2$. 
Thus, hypersurfaces at fixed $s$ are asymptotically locally AdS metrics of radius $s$ in $d+1$ dimensions and we recognise $g_{ij}(x,r)$ as the usual near-boundary Fefferman-Graham expansion. This is a generalization of the AdS slicing  of the embedding space \eqref{EspAdSSlicing}, as sketched in Figure \ref{FigAspaceLCsr}. Note also that the $d-$dimensional manifold is recovered in the limit $r\to0$ and $s\to 0$, keeping fixed $t = \frac{s}{r} = 1$. The homothetic vector  reads $T=s\p_s$ in these coordinates.

This hyperbolic slicing illustrates several interesting properties of the ambient space. First, it tells us that the coefficients in \eqref{AspExpansionEvend} and \eqref{AspExpansionOddd} contain precisely the same information as the corresponding ones in the usual holographic expansion\footnote{Observe that the ambient  coordinate $\rho$ is proportional to the holographic $\rho_{Holo}$ as defined \eg in \cite{Haro2000} according to $\rho = -\frac{1}{2} \rho_{Holo}$.} as in \cite{Haro2000, Skenderis:2002wp}. In particular, $g_{(d)ij}$ is related to the vacuum expectation value (VEV) of the boundary energy-momentum tensor, while the boundary metric $g_{(0)ij}(x)$ plays the role of its source. Finally, $h_{(d)ij}$ is proportional to the metric variation of the boundary Weyl anomaly \cite{Haro2000, Henningson:1998gx}. Therefore the ambient space geometrically encodes both the generic CFT background as well as its possibly non-trivial state, and includes the information about conformal anomalies. Importantly, it does so in a Weyl-covariant way as we will remark later. 

Another important consequence of this slicing is that  exact ambient solutions can be found starting from asymptotically locally Anti-de Sitter (ALAdS) geometries in the Fefferman-Graham gauge by considering the AdS radius as a new coordinate $s$ and fibering it according to \eqref{AspMetricAdSSlicing}. This automatically solves the Ricci-flatness equations.

Note that the change of coordinates \eqref{AspCoordsAdSSlicing} only covers $\rho < 0$. Alternatively we can consider the change of coordinates 
\be \label{AspCoordsdSSlicing}
\rho = + \frac{r^2}{2}, \quad t = \frac{s}{r},
\ee
leading to the metric
\be \label{AspMetricdSSlicing}
\wt{g} =  ds^2 +s^2 \left( \frac{ -dr^2 + g_{ij}(x,r) dx^i dx^j }{r^2} \right).
\ee
As expected from the analogy with Minkowski, here we are covering the $\rho>0$ region of the ambient space with a foliation in terms of $(d+1)-$dimensional asymptotically locally dS (ALdS) spaces. One can move from the positive $\rho$ region to the negative $\rho$ region by taking the analytic continuation $s \to i s$ and $r \to i r$ of the radius and Fefferman-Graham coordinate, recovering the well-known map from Euclidean AdS to dS spaces. Similarly to the negative $\rho$ case, we can find exact ambient geometries in this patch by plugging ALdS metrics within the parentheses of \eqref{AspMetricdSSlicing}.

\subsection{Relation to the embedding space} \label{SSec:RelToESpace}

To illustrate the relationship between the embedding space presented in section \ref{sec:embedding} and the more general ambient metric \eqref{AspGeneralFormTRHO}, we start by rewriting the $(d+2)-$dimensional Minkowski metric $ds^2 = \eta_{MN} dX^M dX^N$ in terms of the Gaussian null coordinates $\X^M = (t, x^i,\rho)$. In view of a comparison with the \Pcr slicing of the embedding space \eqref{EspAdSSlicing} we consider a flat boundary $g_{(0)ij} = \delta_{ij}$. Defining $X^\pm = X^0 \pm X^{d+1}$, a suitable change of coordinates\footnote{A similar set of coordinates for Minkowski was used in \cite{Sonner:2018rmt}.} is
\be \label{EspMApToAsp}
t= X^+, \qquad \rho = \frac{\eta_{MN}X^M X^N}{2 (X^{+})^2}, \qquad x^i = \frac{X^i}{X^+},
\ee
with inverse map
\be \label{AspMApToEsp}
X^0 = \frac{t}{2} \left( 1-2\rho + x^2 \right), \qquad X^i = t x^i, \qquad X^{d+1} = \frac{t}{2} \left( 1+2\rho - x^2 \right).
\ee
The resulting ambient metric is 
\be \label{AspEspaceMetric}
\tilde{g} = 2 \rho dt^2 + 2 t dt d\rho + t^2 \delta_{ij} dx^i dx^j,
\ee
which is simply Minkowski space in Gaussian null coordinates. From the map \eqref{AspMApToEsp} it is clear that fixing a value of $t$ determines a single slice of the nullcone, where the boundary directions $x^i$ play the role of the projective coordinates \eqref{EspProjCoords}, being independent of the section one picks. The coordinate $\rho$ tells us how far from the nullcone we are, with $\rho > 0$ the region with a timelike separation from the origin, and $\rho < 0$ spacelike. As a consistency check, note that the map \eqref{AspMApToEsp} reduces to the embedding \eqref{EspLCembedding} taking $ \rho=0$ and matches the AdS slicing \eqref{EspAdSSlicing} using the change of coordinates \eqref{AspCoordsAdSSlicing}.

Comparing the ambient metric \eqref{AspEspaceMetric} to the general expansion at small $\rho$, we immediately recognise in which sense the embedding formalism can be  generalised via the ambient space. The latter can describe non-trivial states at the boundary in case of a non-vanishing energy-momentum tensor VEV turned on. In addition to this, we remark that a similar map from the embedding space to the ambient formulation as in \eqref{EspMApToAsp} can be found for any conformally flat boundary metric, not only for $g_{(0)ij}(x) = \delta_{ij}$. Assuming the boundary data $g_{(d)ij}$ vanishes, one can check that the ambient space is locally Minkowski ({\it i.e.} its Riemann tensor vanishes) if and only if the boundary metric $g_{(0)ij}(x)$ is conformally flat. In the special case of $d=2$, where all boundary metrics are conformally flat, one can show \cite{Fefferman:2007rka} that the 4-dimensional ambient space is automatically Riemann-flat, even for non-vanishing energy-momentum tensor VEVs.

As an illustration of this fact, let us consider a CFT on a Euclidean AdS$_d$ background, with metric
\be
g_{(0)ij}dx^i dx^j = \frac{dz^2 + dx_a dx^a}{z^2},
\ee
with $a = 2\dots d$. When the boundary metric is an Einstein metric\footnote{By Einstein metric we mean that the Ricci tensor is proportional to the metric. } the ambient expansion truncates at order $O(\rho^2)$. In this case, the ambient metric reads
\be \label{AspForAdS Bdry}
\wt{g} = 2 \rho dt^2 + 2 t dt d\rho + t^2 \left(1- \frac{\rho}{2}\right)^2 \, \frac{dz^2 + dx_a dx^a}{z^2}\, ,
\ee
 and  the change of coordinates mapping this space to Minkowski is 
\be
X^0 = \frac{t}{2 \sqrt{2} z} \left( 2 + z^2 + x^2\right)\left(1-\frac{\rho}{2} \right), \qquad X^1 = t \left(1+\frac{\rho}{2} \right),  
\ee
\begin{equation*}
    X^a = \frac{t}{z} x^a \left(1-\frac{\rho}{2} \right), \qquad X^{d+1} = \frac{t}{2 \sqrt{2} z} \left( 2 - z^2 - x^2\right)\left(1-\frac{\rho}{2} \right).
\end{equation*}

Let us now turn our attention to embedding space correlation functions. Focusing on scalar 2-point functions on $\R^d$, their extension to the ambient space must also be a scalar and this entails that one has to find building blocks which are scalars under ambient diffeomorphisms.\footnote{We will explain in detail this statement in section \ref{SubsecWeylTrAndIsometries}, where in particular it will be shown which class of diffeomorphisms is relevant in this context and why we are requiring full diffeomorphism invariance.} 

In embedding space one can simply use insertion coordinates $X_i^M$ to construct scalars as in \eqref{EspXij}, since Minkowski space is in fact locally isomorphic to its tangent space. In ambient space however, $(t,x^i,\rho)$ are merely coordinates on a curved manifold, and cannot be directly contracted at different insertion points to construct scalars, as these belong to different tangent spaces. Fortunately, by definition, the ambient space comes equipped with the homothetic vector $T=t \p_t$. Since $T$ coincides with $X= X^M \p_M$ in the flat case, it is natural to replace the positions of the insertions in the ambient space is the vector field $T$ evaluated at the insertion points.

Consider two ambient insertion points $\X_{i}, \X_j$. In order to construct a scalar quantity under ambient diffeomorphisms,  we parallel transport $T(\X_i)$ to $\X_j$, and contract it with $T(\X_j)$. Since the two vectors belong to the same tangent space their contraction with the ambient metric at that point results in a well-defined scalar. 

This prescription allowing one to generalise $X_{ij}$ is valid for any ambient space, and we will discuss it at length in section \ref{Sec:Acorrelators}. For now let us check it reproduces the known embedding space invariant in \eqref{EspXij} in the case of a flat boundary with ambient metric \eqref{AspEspaceMetric}.  Given two points not necessarily on the lightcone
\be \label{FlatAspace_EndPoints}
\X_0 = (t_0, x_0, \rho_0), \qquad \quad \X_1 = (t_1, x_1, \rho_1),
\ee
as a first step parallel transport requires solving the geodesic equations from $\X_0$ to $\X_1$ 
\begin{equation}\label{}
	\ddot{\X}{}^M(\lambda) + \Gamma^M_{AB}(\lambda) \dot{\X}{}^A(\lambda) \dot{\X}{}^B(\lambda) =0,
\end{equation}
where $\Gamma^M_{AB}$ here refers to the ambient connection. 
On the flat ambient space \eqref{AspEspaceMetric} they can be easily solved, 
\begin{subequations}
	\begin{align}
		t(\lambda) & = A \lambda + B, \\
		x^m(\lambda) &= \frac{E^m}{A \lambda + B} + F^m,\\
		\rho(\lambda) &= \frac{E^m E_m}{2(A \lambda + B)^2} + \frac{G}{A \lambda + B} + H,
	\end{align}
\end{subequations}
where $A, B, E^m, G, H$ are a total of $2(d+2)$ integration constants accounting for the components of the initial and final point \eqref{FlatAspace_EndPoints} of the geodesic. We set the endpoints to correspond to the values of the affine parameter $\lambda=0$ and $\lambda=1$ respectively, fixing the integration constants to
\begin{subequations}
		\begin{equation}\label{}
			A = - t_0 +t_1, \qquad B=t_0, \qquad E^m = t_0 t_1\frac{x^m_1-x^m_0}{t_0 - t_1},	
		\end{equation}
	\begin{equation}\label{}
		F^m = \frac{t_1 x^m_1 - t_0 x^m_0}{t_1-t_0} \qquad \quad 
		H = - \frac{E^m E_m + 2 G t_0}{2 t_0^2},
	\end{equation}
\begin{equation}\label{}
	G= - \frac{t_0 t_1}{(t_0-t_1)^2}\left[
	(t_0 +t_1)
	(x_1-x_0)^2 +2 (t_0-t_1)(\rho_0-\rho_1)
	\right].
\end{equation}
\end{subequations}
These geodesics are of course simply straight lines on Minkowski in disguise. We now have to evolve the initial condition $T(\lambda=0) = T_0 = (t_0,0,0)$ at $\X_0$ to the point $\X_1$ along these geodesics using the parallel transport equations 
\begin{equation}\label{}
	\dot{\X}{}^M(\lambda) \, \wt{\nabla}_M T^{A}(\lambda) =0.
\end{equation}
In this case also these equations can be solved exactly and after imposing the boundary conditions at the endpoints, at $\lambda=1$ one finds
\begin{equation}\label{eq:TransportedTOnMink}
	\hat{T}_0 \equiv T_0(1) 
	= 
	\left(
	t_0,
	-\frac{t_0}{2t_1} \left[
	(x_0^i-x_1^i)^2 - 2 (\rho_0 - \rho_1)
	\right], \frac{t_0}{t_1}(x_0^i-x_1^i)
	\right).
\end{equation} 
We define the ambient analogue $\wt{X}_{ij}$ of the embedding space invariant $X_{ij}$ as the contraction of $T_1 = (t_1,0,0)$ with $\hat{T}_0$ using the ambient metric evaluated at $\X_1$. This leads to
\begin{equation}\label{AspFlatCaseT01Invariant}
	\wt{X}_{01} \equiv -2 \, \hat{T}_0 \cdot T_1 = t_0 t_1 \left(
	(x_0^i-x_1^i)^2 - 2 (\rho_0 + \rho_1)
	\right).
\end{equation}
Placing the insertions on the lightcone section $\rho_{0,1}=0, t_{0,1}=1$, one recovers the expected value of the embedding space invariant $X_{01}$. 
For more general ambient spaces, we treat the calculation of this invariant in more details in the upcoming sections and in appendix \ref{AppendixGeodPTranspT}. 

As already discussed when constructing the ambient space, and as we will see in more detail in section \ref{SubsecWeylTrAndIsometries}, the conformal dimension under Weyl transformations of an ambient object coincides with minus its weight in $t$. This fixes scalar 2-point functions of dimension $\Delta$  and  bulk-to-boundary propagators to the known forms,
\be \label{eq:EspCorrInAsp}
\braket{O_1(\X_1) O_2(\X_2)} = \left. \frac{C_\Delta}{(\wt{X}_{12})^\Delta} \right|_{\substack{t_{1,2} \to 1 \\ \rho_{1,2} \to 0}} = \frac{C_\Delta}{(x_{12})^{2\Delta}}
\ee
\be
K_{\Delta}(r_0,x_1;x_2) = \left. \frac{C'_\Delta}{(\wt{X}_{12})^\Delta} \right|_{\substack{t_{1} \to 1 \\ \rho_{1} \to 0}} =  \frac{C'_\Delta}{R^\Delta} \left(  \frac{r_0}{(x_{12})^{2} +r_0^2} \right)^\Delta,
\ee
where we set $t_1= \frac{R}{r_1}$ and $\rho_1 = - \frac{r_1^2}{2}$, and $R$ is the AdS radius.

\subsection{Weyl invariance and the Weyl connection} \label{SubsecWeylTrAndIsometries}

The ambient construction reduces to the embedding space for CFTs on conformally flat $d$-dimensional backgrounds in the vacuum state. Correlators must be invariant under conformal transformations, which are conveniently realised as isometries in the embedding space. As detailed in appendix \ref{AppendixBIsometries}, the same happens in the ambient space. In particular, conformal Killing vectors on $g_{(0)}$ are lifted to near-lightcone isometries on the ambient space.\footnote{Such a feature is already present in standard holography as one relates asymptotic symmetries in the bulk to conformal transformations on the boundary \cite{Papadimitriou2005}. This property of the ambient space can be thought of as inherited from the ALAdS realization \eqref{AspMetricAdSSlicing}, where asymptotic symmetries on the ALAdS slices are to be understood as near-nullcone isometries on the ambient space.} The corresponding Ward identity in the CFT constrains ambient correlators in the same way as embedding correlators. For each such near-lightcone isometry $K$ in $d+2$ dimensions, ambient correlators $F$ of quasi-primary operators must satisfy  
\be
\sum_{i=1}^n \mathcal{L}^{(i)}_K F(\X_1, \X_2 \dots \X_n) =0 \, ,\label{ambient_ward}
\ee
where $\mathcal{L}^{(i)}$ is the Lie derivative operator acting on the $i$-th insertion point and where $F$ is a tensor on the ambient space, in general with  different tensorial transformation properties for each insertion.

Since we are interested in CFT backgrounds and states that may break all near-lightcone isometries $K$ that so usefully constrain correlators through \eqref{ambient_ward}, how then is the ambient space formalism useful? The answer is \emph{Weyl covariance}, which represents the universal kinematical constraint on correlators. For a CFT on a generic background $g_{(0)}$ it reads as in \eqref{WeylCovAspace}.\footnote{We are considering all insertions at separated points, so local contributions from Weyl anomalies do not contribute.}

Assume we have an ambient space $\tilde{g}$ of the form \eqref{AspGeneralFormTRHO} constructed from the CFT background metric $g_{(0)ij}$; we wish to construct another one compatible with the metric $\hat{g}_{(0)ij} = e^{2 \Omega(x)} g_{(0)ij}$. It turns out these two ambient spaces are locally diffeomorphic, so that in a new set of coordinates $(\hat{t}, \hat{x}^i, \hat{\rho})$ the ambient metric $\tilde{g}$ reads
\be \label{eq:WeylAspace}
\tilde{g}= 2 \hat{\rho} d\hat{t}^2 + 2 \hat{t} d\hat{t} d\hat{\rho} + \hat{t}^2 \hat{g}_{ij}(\hat{x},\hat{\rho}) d\hat{x}^i d\hat{x}^j \, ,
\ee
which induces the metric $\hat{g}_{(0)ij}(\hat{x})$ when taking $\hat{\rho}=0, \, \hat{t}=1$. Formally \eqref{eq:WeylAspace} is an ambient space constructed from the metric $\hat{g}_{(0)ij}$. One can interpret this fact as the statement that an ambient space is canonically related not only to $g_{(0)ij}$ but to the whole conformal class of metrics $[g_{(0)}]$, all equivalent to $g_{(0)}$ modulo a Weyl transformation.\footnote{This parallels the case of $(d+1)$-dimensional ALAdS spaces, where Weyl transformations are induced onto the boundary by a special class of bulk diffeomorphisms (see \eg \cite{Imbimbo2000, Skenderis2000}).}

The coordinate transformation from $(t, x^i, \rho)$ to $(\hat{t}, \hat{x}^i, \hat{\rho})$ can be easily found by working perturbatively in $\rho$.\footnote{We proceed in this way to keep the discussion as general as possible. Given an exact ambient solution to all orders in $\rho$, this diffeomorphism may be found in closed form.} Algorithmically, one imposes order by order that  the background metric induced at  $\hat{\rho}=0, \, \hat{t}=1$ is the Weyl-rescaled $\hat{g}_{(0)ij}$, as well as that the ambient gauge is preserved (i.e. $\hat{t}$ and $\hat{\rho}$ are Gaussian null coordinates). For what follows we are interested only in the first few orders,
\begin{subequations} \label{AmbientDiffeomAfterBdryConformal}
    \begin{align} 
\hat{t} &= e^{-\Omega(x)} t \left[ 1 - \frac{1}{2} \Omega^i \Omega_i \rho   + O(\rho^2)\right], \\
\hat{x}^i &= x^i +  \Omega^i \rho + O(\rho^2) \, , \\
\hat{\rho} &=e^{2\Omega(x)} \rho + O(\rho^2)\, ,
\end{align}
\end{subequations}
with $\Omega_i = \partial_i \Omega$ and where indices are raised and lowered using $g_{(0)ij}$.  

As anticipated on the nullcone $\rho=0$ this diffeomorphism reduces to a local rescaling of the coordinate $t$. This is the analogue of the local rescaling of the projective section in the embedding space in \eqref{EspWeylRescaledSection} which leads to a Weyl-rescaled background. This agrees with the intuition that $t$ measures the engineering dimensions of ambient quantities. In particular, the scalar invariant $\wt{X}_{ij}$ defined in equation \eqref{AspFlatCaseT01Invariant} for conformally flat backgrounds is manifestly homogeneous in $t$ in both insertions $\wt{X}_{ij} \propto t_i t_j$, hence transforms homogeneously under Weyl transformations with dimension $-2$. Analogously, $X_{12}$ is a dimension $-2$ invariant in the embedding space.

The fact that Weyl transformations are induced by ambient diffeomorphisms represents the key property of the ambient space and it has been the main motivation for its use in conformal geometry, allowing one to find and classify Weyl-invariant objects on arbitrary $d$-dimensional manifolds 
\cite{Fefferman85,Fefferman:2007rka, Cap2002,Graham1992a,graham,Bailey1994,Fefferman2003AmbientMC,branson2005conformally,Gover2004ConformalDR,GOVER2006450,Graham2004TheAO,Graham2008ExtendedOT}. 
Our goal is to use the ambient space to study correlators, meant as multi-local tensorial objects living on the ambient nullcone. To impose their Weyl-covariance, one has to study the precise action of the diffeomorphisms \eqref{AmbientDiffeomAfterBdryConformal} on ambient tensors when restricted to the nullcone \cite{Fefferman:2007rka, Bailey1994, Cap2002}.

Let us focus on vector fields on the ambient space for simplicity. It is straightforward to generalise the discussion to any other ambient tensor. When we restrict an ambient vector field to the nullcone, its components will only depend on $t$ and $x^i$. There turns out to be a privileged class of ambient vectors whose components can be written in the form
\be \label{eq:ATractorVector}
V^M = \left( v^0(x), \frac{v^i(x)}{t} , \frac{v^\rho(x)}{t} \right)
\ee
once restricted to the CFT background.\footnote{By doing this we are effectively constructing a $(d+2)$-dimensional vector bundle on the $d$-dimensional background.} Here $t$ should be thought of as a formal parameter, which keeps track of the weight under Weyl transformations of each component. These expressions should be thought of as evaluated at $t=1$.
In conformal geometry, the vector \eqref{eq:ATractorVector} is known as a (weight 0) \emph{tractor}.  

More specifically, under the ambient diffeomorphisms \eqref{AmbientDiffeomAfterBdryConformal} the components of any such vector restricted to $\rho=0$ transform according to 
\begin{equation}\label{}
\wh{V}^M = \left. \frac{\partial \widehat{\X}{}^M}{\partial \X^N} V^N \right|_{\substack{\rho=0}} =
\begin{pmatrix}
e^{-\Omega} \left[
v^0 - \Omega_i v^i - \frac{1}{2} \Omega_i \Omega^i v^\rho \right]\\
\frac{1}{t} \left[v^i + \Omega^i v^\rho\right]\\
 \frac{e^{2\Omega}}{t}   v^\rho 
\end{pmatrix}.
\end{equation}
The resulting components $\wh{V}^M$ have  the same weight in $t$ as the initial vector \eqref{eq:ATractorVector} and thus this transformation preserves the class of tractor fields. We can rewrite this action in terms of a linear transformation on the $(d+2)$-dimensional space of such vector fields,
\be \label{WeylUmatrix}
U(\Omega)^M_{\,\;N}  = \begin{pmatrix}
 e^{-\Omega}  &
 - \Omega_n e^{-\Omega}  &  -\frac{1}{2} \Omega_i\Omega^i e^{-\Omega}\\
 0 & \delta^m_n  & \Omega^m    \\
 0 & 0 & e^{2\Omega}
\end{pmatrix},
\ee 
parametrised by a function $\Omega(x)$ on the CFT background, and where we set $t=1$. 

One can analogously define weight $w$ tractors as the restriction to a $d$-dimensional nullcone section of ambient vectors with an additional overall homogeneity of $t^w$ with respect to \eqref{eq:ATractorVector}. Weighted tractors can be simply thought of as the restriction of ambient vectors $V=V^M \p_M$ with homogeneity $w-1$ in $t$ to the section $t=1$ of the nullcone. They transform as
\be \label{WeightedTractorsWeylTransf}
\wh{V}^M = e^{w \, \Omega} \, U(\Omega)^M_{\,\;N} V^N
\ee
under a Weyl transformation, where $U(\Omega)^M_{\,\;N}$ is the same matrix as in \eqref{WeylUmatrix}.

If we further inspect the components of the ambient connection, the action of the ambient covariant derivative along the $x^i$ directions (once restricted to the $d$-dimensional section) $\mathcal{D}_k \equiv \left. \wt{\nabla} \right|_{\substack{\rho=0 \\ t=1}}$ can be split as
\begin{equation}\label{}
\mathcal{D}_k = 
\nabla_k +  \mathcal{A}_k, \qquad \quad \mbox{ with } \quad \left(\mathcal{A}_k\right)^M_{\,\,\;N} = 
\begin{pmatrix}
0 & -P_{kn} & 0 \\
P^{\;\,m}_{k} & 0 & \delta^m_{k} \\
0 & -g_{kn} & 0 
\end{pmatrix},
\end{equation}
where the first piece is simply the covariant derivative compatible with the background metric $g_{(0)}$, under which $v^0$ and $v^\rho$ are scalars. Thus the ambient connection acts on a tractor (of any weight) as
\begin{equation}\label{}
\mathcal{D}_k V^M =
\partial_k V^M + \delta^M_m \Gamma^m_{kn}v^n
+ \left(\mathcal{A}_k\right)^M_{\,\,\;N} V^N.
\end{equation}
The additional piece $\mathcal{A}_k$ that the ambient connection induces onto the CFT background is what makes $\mathcal{D}_k$ covariant under Weyl transformations when acting on tractors. This type of connection was originally constructed in \cite{Thomas26} and independently studied in physics in the context of conformal supergravity \cite{Kaku77,Kaku:1978nz,FRADKIN1985233}. In particular one can check that $\mathcal{D}_k$ commutes with Weyl transformations, {\it i.e.} $\mathcal{D}_k V^M$ transforms in the same way as $V^M$,
\begin{equation} 
 (\wh{\mathcal{D}}_k \wh{V})^M = e^{w \, \Omega} \, U(\Omega)^M_{\,\;N} \, \left(\mathcal{D}_k V\right)^N      \, ,
\end{equation}
where $\wh{\mathcal{D}}_k$ indicates the covariant derivative compatible with the ambient metric \eqref{eq:WeylAspace}.
This shows that the ambient connection  canonically induces a Weyl connection on the boundary.  Finding Weyl covariant objects in $d$ dimensions (such as CFT correlators) boils down to the study of multiplets under Weyl transformations given by the matrix $U(\Omega)$. 
It is in this sense that Weyl transformations are linearly realised  on the ambient nullcone, similarly to what happens for conformal symmetries in the embedding space. This is the perspective adopted in the so-called \emph{tractor calculus} \cite{Bailey1994, GOVER2001206,Curry:2014yoa}. In section \ref{Ssec:SpinningCorr} we discuss the implications for the computation of spinning correlators in general backgrounds and states.

\section{Ambient correlators} \label{Sec:Acorrelators}

Given a CFT in a state defined by the VEVs $\{\braket{O_i}\}$ and on the metric background $g_{(0)}$, we must find a prescription to associate a specific ambient space to it.
As discussed in section \ref{sec:ambient_construction} the data $g_{(0)}$ are not enough to specify the ambient metric, since one must also provide additional near-nullcone data $g_{(d)ij}$. Once this data is specified, the construction proceeds by fulfilling the Ricci-flatness condition. 

It is natural to associate $g_{(d)ij}$ with the VEV of the energy-momentum tensor. 
Recall that according to AdS/CFT any hyperbolic slice of an ambient space in the form \eqref{AspMetricAdSSlicing} encodes the dynamics of a CFT on the background  $g_{(0)}$ and in a precise state. We propose to associate a CFT in the state $\{\braket{O_i}\}$ and background $g_{(0)}$ to the ambient space constructed with the corresponding ALAdS slices according to AdS/CFT. Other states where additional VEVs are turned on would require an extension of the ambient space to accommodate for such additional data. In such case one should include other matter fields and a modification of the Ricci-flatness condition.

While we use holography to justify the connection with a non-trivial state, our results do not necessarily apply only to holographic CFTs. Recall that the embedding space solves the kinematics of CFTs in the vacuum state on conformally flat backgrounds, and its hyperbolic slices are pure (A)dS spaces. 
Nonetheless, we know that using the embedding space we can solve the symmetry constraints on correlators not only for theories which are strictly-speaking holographically dual to pure AdS, but also for free or weakly coupled CFTs in the vacuum state. 
The ambient space will be treated in a similar way: although we use the AdS/CFT dictionary to construct it, we expect it to allow one to solve the kinematical constraints of any CFT in that background and state, also non-holographic ones. We will explicitly see this  in the example of thermal CFTs discussed in section \ref{sec:thermal}.

In this section we construct Weyl-covariant building blocks that can appear in correlators of a CFT on the metric background $g_{(0)}$ with a given energy-momentum tensor VEV $\braket{T_{ij}}$, using the corresponding ambient space as prescribed. The focus will be on scalar $n$-point functions as a first test of the formalism. The case of flat ambient space correlators described in section \ref{SSec:RelToESpace} will guide our steps. There, for scalar $n$-point functions the only available building block is $X_{ij}$. Since the ambient space accounts for setups with less symmetry, we expect a larger number of independent invariants than in the embedding space. After assembling these Weyl-covariant building blocks into correlators on the ambient space, the CFT correlators are obtained by taking the projection onto a section of the nullcone.  We take the section to be at $t=1$; through a Weyl transformation one can move to a conformally-related section. In section \ref{Ssec:SpinningCorr} we describe how to generalise this discussion to spinning correlators.

\subsection{The ingredients} \label{SSec:Ingredients}
The objects at hand are the ambient space metric and covariant derivative. Whilst in some sense these objects survive in the embedding space limit as the Minkowski metric and partial derivative, the ambient Riemann tensor $\wt{R}_{ABCD}$ does not. Thus, $\wt{R}_{ABCD}$ and its ambient covariant derivatives form natural ingredients that embody departures from embedding space results.

For correlation functions the other important ingredient is the homothetic vector, $T$. As discussed in section \ref{SSec:RelToESpace}, $T$ provides the ambient space generalisation of the embedding space insertion points $X_i$ for correlation functions. For $n$-point functions we have multiple distinct insertion points and need to parallel transport all relevant quantities to the same point, so that everything lives in the same tangent space and contractions can be made.
Typically the geodesics along which we transport leave the ambient nullcone and explore the bulk of ambient space. This means that transported quantities get affected by the non-trivial $(d+1)$-dimensional ALAdS geometry. The ambient curvature itself contains information about the state. Explicitly, \begin{align}
 \label{generalRiemann}
\text{even $d$: }&\qquad (\nA_\rho)^{\frac{d}{2}-2} \wt{R}_{\rho ij \rho}= \frac{t^2}{2} 
\left( \frac{d}{2}\right)! \; g_{(d)ij} + F[g_{(0)}] + O(\rho) \, , \\
\label{generalRiemannOddD}
\text{odd $d$: } &\qquad (\nA_\rho)^{\frac{d+1}{2}-2} \wt{R}_{\rho ij \rho}= \frac{t^2}{2\sqrt{\pi}} \, \Gamma\left(\frac{d}{2}+1\right) \, \frac{g_{(d)ij}}{\sqrt{\rho}} +  O(\rho^0) \, ,
\end{align}
where $F[g_{(0)}]$ is a local functional of $g_{(0)}$, while $g_{(d)ij}$ is related to the state in the prescription outlined above. This is one of the main ways the CFT state enters the building blocks that we are constructing.

Note that in more general settings where other operators take non-vanishing VEVs these must be added to the legitimate ingredients. If residual conformal Killing vectors are present, the corresponding ambient isometries and their parallel transport may also enter the list of  ingredients necessary to construct a complete set of invariants.

Finally, we close this subsection by listing some useful properties of the ambient Riemann tensor, $\wt{R}_{ABCD}$. The Weyl, Cotton and Bach tensors can be obtained as the restriction of the ambient curvature to the $d$-dimensional background \cite{Fefferman:2007rka}\footnote{Here we are assuming that $d\neq4$. If $d=4$, as we will remark later, the energy-momentum tensor VEV contained in $g_{(d)ij}$ enters the $\left.\wt{R}_{\rho jk\rho}\right|_{\rho=0, t=1}$ components, and thus these cannot be written only in terms of the boundary metric $g^{(0)}$. As a consequence, the expression in \eqref{eq:AcurvBdryConfCurv} is no longer valid in $d=4$.},
\be \label{eq:AcurvBdryConfCurv}
\left. \wt{R}_{ijkl}\right|_{\rho=0, t=1} = W_{ijkl}\,, \quad \quad
\left. \wt{R}_{\rho jkl}\right|_{\rho=0, t=1} = C_{jkl}\,, \quad \quad
\left. \wt{R}_{\rho jk\rho}\right|_{\rho=0, t=1} = -\frac{B_{ij}}{d-4}\,.
\ee 
Working perturbatively at small $\rho$ one can obtain expressions in closed form for the components of the ambient Riemann tensor. For conformally flat $g_{(0)}$,
\begin{align} \label{ARiemannCompg0ConfFlat}
	\wt{R}_{\rho jk\rho} &= 
	\frac{d}{4} \left(\frac{d}{2}-1\right)
	g_{(d)jk} \, \rho^{\frac{d}{2}-2} t^2 + O(\rho^{\frac{d}{2}-1})
	\,, \\
	\wt{R}_{\rho jkl} &= \frac{d}{4}  \left[
	\nabla_l g_{(d)jk} - \nabla_k g_{(d)jl}
	\right] \rho^{\frac{d}{2} - 1} t^2 +O(
	\rho^{\frac{d}{2}}
	)\,, \\\label{ARiemannCompg0ConfFlatE3}
	\wt{R}_{i jkl} &= \frac{d}{4} \left(
	g_{(0)il}g_{(d)jk} +g_{(0)jk} g_{(d)il} - g_{(0)ik} g_{(d)jl} - g_{(0)jl} g_{(d)ik}
	\right) \rho^{\frac{d}{2}-1} t^2 + O(\rho^{\frac{d}{2}})\,,
\end{align}
while for generic $g_{(0)}$ in $d=3$ the components take the same form as above except for  
\begin{equation}  \label{ARiemannCompd=3}
    \wt{R}_{\rho jkl} = \left[ \nabla_l P_{jk} - \nabla_k P_{jl} \right] t^2 +\frac{d}{4}  \left[
	\nabla_l g_{(d)jk} - \nabla_k g_{(d)jl}
	\right] \rho^{\frac{d}{2}-1} t^2 +O(
	\rho
	)\,.
\end{equation}
where $P_{ij}$ is the boundary Schouten tensor, \eqref{Schouten_def}.
We will make use of these expressions later when studying CFTs at finite temperature and on squashed sphere backgrounds.

\subsection{The building blocks} \label{SSec:BuildingBlocks} 

We now construct building blocks on the ambient space that can enter CFT correlators based on Weyl covariance. Following the previous discussion, using parallel transport we must combine the local quantities
\be \label{AmbientIngredientScalarInvariants}
T, \quad \tilde{g}, \quad (\nA)^k \Riem \qquad (k=0,1\dots),
\ee
evaluated at the different insertion points.\footnote{Note that gradients of $T$ do not need to be considered since $\wt{\nabla}_A T_B = \tilde{g}_{AB}$.} We first focus on scalar invariants, turning to invariants with spin in section \ref{Ssec:SpinningCorr}.

The simplest scalar invariant is $\X_{ij}$, the ambient space analogue of the square-distance between insertions. We construct it as prescribed in the flat case in section \ref{SSec:RelToESpace}: we parallel transport $T_i = T(\X_i)$ to $\X_j$ yielding $\T_i$, which we then contract with $T_j=T(\X_j)$ at $\X_j$. In appendix \ref{AppendixGeodPTranspT} we discuss in detail how to find geodesics between two points lying on a section of the ambient nullcone, how to perform the parallel transport and finally obtain $\hat{T}_i$. The key result is that
\be \label{RelationXij-AGeodDistance}
\X_{ij} = -2 \, \T_i \cdot T_j = \ell (\X_i, \X_j)^2,
\ee
where $\ell (\X_i, \X_j)$ is the geodesic distance between the two insertion points on the ambient space.
This generalises the result we found earlier for the flat background, \eqref{AspFlatCaseT01Invariant}. Note that it does not matter which insertion we parallel transport, the result is symmetric under $i 
\leftrightarrow j$.

Note that the invariant \eqref{RelationXij-AGeodDistance} relies on the existence of an ambient geodesic between the two insertions. It is conceivable that in some cases no such geodesic exists, in which case this building block doesn't exist. It is also possible that there is more than one geodesic, in which case there will be an enhancement in the number of invariants available to build correlators.\footnote{This latter possibility occurs for states described by thermal AdS spaces, where there is an infinite number of geodesics connecting any two nullcone points, enumerated by the number of windings around the thermal circle. Indeed, a sum over such contributions is required to reproduce the corresponding thermal correlator, as we show later in section \ref{SSSec_Double_Twists}.}  However, under mild assumptions given any two points on the ambient nullcone there is one and only one geodesic connecting them \cite{Mazzeo86, Graham:2017fmz}.

In addition to $\X_{ij}$ we can construct new bi-local scalar invariants by directly using the ambient curvature and its covariant derivatives. Assembling these ingredients one immediately discovers that not all such invariants are independent, due to a number of identities: the contractions of $T$ with Riemann are trivial \cite{Fefferman:2007rka},
\begin{equation}\label{TR0}
	T_j^D \wt{R}_{ABCD} =0,
\end{equation}
and contractions with gradients of Riemann are redundant since,
\begin{subequations}
    \begin{align}
     \label{TR1}
		T_j^D \wt{R}_{ABCD;M_1\dots M_r} &= - \sum_{s=1}^r
	\wt{R}_{ABCM_s;M_1\dots \wh{M}_s\dots M_r}, \\
	\label{TR2}
		\begin{split}
		    T_j^P \wt{R}_{ABCD;M_1\dots M_s P M_{s+1} \dots M_r} &= - (s+2) \wt{R}_{ABCD;M_1\dots M_r}\\ 
		& \quad \,   - \sum_{t=s+1}^r
		\wt{R}_{ABCD;M_1\dots M_s M_t M_{s+1} \dots \wh{M}_t \dots M_r},
		\end{split}
    \end{align}
\end{subequations}
where semicolons denote covariant derivatives and hatted indices are understood as removed. These properties, along with  Ricci-flatness $\wt{R}_{AB}=0$, reduce the number of independent scalar invariants.

Based on these observations, in what follows we restrict our attention to the following set of scalar invariants constructed at $X_j$, the \emph{weighted curvature invariants}:
\be \label{eq:CurvatureInvariantsForm2} 
W_{ij}^{(k, n)} \sim \mbox{contr} \left[ \T_i \otimes \dots  \otimes (\wt{\nabla})^{r_1} \mbox{Riem} \otimes \dots \otimes (\wt{\nabla})^{r_k}
 \mbox{Riem} \right],
\ee
where $\mbox{contr}$ indicates the full contraction of all the indices using the ambient metric at $X_j$. They are diffeomorphism invariants in $d+2$ dimensions, while displaying a precise weight under Weyl transformations (hence being Weyl covariant quantities).  Since $\T_i$ are obtained by parallel transport, one can build a distinct set of invariants for each corresponding geodesic.

We have labelled the $W_{ij}^{(k, n)}$ by the number of Riemann's they contain, $k$. This is a good label once we fix some redundancies. The first redundancy is associated to use of the identity $\wt{\nabla}^D \wt{R}_{ABCD} = 0$ which follows from the second Bianchi identity and Ricci-flatness. Because of this, we require that none of the covariant derivatives within each factor  $(\wt{\nabla})^{r} \mbox{Riem}$ in \eqref{eq:CurvatureInvariantsForm2} are contracted with the Riemann tensor itself, regardless of the ordering. This is because by repeated commutation of the covariant derivatives one can eventually reach a form where $\wt{\nabla}^D \wt{R}_{ABCD} = 0$ can be applied to one term; all remaining terms then take the form of other terms appearing in \eqref{eq:CurvatureInvariantsForm2} with higher $k$. The second redundancy is the remaining ordering ambiguity of the covariant derivatives within each factor $(\wt{\nabla})^{r} \mbox{Riem}$, which we fix by symmetrisation, as a matter of convention. The remaining label $n$ enumerates all possible invariants with that $k$.

As a note of caution, the weighted curvature invariants \eqref{eq:CurvatureInvariantsForm2} do not necessarily include all possible invariants. For example, we have not considered covariant derivatives of $\T_i$ at $\X_j$, nor do we consider parallel transport of the ambient curvature and its covariant derivatives from $\X_i$ to $\X_j$. In what follows we assume that \eqref{eq:CurvatureInvariantsForm2} constitute a basis without including such contributions. Evidence in support of these assumptions is brought by the results presented in the explicit examples in sections \ref{sec:thermal} and \ref{Sec:SqSphere} where we show that the invariants of the form \eqref{eq:CurvatureInvariantsForm2} under these assumptions constitute a basis.

Let us discuss invariants with $k=0,1,2$.
There are no non-trivial $k=0$ weighted curvature invariants, since without Riemanns in \eqref{eq:CurvatureInvariantsForm2} there are only contractions of $\T_i$ which are zero; there is just the identity.
There are also no $k=1$ weighted curvature invariants, and a proof of this result proceeds as follows. At most two of the indices of Riemann can be contracted with $\hat{T}_i$ due to the antisymmetry of Riemann indices. Thus at least two of the four indices of the ambient Riemann are to be contracted with either the inverse metric or covariant derivatives. Any contraction with an inverse metric yields zero by Ricci-flatness. Any contraction with covariant derivatives is a term that is not a member of the $k=1$ set of invariants, according to the definition given above. Later, in the examples discussed in sections \ref{sec:thermal} and \ref{Sec:SqSphere} we provide explicit examples of $k=2$ building blocks, which play an important role in constructing ambient 2-point functions. 

As explained in section \ref{SubsecWeylTrAndIsometries} the engineering dimension, $\Delta$, of an ambient scalar is minus its overall weight in $t$. It can be easily computed with the same rules used in the embedding space\footnote{For ease of comparison with the mathematical literature, we observe this is not the perspective typically adopted in conformal geometry. There the metric $\tilde{g}$ and the Riemann (meant as tensors) both have  dimensions --2 following from their homogeneity in $t$, while $T$ and the ambient derivative $\wt{\nabla}_M$  have dimension zero. Their components have of course different weight, and this is what one considers in the embedding formalism instead. For example, the vector $T=t\p_t$ has weight zero in $t$, while its components $T^M= t \delta^M_0$ clearly have dimension --1. In practice either perspective lead to the same answer \eqref{eq:DimensionAinvariants}, hence in the main discussion we stick to the component-based picture, which is rather unnatural from the perspective of conformal geometry but very common in the QFT literature.}, by viewing $T^M$ and $\nabla_M$ as dimension $-1$ and 1 quantities respectively. The Riemann tensor contains two derivatives of the metric and we conclude it has dimension 2. For weighted curvature invariant \eqref{eq:CurvatureInvariantsForm2} with $k$ Riemann tensors, $r$ covariant derivatives and $\ell$ $\T_i$ vectors,
\be \label{eq:DimensionAinvariants}
\Delta = 2k +r -\ell.
\ee
Note that $r+\ell$ must be even in order to be able to build a scalar with an integral number of inverse metrics. From \eqref{eq:DimensionAinvariants} this entails that all such invariants have even dimensions. 

If an invariant of the form \eqref{eq:CurvatureInvariantsForm2} has $\Delta \neq 0$ we can easily construct a $\Delta = 0$ invariant from it by multiplying by an appropriate power of $\X_{ij}$. However, a useful class of $\Delta = 0$ invariants are those of the form \eqref{eq:CurvatureInvariantsForm2} with $2k +r = \ell$. Due to the symmetries of the Riemann tensor their structure is completely fixed and one can list them in full generality. If we define the partial contraction
\begin{equation}\label{mathcalRweight0inv}
	\mathcal{R}^{(\hat{r})}_{AC} = \T_i^{M_1} \dots \T_i^{M_{\hat{r}}} \, \T_i^U \T_i^V \, \wt{\nabla}_{M_1} \dots \wt{\nabla}_{M_{\hat{r}}} \wt{R}_{AUCV} \,,
\end{equation}
any $\Delta = 0$ curvature scalar constructed out of $k$ Riemann's and $r$ derivatives can be written as a linear combination of chains of the form
\be \label{eq:Generic0weightCurvInv}
\mathcal{R}^{(r_1)\; M_2}_{\;\;\; M_1} \,
\mathcal{R}^{(r_2)\; M_3}_{\;\;\; M_2}
\dots \mathcal{R}^{(r_k)\; M_1}_{\;\;\; M_k}
,
\ee
where each such chain is constrained to have  $\sum_i r_i = r$. We will utilise invariants from this class in sections \ref{sec:thermal} and \ref{Sec:SqSphere}.

A caveat to be aware of concerns the limit of expressions of the form \eqref{eq:CurvatureInvariantsForm2} to a section of the nullcone $\rho=0$, $t=1$. In particular this involves the behaviour of the ambient Riemann tensor when approaching the nullcone, some examples of which are given in \eqref{generalRiemann}-\eqref{generalRiemannOddD} and \eqref{ARiemannCompg0ConfFlat}-\eqref{ARiemannCompd=3}. From the metric expansion one can show that in even $d$ only non-negative integer powers of $\rho$ appear in components of the ambient Riemann tensor, while in odd $d$ fractional powers of $\rho$ appear when a non-vanishing $g_{(d)}$ is present. For odd $d$ the RHS of equation \eqref{generalRiemannOddD} diverges for $\rho\to0$.
By taking more derivatives such divergences become stronger. For the purpose of constructing ambient invariants this means that scalars constructed using curvature terms $(\wt{\nabla})^r \wt{R}$ with high enough $r$ may be singular in odd $d$ when restricted to the boundary. Such terms must either be discarded, or combined into linear combinations to cancel such infinities. Despite these apparent complications for odd $d$, we were able to find a complete basis of curvature invariants for the $d=3$ example of a CFT on a squashed 3-sphere in section \ref{Sec:SqSphere}.

Analogously to the embedding formalism, correlators in $d+2$ dimensions must be invariant under the (near-nullcone) isometries encoding $d$-dimensional conformal symmetries. Geodesics and geodesic transport preserve the symmetries of the geometry and hence ambient building blocks constructed out of the ambient metric and covariant derivatives of the curvature automatically satisfy the constraints imposed by the near-nullcone symmetries. One can explicitly see this for instance in the invariants constructed in section \ref{SSec:BBAinv} in the case of thermal CFTs -- they are invariant under the residual symmetries of the CFT. To conclude, the prescription for the ambient building blocks discussed here automatically implements the residual conformal Ward identities, leaving Weyl covariance as the only non-trivial kinematic constraint to be imposed.

\subsection{Scalar 2-point functions} \label{SSec:ProposalA2ptf}
In the previous subsections we constructed a class of  ambient invariants -- namely, $\wt{X}_{ij}$ \eqref{RelationXij-AGeodDistance} and $W_{ij}^{(k,n)}$ \eqref{eq:CurvatureInvariantsForm2} -- that enter CFT correlators on general backgrounds and states based on Weyl covariance. We now propose a general form of ambient scalar 2-point functions that arranges those invariants so as to exhibit the required properties,
\begin{equation}\label{eq:ProposalForm1}
	\braket{O(\X_1) O(\X_2)}
	= 
	\frac{C_\Delta}{(\wt{X}_{12})^{\Delta}}\, \lim_{\substack{\rho\to 0\\ t\to 1}}\left[1+\sum_{k=2}^\infty\mathcal{I}^{(k)}_2\right],
\end{equation}
where
\be
{\cal I}_{2}^{(k)} = \sum_{n} c_n \X_{12}^{\Delta_n/2}W_{12}^{(k,n)},
\ee
and $\Delta_n$ denotes the dimension of $W_{ij}^{(k,n)}$ given by \eqref{eq:DimensionAinvariants}.
The constant coefficients $c_n$ are determined by the dynamics of the CFT. The sum over $k$ in \eqref{eq:ProposalForm1} starts from terms of order $O(\wt{\text{R}}\text{iem})^2$ since in section \ref{SSec:BuildingBlocks} we proved that $\mathcal{I}^{(1)}_2=0$, while $\mathcal{I}^{(0)}_2$ is just the identity, already accounted for as the first term in \eqref{eq:ProposalForm1}. The overall scaling dimension is $-2\Delta$, as required by Weyl covariance. The correlator is analytic in curvatures and continuously connected to the flat space limit in which $\wt{X}_{12} \to X_{12}$ and $\mathcal{I}^{(k)}_2\to 0$, where we recover the embedding space 2-point function \eqref{eq:EspCorrInAsp} with the same constant $C_\Delta$.

As discussed in section \ref{SSec:BuildingBlocks} there may be more than one geodesic path connecting the two insertion points. Parallel transporting along each of them can generate independent invariants and thus an implicit sum over all the ambient geodesics connecting $\X_1$ and $\X_2$ is understood in the RHS of \eqref{eq:ProposalForm1}.

Let us now discuss which states we expect to be able to describe using \eqref{eq:ProposalForm1}. 	For a CFT in any background $g_{(0)}$ and state, at short distances the background becomes approximately flat and as such we should have a convergent OPE (see for example \cite{Zinn-Justin:1989rgp,Fredenhagen:1986jg,Hollands:2006ag,Hollands:2023txn} for a general discussion). We can use it to reduce a 2-point function of a scalar operator $O$ of scaling dimension $\Delta$ to a sum of 1-point functions of exchanged operators,
\begin{equation} \label{2-ptfGeneralOPE}
    \braket{O(x_1) O(x_2)} \simeq  
		\frac{1}{|x_{12}|^{2\Delta}} \sum_{\phi\in O\times O} h_{\phi}(x_i,\p_i) \braket{\phi(x_2)} \,,
\end{equation}
where the $\simeq$ is understood as an equality modulo contact terms. 
Since $\Riem \sim \braket{T}$ (see \eqref{generalRiemann} and \eqref{generalRiemannOddD}) schematically we have that ${\cal I}^{(k)}_2 \sim \braket{T}^k$. Therefore we expect \eqref{eq:ProposalForm1} to account for the multi-energy-momentum tensor contributions in \eqref{2-ptfGeneralOPE}, at least for large-$N$ theories where multi-energy-momentum tensor 1-point functions factorise. However, we conjecture that our curvature invariants provide a basis for multi-energy-momentum tensor contributions also for theories which are not at large-$N$. Operators other than the multi-energy-momentum tensors contributing to the RHS of \eqref{2-ptfGeneralOPE} must be captured using other classes of ambient invariants, and we comment on this issue in section \ref{sec:outlook}. We stress that the multi-energy-momentum tensors are universal contributions in any CFT correlator, and this is what the ambient geometry captures through \eqref{eq:ProposalForm1}.

For holographic CFTs the ambient 2-point function \eqref{eq:ProposalForm1} has an additional interpretation, providing multi-energy-momentum tensor corrections to the well-known geodesic approximation of 2-point functions in the context of AdS/CFT \cite{Graham:1999pm,Balasubramanian:1999zv,Louko:2000tp}. In appendix \ref{AppendixGeodPTranspT} we discuss how the presence of the homothetic vector $T$ on the ambient space fully fixes the component of a particle trajectory along that direction. As we show in appendix \ref{AppAmbientAdSGeods} if we focus on geodesics connecting points on the ambient nullcone, the remaining $d+1$ equations for the unknown components of the geodesic path turn out to be the geodesic equations on the ALAdS$_{d+1}$ section associated to that ambient space in a non-affine parametrisation.  In this picture, the endpoints of the geodesic are points on the conformal boundary of ALAdS$_{d+1}$. In appendix \ref{AppAmbientAdSGeods} we further prove that the square-geodesic distance on the ambient space $\X_{12}$ is related to the (renormalised) geodesic distance on the associated ALAdS$_{d+1}$ space. Through \eqref{RelationXij-AGeodDistance} we can write their relation as
\begin{equation} \label{GeodApproxAdSAmbient}
    \frac{1}{(\X_{12})^\Delta} = \left. \frac{r^{-2\Delta}}{(t_1 t_2)^\Delta} e^{-\Delta L_{AdS}}\right|_{r=0}.
\end{equation}
for an arbitrary real $\Delta$, where $r$ is the Fefferman-Graham coordinate on the ALAdS space as in the metric \eqref{AspCoordsAdSSlicing}, while $t_1$ and $t_2$ are the $t$-components of $\X_1$ and $\X_2$ respectively. Here $L_{AdS}$ indicates the (divergent) length of the corresponding geodesic on the ALAdS$_{d+1}$ section.
The RHS of \eqref{GeodApproxAdSAmbient} coincides with the geodesic approximation for a scalar 2-point function of an operator of dimension $\Delta$ in the context of AdS/CFT. It can be argued to follow from the saddle-point approximation of the first-quantised path integral for a massive particle and consequently its validity is restricted to the large-$\Delta$ regime. We can thus interpret the ambient curvature invariants in \eqref{eq:ProposalForm1} as encoding the quantum corrections due to multi-energy-momentum tensor contributions at finite $\Delta$ beyond the semi-classical approximation provided by $(\X_{12})^{-\Delta}$.

Given that $\mathcal{I}^{(1)}=0$, the coefficient in the ambient expansion at order $O(\wt{\text{R}}\text{iem})$ predicted by the geodesic approximation is exact. Therefore a universal prediction from the ambient formalism is that correlators in the geodesic approximation are exact up to order $O(\wt{\text{R}}\text{iem})^2$ corrections if  no other operator with scaling dimension $\Delta <2d$ acquires a VEV.

As mentioned in the introduction, one may consider 2-point functions on a non-trivial background/state as a flat-space higher-point function in vacuum. For example for a non-trivial state, the 2-point function is equivalent to a 4-point function, whose form is fixed up to a function of cross-ratios, and with the in-state represented by an operator inserted at the origin, and the out-state at infinity. We thus expect that 2-point functions on a non-trivial state are fixed by Weyl-invariance up to a function of one variable (the two cross-ratios are degenerate in this limit). Indeed, one can think of the free coefficients appearing in our proposal \eqref{eq:ProposalForm1} as being related to the Taylor series coefficients of the function of cross-ratios in the single remaining variable.\footnote{We thank the anonymous referee for raising this point.}

\subsection{Scalar higher-point functions} \label{Ssec:Higher-ptf}

Similar expressions to \eqref{eq:ProposalForm1} can be written for scalar higher-point functions. In the ambient formalism scalar 3-point functions read
\begin{equation}\label{AspScalar3pt}
	\braket{O_1 O_2 O_3} 
= \frac{C_{123}}{ 
	( \X_{12})^{\alpha_{123}}
	( \X_{13})^{\alpha_{132}}
	( \X_{23})^{\alpha_{231}}
}
\, \lim_{\substack{\rho\to 0\\ t\to 1}}
\left[ 
1 + \sum_{k=2}^\infty\mathcal{I}^{(k)}_3
\right]
,
\end{equation}
where the $\alpha_{ijk}$ coefficients are the same as those defined in \eqref{EspScalar3pt} to ensure the correct scaling properties. To recover the expression on the embedding space in the flat limit, $C_{123}$ must be the same as in \eqref{EspScalar3pt}.
Here $\mathcal{I}^{(k)}_3$ denote linear combinations of weight-0
curvature invariants containing $k$ ambient Riemanns and constructed with the pairwise parallel transport of tensors from the three insertions $\X_1, \X_2, \X_3$.
The fact that bi-local invariants provide a basis for 3- and higher-point functions (with no need to resort to $n$-local invariants)  is justified by the following remarks. First, this is what happens in embedding space correlators with arbitrary spin and with an arbitrary number of insertions. Second and more fundamental, as stressed above the OPE is expected to converge at short enough distances in general backgrounds and states, and OPE contractions are pairwise. 

The linear combinations $\mathcal{I}^{(k)}_3$ are thus products of bi-local invariants from the three insertion points and as such they can be decomposed in terms of the 2-point linear combinations $\mathcal{I}^{(m)}_2$ with generic coefficients. From $\mathcal{I}^{(1)}_2 = 0$ it follows that $\mathcal{I}^{(1)}_3 = 0$; furthermore one can check explicitly that 
\begin{align} 
\mathcal{I}^{(2)}_3 (\X_1, \X_2, \X_3) &= P_3^{(2)}(\X_1, \X_2, \X_3) \, , \label{I32}\\
\mathcal{I}^{(3)}_3 (\X_1, \X_2, \X_3) &= P_3^{(3)}(\X_1, \X_2, \X_3)  \, ,  \label{I33}
\end{align}
where we defined 
\begin{equation} \label{Higherptf-P3k}
P_3^{(k)} (\X_1, \X_2, \X_3) = \mathcal{I}^{(k)}_2 (\X_1,\X_2) + \mathcal{I}^{(k)}_2 (\X_1,\X_3) + \mathcal{I}^{(k)}_2 (\X_2,\X_3),
\end{equation}
and where each $\mathcal{I}^{(k)}_2$ is thought of as containing generic different constant coefficients.
Turning to fourth order invariants, the most general linear combination is of the form
\begin{equation}
    \mathcal{I}^{(4)}_3 (123) = P_3^{(4)}(123) + 
    \mathcal{I}^{(2)}_2(12) \,\mathcal{I}^{(2)}_2(13) + \mathcal{I}^{(2)}_2(12) \,\mathcal{I}^{(2)}_2(23) +
    \mathcal{I}^{(2)}_2(13) \,\mathcal{I}^{(2)}_2(23)\, ,
\end{equation}
where we adopted a convention where one denotes $X_\ell$ by $\ell$. The last three terms can be rewritten as $[P^{(2)}_3(123)]^2$. The latter also includes $[\mathcal{I}^{(2)}_2(12)]^2$, $[\mathcal{I}^{(2)}_2(13)]^2$ and $[\mathcal{I}^{(2)}_2(23)]^2$. Note that these three terms are already contained in $P_3^{(4)}(123)$ and their appearance in $[P^{(2)}_3(123)]^2$ can be reabsorbed by shifting the corresponding arbitrary coefficients in $P_3^{(4)}(123)$. All in all we can rewrite the linear combination of fourth order 3-point invariants as 
\begin{equation}
    \mathcal{I}^{(4)}_3 (123) = P_3^{(4)}(123) + 
    \left(P^{(2)}_3(123)\right)^2\,.
\end{equation}
Studying higher orders one finds the  following recursive relation between order $k$ and order $k-2$ invariants,
\begin{equation}
    \mathcal{I}^{(k)}_3 (123) = P_3^{(k)}(123) + 
    P^{(2)}_3(123) P^{(k-2)}_3(123)
    \,.
\end{equation}
Using these recursion relations one finds the  expression for the general linear combination of curvature invariants of order $k$  in terms of the bi-locals $\mathcal{I}^{(k)}_2(\X_i,\X_j)$ involved in 2-point functions,
\begin{align}
\text{$k$ even:} \qquad &\mathcal{I}^{(k)}_3 (\X_1, \X_2, \X_3) =  
\sum_{\ell=1}^{k/2} \left( P_3^{(2)}\right)^{\frac{k}{2}-\ell} P_3^{(2\ell)}
\, , \label{I3kEven} \\
\text{$k$ odd:} \qquad &\mathcal{I}^{(k)}_3 (\X_1, \X_2, \X_3) = 
\sum_{\ell=3/2}^{k/2} \left( P_3^{(2)}\right)^{\frac{k}{2}-\ell} P_3^{(2\ell)}
\, , \label{I3kOdd}
\end{align}
where the sum in the odd case is over half-odd $\ell$. These expressions are manifestly symmetric (modulo the different coefficients in the linear combinations) under permutations of the insertion points $\X_i$ and are of the appropriate order in the ambient Riemann. The full 3-point function is not invariant under permutations of the three insertion points for different scaling dimensions $\Delta_i$. However this different behaviour under Weyl transformations is accounted for by the overall factor in front of \eqref{AspScalar3pt}.

Let us now turn to scalar $n$-point functions. Based on our assumptions and on a consistent reduction to \eqref{EspHigherptf} in the flat limit, their form on the ambient space is fixed to
\begin{equation} \label{AspHigherptf}
    \braket{O_1(\X_1) \ldots O_n(\X_n)} =  \left(\prod_{i<j} \widetilde{X}_{ij}^{\alpha_{ij}}\right) \, \lim_{\substack{\rho\to 0\\ t\to 1}} \left[f\left(u\right)+\sum_{k=2}^\infty\mathcal{I}^{(k)}_n\right],
\end{equation}
where the cross-ratios $u$ are now in terms of the ambient geodesic distances 
\begin{equation}
    u_{[pqrs]} = \frac{\X_{pr}\X_{qs}}{\X_{pq}\X_{rs}} \,,
\end{equation}
and $f$ is the same function of the cross-ratios present in the corresponding correlator for the same CFT in vacuum on flat space.

Using combinatorial arguments similar to those for 3-point functions one can straightforwardly generalise \eqref{I3kEven}-\eqref{I3kOdd} to any $n$,
\begin{align}
\text{$k$ even:} \qquad &\mathcal{I}^{(k)}_n (\X_1\dots \X_n) =  
\sum_{\ell=1}^{k/2} \left( P_n^{(2)}\right)^{\frac{k}{2}-\ell} P_n^{(2\ell)}
\, , \label{InkEven} \\
\text{$k$ odd:} \qquad &\mathcal{I}^{(k)}_n (\X_1\dots \X_n) = 
\sum_{\ell=3/2}^{k/2} \left( P_n^{(2)}\right)^{\frac{k}{2}-\ell} P_n^{(2\ell)}
\, . \label{InkOdd}
\end{align}
Here we defined 
\begin{equation}
P_n^{(k)}(\X_1 \dots \X_n) = \sum_{(Y,Z) \in \,C^n_2 (\X_1 \dots \X_n)} \mathcal{I}^{(k)}_2 (Y,Z) \, ,
\end{equation}
where the sum is over the $\binom{n}{2}$ pairwise combinations $(Y,Z)$  of the points $(\X_1 \dots \X_n)$. This definition reduces to \eqref{Higherptf-P3k} for $n=3$.

These combinatorial relations show that knowing the ambient curvature invariants that enter the scalar 2-point function up to a certain order $k$ allows one to straightforwardly write  the form of generic ambient scalar $n$-point functions to the same order $k$.
In particular, $\mathcal{I}^{(1)}_2 = 0$ implies $\mathcal{I}^{(1)}_n = 0$ for any $n$. This entails that the universal validity of the geodesic approximation up to $O(\Riem^2)$ corrections extends to any scalar $n$-point function in any CFT on generic backgrounds and states, as long as no operator with scaling dimension lower or equal than $2d$ has a non-vanishing VEV. Here by geodesic approximation of a generic $n$-point function we mean the first term $\left(\prod \widetilde{X}_{ij}^{\alpha_{ij}}\right) f(u)$ in \eqref{AspHigherptf}.

\subsection{Correlators with spin} \label{Ssec:SpinningCorr}
In this section we provide some comments on how to generalise the scalar correlators discussed above to those with spin. As before, embedding space is generalised to ambient by adopting the homothetic vector $T^M$ in lieu of the position on Minkowski $X^M$, and the ambient metric $\tilde{g}_{MN}$ instead of $\eta_{MN}$. On top of this, one considers corrections that depend on ambient curvature invariants which vanish in the flat limit.

The building blocks to be used in this case must have free indices on the ambient space, meaning that curvature invariants will be of the same form as \eqref{eq:CurvatureInvariantsForm2}, where contractions are understood as partial contractions so as to end up with the appropriate spin. Such free ambient indices transform as generic ambient tensors ({\it i.e.} weighted tractor tensors when restricted to the nullcone) under Weyl transformations through the matrix $U^M_{\;\;\;N}(\Omega)$ defined in equation \eqref{WeylUmatrix}. It is natural to expect that a set of ambient curvature invariants with such partial contractions would form a basis for multi-energy-momentum tensor contributions to $n$-point functions of general spin.

Following corresponding discussions in the embedding space \cite{Costa2011}, it is convenient to reduce the problem of classifying ambient spinning structures to finding scalar structures by considering ambient polarisation vectors $Z_{(i)}^M$, one at each insertion point, and treating them as additional local ingredients to be used to construct scalar bi-local invariants besides \eqref{AmbientIngredientScalarInvariants}. The number of $Z_{(i)}$'s that such invariants must contain is fixed by the spin of the inserted operators. To retrieve the spinning expression on the ambient space one would then use appropriate differential operators acting on the $Z_{(i)}$'s.  

There also exists an alternative path for spinning ambient building blocks. The relationship described in section \ref{SubsecWeylTrAndIsometries} between tractor and ambient connections allows one to generalise the so-called weight- and spin-shifting operators on the embedding space introduced in \cite{Karateev2017} to the ambient space. These differential operators act on tensor structures modifying their scaling dimension and spin, and by leveraging the many results of tractor calculus one may be able to generalise them to the ambient space. 

More precisely, given such operators on the embedding space one would perform the map
\begin{equation}
X^M \to T^M, \qquad \p_M \to \wt{\nabla}_M, \qquad \eta_{MN} \to \tilde{g}_{MN} \,,
\end{equation}
giving local weight- and spin-shifting operators on the ambient space. Operators obtained in this way satisfy all the required properties of a weight- or spin-shifting operator as put forward in the flat space case \cite{Karateev2017}. Note that to use these operators requires pairwise contractions to obtain bi-local differential operators acting on two distinct insertions, and on the ambient space this would involve parallel transport of differential operators. It would be interesting to work out the details and we leave this to future work.

\section{Finite temperature CFTs} \label{sec:thermal}

In this section we apply the ambient space formalism to the case of finite temperature CFTs. This example allows for explicit checks of the ambient predictions on correlators by matching them with results from thermal OPEs and holography. The agreement we find represents a non-trivial test of the formalism.

Euclidean thermal CFTs in $d$ dimensions on flat space live on the thermal cylinder $S^1_\beta \times \R^{d-1}$, where $\beta$ is the inverse temperature. We parametrise this background with coordinates $x^i=(\tau, x^a)$ where $0\leq \tau <\beta$.  This geometry breaks conformal invariance because of the length scale $\beta$; the only global symmetries remaining are translations along the $\tau$ and $x^a$ directions, as well as rotations on $\R^{d-1}$. We restrict our analysis to states which respect these spacetime symmetries and do not spontaneously break them further.

These residual symmetries constrain 1-point functions. By translational invariance they are non-vanishing only for primary operators, and rotational symmetry implies they must be constant tensors of the form 
\begin{equation} \label{eq:Th1point}
    \braket{O^{i_1 \dots i_J}}^{(\beta)}_\Delta = \frac{b_O}{\beta^{\Delta}} \left( e^{i_1} \dots e^{i_J} - \mbox{traces} \right),
\end{equation}
where $b_O$ is a theory-specific constant (which may be zero) and 
$e^i$ is the unit vector along $\tau$ \cite{Iliesiu2018}. In particular, for the energy-momentum tensor VEV we have
\begin{equation} \label{BBthermalTij}
    \braket{T_{ij}}^{(\beta)} = \dfrac{c_{(d)}}{\beta^d}\,\, \text{diag}\left(
1-d,1, \dots,1
\right)\,,
\end{equation}
which is traceless, as expected on the thermal cylinder.
From now on (and unless stated otherwise) we restrict our attention to $d=4$; we will comment later on the generalisation to any $d$. Given the state specified by \eqref{BBthermalTij}, the ambient space to be used has  the AdS planar black hole as ALAdS$_{5}$ slices. The 6-dimensional ambient geometry  relevant for this problem then reads
\begin{equation}\label{eq:AmbientSpaceBrane}
	\tilde{g} = - ds^2 + \frac{s^2}{z^2}\left[
	\frac{dz^2}{1-\frac{z^4}{z_H^4}}
	+ \left(1-\frac{z^4}{z_H^4}\right) d\tau ^2 + \delta_{ab}dx^a dx^b
	\right],
\end{equation}
with $a,b = 1,2,3$, horizon scale $z_H= \pi / \beta$ and compact time direction $0\leq \tau < \beta$. 
This choice of AdS bulk metric corresponds to $c_{(4)}=2 \pi^4$ in \eqref{BBthermalTij}.
Finally note that \eqref{eq:AmbientSpaceBrane} is not in the usual Fefferman-Graham ambient gauge \eqref{AspMetricAdSSlicing}, which can be reached with the transformation $z=r/\sqrt{1+\frac{r^4}{4 z_H^4}}$.

Our aim is to find the expression for scalar 2-point functions in such a CFT using the ambient space formalism. This translates into finding the ambient building blocks that account for the multi-energy-momentum tensor contributions. Following the prescription in \eqref{eq:ProposalForm1} we set up the problem so as to identify these invariants order by order in the ambient Riemann. In this specific case, since $\beta$ is the only scale in the CFT, we have that
\begin{equation}
\Riem \sim \beta^{-4}. \label{thermal_riem_beta}
\end{equation}
Thus the Riemann expansion in \eqref{eq:ProposalForm1} can be viewed either as an expansion in small temperature, or as an expansion in small distance between insertions. The former allows us to use $\beta$ power-counting to organise the number of Riemann tensors in \eqref{eq:ProposalForm1}. The latter allows us to make contact with the thermal OPE, presented in section \ref{SSec:AcorrBB}.

As a first step we find the relevant ambient invariants up to second order in the Riemann tensor.

\subsection{Ambient geodesics and geodesic transport}
\label{SSec_B_Geods}

The first step is to identify the geodesics between the two insertion points and compute the corresponding geodesic distance. As we showed in section \ref{SSec:BuildingBlocks} this  yields the invariant $\X_{12}$. Adopting the ambient parametrisation $\X^M = (t,z,\tau,x^a)$ and using the residual rotational and translational symmetries of the problem, we can move the two insertions to lie at $\X_1=(t_1, 0, 0,0,0,0)$ and $\X_2=(t_2, 0, \tau_f,x_f,0,0)$.  

The strategy to solve the geodesic equations is the following. Because of the presence of the homothetic vector $T=s\p_s$, the expression for the trajectory along $s = r\, t$ is automatically fixed up to an integration constant, which is the square geodesic length $C$, as derived in appendix \ref{AppendixGeodPTranspT}, see \eqref{eq:sLambdaGeods}. The second order geodesic equation for $s$ then becomes a first order equation involving $z, \tau$ and $x_1$ and their derivatives. One can get rid of $\dot{\tau}$ and $\dot{x_1}$ using the equations for the integrals of motion related to translations along $\tau$ and $x_1$, 
\begin{equation}\label{eq:BBGeodEqtaux1}
	\dot{\tau} = \frac{A_1 z^2}{\lambda (1-\lambda)  \left(1-\frac{z^4}{z_H^4}\right)}, \quad \qquad
	\dot{x_1} = \frac{A_2 z^2}{\lambda (1-\lambda)  },
\end{equation}
with $A_{1,2}$ constants of motion. The geodesic equation for $s$ thus becomes a non-linear first order equation in $z$ only,
\begin{equation}\label{eq:BBGeodEqz}
	4 \lambda^2 (1-\lambda)^2  \dot{z}^2 -\frac{4 A_2^2 z^8}{z_H^4}+\frac{z^6}{z_H^4}+4 \left(A_1^2+A_2^2\right) z^4-z^2 =0.
\end{equation}
The three equations \eqref{eq:BBGeodEqtaux1} and \eqref{eq:BBGeodEqz} are the only independent equations left. 

We are interested in computing ambient correlators, which are expressed as expansions in terms of the ambient Riemann. Given \eqref{thermal_riem_beta}, it is sufficient to solve the geodesic equations perturbatively, considering the distance between the insertions as small compared to the inverse temperature. Denoting the distance between the insertions on the thermal cylinder by $|x| = \sqrt{\tau_f^2 +x_f^2}$, this corresponds to the regime $|x|/\beta \ll 1$. We solve the equations by expanding the trajectory $z, \tau, x_1$ and the integration constants $C, A_1, A_2$ as, 
\begin{equation}
        z(\lambda) = \sum_{k=0}^\infty \frac{z^{(k)}(\lambda)}{z_H^{4k}}, \qquad \quad
        A_i = \sum_{k=0}^\infty \frac{A_i^{(k)}}{z_H^{4k}},
\end{equation}
and analogously for $\tau, x_1$ and $C$.
This is a consistent expansion since in this perturbative scheme we intend to capture the corrections to geodesics on $(d+2)$-dimensional Minkowski provided by the non-trivial geometry on the ALAdS slices, where only powers of $z_H^4$ appear. We start by solving equation \eqref{eq:BBGeodEqz} in $z(\lambda)$ order by order. By subsequently feeding the $z^{(k)}(\lambda)$'s into \eqref{eq:BBGeodEqtaux1} one finds the coefficients in the expansion of $\tau$ and $x_1$.

At each perturbative order, the solution just obtained  contains six integration constants, that is $A_1^{(k)}, A_2^{(k)}, C^{(k)}$ as well as the three following from the integration of the first order equations \eqref{eq:BBGeodEqtaux1}-\eqref{eq:BBGeodEqz}. These can be fixed order by order imposing the boundary conditions\footnote{Note that $z$ differs from the Fefferman-Graham coordinate $r$ by $O\left(z_H^{-4}\right)$ corrections. However, close to the boundary $\lambda \to 0,1$ the behaviour in $\lambda$ of $z(\lambda)$ and $r(\lambda)$ is the same and this ensures we can use \eqref{eq:BBsmallTt01BC} as boundary conditions.}
\begin{subequations}\label{eq:BBsmallTt01BC0}
	\begin{align}
		\tau(0) &= 0\,, \qquad \qquad \tau(1) = \tau_f\,, \\
		x_1(0) &= 0\,, \qquad \qquad \!\!\! x_1(1) = x_f\,, \\
		\label{eq:BBsmallTt01BC}\lim_{\lambda\to0}  \frac{s(\lambda)}{z(\lambda)} &= t_1\,, \quad\;\; 
		\lim_{\lambda\to1}  \frac{s(\lambda)}{z(\lambda)} = t_2\,.
	\end{align}
\end{subequations}

The leading order of both the trajectory and the integration constants coincide with the corresponding Minkowski expressions shown in section \ref{SSec:RelToESpace}. Following this integration scheme and renaming $\tau_f \to \tau$ and $x_f \to x$, to second order in the perturbative parameter the invariant $\X_{12}$ reads, 
\begin{equation}\label{eq:BBXij}
	\X_{12} = t_1 t_2 |x|^2  \left[
	1+\frac{|x|^2 \left(x^2-3 \tau ^2\right)}{120 z_H^4}-\frac{|x|^4 \left(91 \tau ^4-98 \tau ^2 x^2+19 x^4\right)}{201600 z_H^8} +  O\left(\frac{|x|^{12}}{z_H^{12}}\right) 
	\right].
\end{equation}
One can straightforwardly proceed to arbitrarily high order. Through the relation between the ambient and AdS geodesic lengths \eqref{GeodApproxAdSAmbient} this result matches the geodesic distance on the AdS planar black hole found in \cite{RodriguezGomez2021,RodriguezGomez2021a}.

\subsection{The ambient 2-point function} \label{SSec:BBAinv}  \label{SSec:AcorrBB}

After finding the geodesic trajectories and $\X_{12}$ we turn to the curvature invariants. As a first step we are interested in writing the ambient 2-point function \eqref{eq:ProposalForm1} up to second order in the ambient Riemann. The homothetic vector $T$ can be parallel transported along the perturbative geodesics we are considering order by order in $z_H^{-4}$ taking the form
\begin{equation}
    \hat{T}=  \hat{T}^{(0)} + \sum_{n=1}^{\infty} \frac{\hat{T}^{(n)}}{z_H^{4n}}\, ,
\end{equation}
where $\hat{T}^{(0)}$ is the homothetic vector  \eqref{eq:TransportedTOnMink} transported on the flat ambient space. 
Since $\mathcal{I}_2^{(1)}=0$, it is sufficient to use $\hat{T}^{(0)}$ for invariants up to second order in the Riemann since higher $\hat{T}^{(n)}$'s contribute at order $O(\Riem)^3$ in contractions of the form \eqref{eq:CurvatureInvariantsForm2}.

We now turn to determining a basis of ambient invariants quadratic in the curvature. In principle one could pick them to be of any scaling dimension and then multiply them by the appropriate power of $\X_{12}$. As we discussed in 
section \ref{SSec:BuildingBlocks} invariants with vanishing scaling dimension are particularly rigid in their structure and thus easy to completely classify. Their general form is given in equation \eqref{eq:Generic0weightCurvInv} and if we restrict to $k=2$ curvature tensors, one can show that in the present setup there are only three independent such invariants of order $O(z_H^{-8})$. One possible choice is 
\begin{subequations}\label{BB_2ndOrderCurvatureInvariants}
	\begin{align}
		e_0 &= \mathcal{R}^{(0)}_{AC} \, \mathcal{R}^{(0) AC}   = \frac{3}{4} \frac{|x|^8}{z_H^8} + O \left( \frac{|x|}{z_H} \right)^{12},\\
		e_1 &=  \mathcal{R}^{(1)}_{AC}\, \mathcal{R}^{(0)AC}   = - \frac{|x|^6}{z_H^8} (3\tau^2+7x^2) + O \left( \frac{|x|}{z_H} \right)^{12}, \\
		e_2 &=  \mathcal{R}^{(1)}_{AC}\, \mathcal{R}^{(1)AC}  =
		4 \frac{|x|^4}{z_H^8} (3\tau^4 + 16 \tau^2 x^2 + 17x^4) + O \left( \frac{|x|}{z_H} \right)^{12},
	\end{align}
\end{subequations}
where the superscripts refer to the number of covariant derivatives required to construct them, with the tensors $\mathcal{R}^{(r)}$ defined in \eqref{mathcalRweight0inv}. In these expressions we have already taken the limit from generic ambient points to the CFT background on the nullcone. As we detail in appendix \ref{App:DetailsCurvatureInvFiniteT} any curvature invariant quadratic in the ambient Riemann can be obtained as a linear combination of the form
\begin{equation}
    {\cal I}_2^{(2)} = c_0 e_0 + c_1 e_1 + c_2 e_2 \,.
\end{equation}

Putting all together, we assemble the ambient scalar 2-point function for operators of scaling dimension $\Delta$ as prescribed by equation \eqref{eq:ProposalForm1},
\begin{align} \label{BBambientPrediction}
&\braket{O(\tau, x) O(0)}^{(\beta)}_{d=4,\Delta} =
\frac{C_\Delta}{|x|^{2\Delta}}\!\Bigg[ 1
-\frac{ \Delta  \left(x^2-3 \tau ^2\right) \!|x|^2}{120 \pi^{-4} \beta^4} +\frac{ |x|^4}{ \pi^{-8}\beta^8} \! \bigg[
\frac{3}{4}\!\left(\!c_0  + \frac{\Delta  (63 \Delta +170)}{30240}\!\right) \! |x|^4  \nonumber
\\
&  - \left(c_1  + \frac{\Delta  (14 \Delta +39)}{25200}\right) |x|^2 (3\tau^2\!+\!7x^2) + 4 \!\left(c_2 + \frac{\Delta  (7 \Delta +20)}{201600}\right)  (3\tau^4 \!+\! 16 \tau^2 x^2\! +\! 17x^4) \bigg] \nonumber \\
& + O \left( \frac{|x|}{\beta} \right)^{12}\Bigg].
\end{align}
Here the constants $c_i$ are to be fixed by the dynamics of the specific thermal CFT, and they quantify the quantum corrections to the semi-classical geodesic approximation as discussed in section \ref{SSec:ProposalA2ptf}.

\subsection{Matching with the thermal OPE} \label{SSec:MatchThOPE}

As reviewed in section \ref{SSec:ProposalA2ptf}, the OPE is expected to converge for CFTs on generic backgrounds and states for short enough distances. It can  thus be used to reduce 2-point functions to a sum over 1-point functions as in equation \eqref{2-ptfGeneralOPE}. Specialising to thermal CFTs on flat space it has been argued in \cite{Katz:2014rla, Witczak-Krempa:2015pia, Iliesiu2018} (see also \cite{Gobeil:2018fzy,Fitzpatrick2019, Li:2019tpf, Karlsson:2021duj,Marchetto:2023fcw} for related discussions) that for a distance between insertions shorter than the thermal radius $|x| < \beta$ one can expand a scalar correlation function of two operators of dimension $\Delta$ as 
\begin{equation} \label{eq:ThOPE}
    \braket{O(\tau,x)O(0)}^{(\beta)}_{d,\Delta} = 
    \sum_{\phi \,\in \,O\times O} \frac{a_\phi}{\beta^{\Delta_\phi}} \, C_J^{(\nu)} (q) |x|^{\Delta_\phi-2\Delta}\,,
\end{equation}
where $J$ and $\Delta_\phi$ are the spin and scaling dimension of the exchanged operator $\phi$, and we defined $\nu = \frac{d}{2}-1$ and $a_\phi = \frac{f_{ O O \phi} b_\phi}{c_\phi}  \frac{J}{2^J (\nu)_{J}}$. Here $c_\phi$ and $f_{O \phi \phi}$ are the 2- and 3-point function coefficients on flat space, while $C_J^{(\nu)} (q)$ are Gegenbauer polynomials of the dimensionless ratio $q = \tau/|x|$.

The products $C_J^{(\nu)} (q) |x|^{\Delta_\phi-2\Delta}$ can be thought of as thermal conformal blocks. The fact that in thermal CFTs 1- and 2-point functions contain non-trivial dynamical data through the coefficients $a_\phi$ mirrors the freedom in the coefficients $c_i$ appearing in the ambient 2-point function \eqref{BBambientPrediction}. In this subsection we would like to make this connection more precise by relating the coefficients $a_\phi$ with the $c_i$. 

As anticipated in section \ref{SSec:ProposalA2ptf} we expect  the ambient curvature invariants to account for the multi-energy-momentum tensor contributions  :$ T^n $:$\,$. These operators are defined as the $n+1$ symmetrised traceless partial contractions of tensor products of $n$ energy-momentum tensors, with scaling dimensions $n d$ in $d$ dimensions and even spins ranging from $J=0$ to $J=2n$. Their contribution takes the form
\begin{equation} \label{ThermalCorrOPETn}
\braket{O(\tau,x)O(0)}^{(\beta)}_{d,\Delta} \supset 
    \sum_{n=0}^\infty \sum_{\substack{J=0\\J\,\text{even}}}^{2n} a^{(T)}_{n,J} \, C^{(\nu)}_{J}(q) \frac{|x|^{n d-2\Delta}}{\beta^{n d}}.
\end{equation}
Comparing \eqref{BBambientPrediction} with \eqref{ThermalCorrOPETn}  to second order in the energy-momentum tensor yields the following dictionary re-expressing the thermal OPE coefficients in terms of the ambient free coefficients for any $\Delta$ in $d=4$,
\begin{align}
a_{0,0}^{(T)} &= C_\Delta, \qquad     a_{1,0}^{(T)} = 0, \qquad
    a_{1,2}^{(T)} = \frac{\Delta  }{120} C_{\Delta }, \label{OPEbeta4}\\
    a_{2,0}^{(T)} &= \left(\frac{3 c_0}{4}-6 c_1+52 c_2+\frac{\Delta  (7 \Delta +18)}{201600}\right) C_{\Delta },\label{OPEbeta8_1}\\
    a_{2,2}^{(T)} &= \left(c_1-15 c_2+\frac{\Delta  (7 \Delta +12)}{201600}\right) C_{\Delta },\label{OPEbeta8_2}\\
    a_{2,4}^{(T)} &= \left(c_2+\frac{\Delta  (7 \Delta +20)}{201600}\right) C_{\Delta }.\label{OPEbeta8_3}
\end{align}
Note once more that the ambient prediction at first order in the energy-momentum tensor is fully fixed by the geodesic distance factor $(\X_{12})^{-\Delta}$ as a consequence of $\mathcal{I}^{(1)}_2=0$. 

The relations \eqref{OPEbeta4}-\eqref{OPEbeta8_3} entail that to this order, ambient curvature invariants and thermal conformal blocks are two equivalent bases to describe multi-energy-momentum tensor contributions. This can be made more precise by mapping the thermal conformal blocks to the basis of curvature invariants $\{e_0,e_1,e_2\}$. After taking the large-$N$ limit in the CFT the multi-energy-momentum tensor VEVs factorise, $\braket{:\!T^n\!:} \sim \braket{T}^n$. Denoting the energy-momentum tensor VEV \eqref{BBthermalTij} by $T_{ij}$ to avoid cluttering, in terms of $T_{ij}$ the double-energy-momentum tensor VEVs with $J=0, 2, 4$ read,
\begin{subequations}
	\begin{align}
		\braket{T^2} &= T^{kl} T_{kl}, \\
		\braket{T^2}_{ij} &= T_{ik} T^{k}_{\;\;\,j} - \frac{1}{4}  T^{kl} T_{kl} \delta_{ij}, \\
		\braket{T^2}_{ijkl} &= \Sigma_{ijkl}
		- \frac{3}{4} \delta_{(ij} \Sigma_{kl)m}^{\;\;\;\;\;\;\;\; m}
		+\frac{1}{16} \Sigma^{m\;\;\;n}_{\;\;\;m\;\;\;n} \, \delta_{(ij} \delta_{kl)}
		,
	\end{align}
\end{subequations}
where we defined $\Sigma_{ijkl} = T_{(ij} T_{kl)}$. In terms of these, the second order curvature invariants can be written as
\begin{subequations}\label{eq:d=4InvTT}
	\begin{align}
	 64	\, e_0 &= 
		\braket{T^2} |x|^8 ,\\
	8 \,	e_1 &= 
		\braket{T^2}_{ij} x^i x^j |x|^6  -
		 \braket{T^2} |x|^8, \\
	4 \,	e_2 & = 
		\braket{T^2}_{ijkl} x^i x^j x^k x^l |x|^4
		-\frac{15}{2} \braket{T^2}_{ij} x^i x^j |x|^6 + \frac{13}{3} \braket{T^2} |x|^8.
	\end{align}
\end{subequations}
Thus the thermal conformal blocks at order $n=2$ in the large-$N$ limit are simply proportional to trace modifications of  the ambient invariants $e_i$. In appendix \ref{App:DetailsCurvatureInvFiniteT} we describe how to extend these conclusions to any order in the ambient Riemann and to other dimensions $d$. In particular we argue that the dimensionless invariants \eqref{eq:Generic0weightCurvInv} constructed as chains of tensors $\mathcal{R}^{(r)}$ form a basis for the contribution of generic multi-energy-momentum tensor operators :$T^n$: for thermal CFTs in even dimensions $d$.\footnote{As we discussed in section \ref{SSec:BuildingBlocks}, in odd $d$ divergences appear in the ambient Riemann in the limit $\rho\to 0$. This implies that some of the weight-0 scalars \eqref{eq:Generic0weightCurvInv} may diverge and other invariants must be used in addition to them. We will see this explicitly in section \ref{Sec:SqSphere}.}

At finite $N$, typically more operators take non-trivial VEVs and contribute in correlators, meaning that usually additional ambient invariants are required. However conformal blocks retain their form independently of the regime the theory is in, since they follow from kinematics and not from dynamics. In the thermal case this means that the thermal conformal blocks describing the multi-energy-momentum tensor contributions in \eqref{ThermalCorrOPETn} are the same Gegenbauer polynomials at any $N$. We have shown that multi-energy-momentum tensor conformal blocks are equivalent to the basis of ambient curvature invariants of the form \eqref{eq:Generic0weightCurvInv} at large $N$. We now conclude that this equivalence extends trivially to finite $N$: the ambient curvature invariants provide a basis for multi-energy-momentum tensor contributions in any thermal CFT. 

This represents evidence for the conjectured validity of the ambient formalism as a tool to solve the kinematics of generic CFTs. In this case it was possible to compare the ambient prediction with OPE computations and we found perfect agreement even for non-holographic CFTs. In section \ref{Sec:SqSphere} we consider CFTs on squashed spheres, where no OPE result is available and the ambient formalism produces genuinely new  predictions.

\subsection{Matching with a holographic correlator}
\label{SSec_BBHoloCorr}

In the previous subsection we showed that ambient 2-point functions account for the multi-energy-momentum tensor contributions to correlators in thermal CFTs (holographic or otherwise). We did this by comparing with the thermal OPE. 
In this section we check this statement through a holographic computation, without relying on the thermal OPE.

To this aim we will study holographic correlators
in the state dual to the Euclidean AdS$_{d+1}$ planar black hole with metric
\begin{equation} \label{eq:BBmetric}
	ds^2 = \frac{1}{z^2}\left[
	\frac{dz^2}{1-\frac{z^d}{z_H^d}}
	+ \left(1-\frac{z^d}{z_H^d}\right) d\tau ^2 + \delta_{ab}dx^a dx^b
	\right] .
 \end{equation}
The dual CFT is in the same background and state as the previous subsections, with inverse temperature $\beta = 4\pi z_H / d$ and energy-momentum tensor expectation value \eqref{BBthermalTij}.
This problem involves solving the free scalar equation 
\be \label{eq:BBKGeq}
\left[- \Box_{d+1} + \Delta(\Delta-d)\right] \Phi(z,\tau,\textbf{x})=0
\ee
on the fixed background \eqref{eq:BBmetric}, subject to Dirichlet conditions at the boundary $z \to 0$ and regularity conditions in the bulk interior $z \to \infty$. Here $\Delta$ is the scaling dimension of the operator whose 2-point function we wish to compute. Given a regular solution of \eqref{eq:BBKGeq} we can extract the 2-point function from its asymptotic expansion:
\begin{equation}
    \Phi(z, \tau, \textbf{x}) = z^{d-\Delta}\phi_{(0)}(\tau, \textbf{x}) \left(1 + \cdots +  \frac{z^{2 \Delta -d}}{(2 \Delta-d)}\braket{O(\tau,\textbf{x})O(0)} + \cdots\right)
\end{equation}
where $\phi_{(0)}(\tau, \textbf{x})$ is the boundary function that specifies the Dirichlet boundary condition (and by AdS/CFT the CFT source that couples to the operator $O$). 

In practice, it is easier to solve \eqref{eq:BBKGeq} after Fourier transforming to momentum space. Translational invariance along the boundary directions and periodicity along $\tau$ allow one to expand the scalar fields in terms of Fourier modes, \begin{equation}\label{BB_CompactFourierTransf}
	\Phi(z,\tau,\textbf{x}) = \sum_{m\in \Z} \int d^{d-1}\textbf{k} \, e^{i (\omega_m \tau + \textbf{k} \cdot \textbf{x})} B(z,\omega_m,\textbf{k}),
\end{equation}
where $\omega_m = 2 \pi m / \beta$ are the Matsubara frequencies. The momentum space correlator is defined similarly as
\begin{equation} \label{eq:FT_FS}
    \braket{O(\tau,\textbf{x})O(0)} = \sum_{m,n} \int d\textbf{k}_1\int d\textbf{k}_2 
    \braket{O(\omega_m,\textbf{k}_1) O(\omega_n, \textbf{k}_2)} e^{i \omega_m \tau + i \textbf{k}_1 \cdot \textbf{x}} \,,
\end{equation}
and translational invariance and orthogonality of Fourier coefficients imply that 
\begin{equation} \label{eq:mom}
    \braket{O(\omega_m,\textbf{k}_1) O(\omega_n, \textbf{k}_2)} = 
    \delta_{m,-n} \delta(\textbf{k}_1+\textbf{k}_2)
    \braket{OO}(\omega_m,\textbf{k})\, ,
\end{equation}
where $\braket{OO}(\omega_m,\textbf{k})$ is defined by this equation \footnote{
When $\tau$ is non-compact, we can Fourier transform also along this direction and the result is the same as in \eqref{eq:FT_FS} and \eqref{eq:mom} with $\omega_m, \omega_n$ replaced by continuous variables $\omega_1, \omega_2$,
the sum over $m, n$ replaced by integrals over $\omega_1, \omega_2$ and $\delta_{m,-n}$ by $\delta(\omega_1+\omega_2)$.}.

Defining $k=\sqrt{\omega_m^2 + \textbf{k}^2}$ and after rescaling the radial coordinate as $r= k z$, $r_H = k z_H$ and redefining $r_{H}=\epsilon^{-\frac{1}{d}}$, equation \eqref{eq:BBKGeq} reads
\begin{equation}
\begin{split}
    	 r \left(\epsilon  r^d-1\right) \left(r B''(r) \left(\epsilon  r^d-1\right)+B'(r) \left(\epsilon  r^d+d-1\right)\right)&\\
      +B(r) \left(\left(\Delta  (\Delta -d)+r^2\right) \left(\epsilon  r^d-1\right)- \frac{\omega_m^2}{k^2} \epsilon  r^{d+2}\right) &=0.
\end{split}
\end{equation}
where prime indicates derivative w.r.t. $r$, and we left the dependence of $B$ on $\omega_m$ and $\textbf{k}$ implicit.
This equation is of Heun type. We will first solve it perturbatively in the limit of short boundary distance  between the insertions (or equivalently at high momenta) with respect to the thermal radius, $\epsilon = (k z_H)^{-d} \ll 1$. In this regime we are able to compare with the expansion in the curvature of ambient correlators (and also double-check results from the thermal OPE). We will then turn to a fully non-perturbative numerical computation to further check the ambient correlator, as well as to study effects that may elude the perturbative analysis.

\subsubsection{Perturbative 2-point function} 
\label{SSSec_BBPert}

We set up the perturbative problem by expanding at $\epsilon \ll 1$ corresponding to large momenta $k \gg (z_H)^{-1}$,
\begin{equation}\label{}
	B(r)= \sum_{n=0}^\infty b_n(r) \epsilon^n \,.
\end{equation}
The equations for the first few orders read
\begin{subequations}
	\begin{align}
		\mathcal{D} \, b_0(r) &=0, \label{BBKG_LOeq}\\
		\mathcal{D} \, b_1(r) &=r^{d-2} \left[b_0(r) \left(\Delta  (\Delta -d)+\left(\eta ^2+1\right) r^2\right)+d r b_0'(r)\right], \label{BBKG_1stOeq}\\
		\mathcal{D} \, b_2(r) &=r^{d-2} \left[b_0(r) r^d \left(\Delta  (\Delta -d)+\left(2 \eta ^2+1\right) r^2\right)+b_1(r) \left(\Delta  (\Delta -d)+\left(\eta ^2+1\right) r^2\right) \right. \nonumber\\ 
		& \quad \left. +d r \left(r^d b_0'(r)+b_1'(r)\right)\right], \label{BBKG_2ndOeq}
	\end{align}
\end{subequations}
where we defined $\eta = \omega_m / k$ and the differential operator
\begin{equation}\label{BBhomogeneousDiffOp}
	\mathcal{D}= \p^2_r +\frac{1-d}{r} \p_r -\frac{\Delta  (\Delta -d)+k^2}{r^2}.
\end{equation}
The perturbative equations at a generic order $n$ reads
\begin{equation}
    \mathcal{D} \, b_n (r) = 
     \sum_{\ell=1}^n r^{\ell d -1}  
     \left[ 
     d b_{n-\ell}'(r) + r \left( \Delta(\Delta-d) + (1+\ell \eta^2) r^2 \right) b_{n-\ell}(r)
     \right].
\end{equation}
The solution to the  leading order equation \eqref{BBKG_LOeq} corresponds to a free scalar on Euclidean AdS$_{d+1}$. Defining $\kappa = \Delta - d/2$, if we assume $\kappa$ is not an integer\footnote{When $\kappa$ is an integer, the CFT correlator has short-distance singularities leading to conformal anomalies \cite{Petkou:1999fv}. On the bulk side, a different choice of basis for the solution space must be made because $u_1=u_2$ when $\kappa$ is an integer. Furthermore, logarithmic terms appear in the Fefferman-Graham near-boundary expansion \cite{Haro2000} and the present analysis must be modified.}, a possible choice for the basis of the solutions space is in terms of modified Bessel functions of the first kind,
\begin{equation}
u_1(r) = \sqrt{\frac{\pi}{2}} \, r^{\frac{d}{2}} \, I_{-\kappa} (r)\,, \qquad \quad  
u_2(r) = \sqrt{\frac{\pi}{2}} \, r^{\frac{d}{2}} \, I_{\kappa} (r)\,.
\end{equation}
Imposing regularity in the interior $r\to \infty$ fixes the leading order solution to a modified Bessel function of the second kind,
\begin{equation}\label{}
	b_0 = u_2-u_1 =  - \sqrt{\frac{2}{\pi}} \cos\!\left(\frac{2\kappa-1}{2} \pi\right)K_{\kappa} (r),
\end{equation}
recovering the expected solution on pure AdS (see e.g. \cite{Skenderis:2002wp}).

Solving the first order equation \eqref{BBKG_1stOeq} is more involved, and we refer the reader to appendix \ref{App:HoloCorrgeneraldD} for the details. The holographic correlator to first order in $\epsilon$ in momentum space results in\footnote{When $\kappa$ is an integer, the correlators develop poles and need renormalization. For example when $d=4$ and $\Delta=3$, so $\kappa=1$, the leading terms are \begin{equation}\label{}
	\braket{OO}^{(\beta)}_{d=4,\Delta=3} = \frac{k^2}{4 (\Delta -3)}+\frac{ k^2}{4} \left(\log \frac{ k^2}{4} +2 \gamma -1 +\frac{4}{5} \left(1-4 \omega_m ^2\right) \epsilon \right)+O\left(\Delta -3\right) + O(\epsilon^2).
\end{equation}
The divergent term in analytic in $k^2$, and thus a contact term in position space and it should be removed them using holographic renormalization \cite{Skenderis:2002wp}. 
As mentioned above, to simplify our presentation we restrict ourselves to cases where $\kappa$ is not an integer and such renormalization is not needed.} 
\begin{align} \label{BBKG_1stSOLGeneraldD}
    &\braket{OO}^{(\beta)}_{d,\Delta}(\omega_m,\textbf{k})	 =
		-\frac{2^{d-2 \Delta } \Gamma \left(\frac{d}{2}-\Delta +1\right) }{\Gamma \left(-\frac{d}{2}+\Delta +1\right)} k^{2 \Delta -d}  \Bigg[\, 1 \, +   \\
		& \quad \frac{\pi ^{3/2+d} (-1)^{d+1}  \cot \left(\frac{\pi  d}{2}\right) \Gamma \left(-\frac{d}{2}-\frac{1}{2}\right) \csc ^2(\pi  \Delta ) \sin \left(\frac{1}{2} \pi  (d-2 \Delta )\right)  \left(k^2-d \omega_m ^2\right)}{4 \Gamma \left(1-\frac{d}{2}\right) \Gamma (-\Delta ) \Gamma (\Delta -d)  \,k^{d+2} \beta^d}  + O(\epsilon^2) \Bigg].\nonumber
	\end{align}
 $\braket{OO}(\omega_m,\textbf{k})$ is defined in \eqref{eq:FT_FS}, \eqref{eq:mom},  
 the superscript $\beta$ indicates that this is a finite temperature correlator and the subscripts are the spacetime dimension $d$ and 
 the dimension of the operator, $\Delta$.

Solving the second order equation \eqref{BBKG_2ndOeq} and higher is particularly involved for general $d$ and $\Delta$. A simplification happens when $2\Delta+d$ is integer ({\it i.e.} $\kappa$ is half-odd). In this case the homogeneous solutions can be written in terms of products of polynomials and exponentials since
\begin{equation}\label{}
\begin{split}
		I_{\kappa} (r) &= \sqrt{\frac{2}{\pi}} \, i^{\kappa-\frac{3}{2}} r^{\kappa} \left( \frac{1}{r} \frac{\text{d}}{\text{d}r}\right) ^{\kappa-\frac{1}{2}} \frac{\sinh r}{r}, \\
         I_{-\kappa} (r) &= \sqrt{\frac{2}{\pi}} \, i^{\kappa-\frac{3}{2}} r^{\kappa+1} \left( \frac{1}{r} \frac{\text{d}}{\text{d}r}\right)^{\kappa+\frac{1}{2}} \frac{\cosh r}{r} \,.
\end{split}
\end{equation}
 This observation allows one to find a case-by-case solution to arbitrarily high order in the inverse temperature.  Considering for simplicity $\frac{d}{2} \leq \Delta \leq d$, the generic form of such momentum space correlators is
 \begin{equation} \label{BB_MomentumSpaceGeneric2pf}
     \braket{OO}^{(\beta)}_{d,\Delta}(\omega_n,\textbf{k}) =      \frac{1}{k^{d-2\Delta}} \sum_{q=0}^\infty \frac{\pi^{qd}}{k^{q(d+2)} \beta^{q d}} \sum_{j=0}^q 
     \alpha_j^{(q)} \, \textbf{k}^{2q-2j} \omega_n^{2j},
 \end{equation}
where the coefficients $\alpha^{(q)}_j$ are given in terms of $d$ and $\Delta$. Up to second order it reads explicitly
 \begin{equation} \label{eq:HoloBBMomentumSpaceGeneric}
	\braket{OO}^{(\beta)}_{d,\Delta}(\omega_m,\textbf{k})	  = \frac{\alpha_0^{(0)}}{k^{d-2\Delta}} + \frac{ \alpha_0^{(1)} \textbf{k}^2+\alpha_1^{(1)}\omega_m ^2}{ \pi^{-d} \, k^{2d-2\Delta +2} \beta^d} + 
	\frac{\alpha_0^{(2)} \textbf{k}^4 + \alpha_1^{(2)} \textbf{k}^2 \omega_m^2 + \alpha_2^{(2)} \omega_m^4}{\pi^{-2d} \, k^{3d - 2\Delta +4}\beta^{2d}} +O\left(k \beta\right)^{-3d}.
\end{equation}
 The coefficients $\alpha^{(0)}_0, \alpha^{(1)}_0$ and $\alpha^{(1)}_1$ for generic $d$ and $\Delta$ can be extracted from  \eqref{BBKG_1stSOLGeneraldD}; in particular, $\alpha^{(1)}_1 = (1-d)\alpha^{(1)}_0$. As an example, for $d=4$ and $\Delta=\frac{3}{2}$ the first few coefficients take the values
\begin{equation}
    \alpha^{(0)}_0=-1, \qquad \alpha^{(1)}_0 = -\frac{3}{16} , \qquad \alpha^{(1)}_1 = \frac{9}{16},  
\end{equation}
\begin{equation}
    \alpha^{(2)}_0 = -\frac{2637}{512}, \qquad \alpha^{(2)}_1 = \frac{11511}{256} , \qquad \alpha^{(2)}_2 = -\frac{10773}{512} .
\end{equation}

Transforming momentum space correlators of the form \eqref{eq:HoloBBMomentumSpaceGeneric} back to position space is subtle since the Fourier transform should be performed over all real momenta and all Matsubara frequencies, while the expression for the correlators we have found is only valid  at large frequencies $\omega_m \gg 1/\beta$. To explicitly perform the Fourier transform one should thus resum the perturbative expansion to assess the full dependence on $k$ and $\omega_m$. 
It is particularly hard to directly compute the Fourier series in \eqref{BB_CompactFourierTransf} for a correlator of the form  \eqref{BB_MomentumSpaceGeneric2pf}. 

To bypass this difficulty we will use the fact that a periodic function $F(\tau)$ may be viewed as the sum of images of an aperiodic function $f(\tau)$, $F(\tau)= \sum_{m\in \Z} f(\tau+m \beta)$, where $\beta$ is the period. Then the Fourier series of $F(\tau)$ may be expressed as a sum of images of the Fourier transform of $f(\tau)$ \footnote{
Manipulations of this type are used in the proof of the Poisson summation formula, see for example \cite{Pinsky2001}.}. 
In our context, 
\begin{align} \label{eq:sum_of_images}
 \braket{O(\tau,x)O(0)}^{(\beta)}_{d,\Delta}	&=  \sum_{m\in \Z} \braket{O(\tau+ m \beta ,x)O(0)}^{(nc)}_{d,\Delta}\, , 
\end{align}
where the thermal correlator 
$\braket{O(\tau,x)O(0)}^{(\beta)}_{d,\Delta}$ is the periodic function, 
and the aperiodic function is the holographic 2-point function,
$\braket{O(\tau ,x)O(0)}^{(nc)}_{d,\Delta}$, obtained by starting from \eqref{eq:BBmetric} but with the $\tau$ coordinate non-compact (the superscript $nc$ stands for ``non-compact'').
This geometry with $\tau$ non-compact is singular at $z=z_H$ (because of lack of periodicity of $\tau$) so the corresponding holographic correlators are not physical. Also note that they are not the zero temperature CFT correlators. The correlator $\braket{O(\tau ,x)O(0)}^{(nc)}_{d,\Delta}$ is useful however because it is computable and as we argue in section \ref{SSSec_Double_Twists} it captures the multi-energy-momentum contributions.
More precisely, the sum over images does not affect the  multi-energy-momentum tensor contributions in position space as long as $\kappa$ is not an integer. Thus knowing the non-compact correlator is sufficient to match the multi-energy-momentum tensor contributions with the predictions from the ambient space.\footnote{In section \ref{SSSec_Double_Twists} we discuss the mixing between the multi-energy-momentum tensor and the so-called double-twist spectra for integer $\kappa$.}

We therefore proceed to compute the  Fourier transform of the momentum space correlator \eqref{eq:HoloBBMomentumSpaceGeneric} (with $\omega_m$ replaced by the continuous variable $\omega$) order by order in the perturbative expansion to obtain the non-compact correlator $\braket{O(\tau ,x)O(0)}^{(nc)}_{d,\Delta}$. 
Recall that the Fourier transform of a spherically symmetric distribution in momentum space reduces to a Hankel transform,
\begin{equation}\label{}
	\begin{split}
	    F(x) &= \int  d^dp f(|p|) e^{i p \cdot x}
	= \frac{(2\pi)^{\frac{d}{2}}}{|x|^{\frac{d}{2}-1}}
	\int^\infty_0 dp f(p) J_{\frac{d}{2}-1}(|x| p) p^{\frac{d}{2}} \\ 
    &= \frac{(2\pi)^{\frac{d}{2}}}{|x|^{\frac{d}{2}-1}} \, 
	H_{\frac{d}{2}-1}\left[p^{\frac{d}{2}-1} f(p)\right](x) \,.
	\end{split}
\end{equation}
The Fourier transform of each order in \eqref{eq:HoloBBMomentumSpaceGeneric} can thus be rewritten as a linear combination of derivatives of Hankel transforms. Defining the integral
\begin{equation}\label{}
	I_{\gamma}(\tau,x)\equiv\int d^dp |p|^\gamma e^{i p \cdot x} = \frac{(2\pi)^{\frac{d}{2}}}{|x|^{\frac{d}{2}-1}} 
	H_{\frac{d}{2}-1}\left[p^{\gamma+\frac{d}{2}-1}\right](x)
	= \frac{\pi ^{d/2} 2^{\gamma +d} \Gamma \left(\frac{d+\gamma }{2}\right) }{\Gamma \left(-\frac{\gamma }{2}\right)} \frac{1}{|x|^{\gamma +d }}\,,
\end{equation}
one can rewrite the first few orders of the correlator in position space as 
\begin{align}
    \braket{O(\tau,x)O(0)}^{(nc)[0]}_{d,\Delta} &=  \alpha^{(0)}_0 \, I_{2 \Delta- d} \, , \label{BBKG_Hankel0}\\
    \braket{O(\tau,x)O(0)}_{d,\Delta}^{(nc)[d]} &=  \alpha^{(1)}_0 \left[
	I_{2\Delta-2d}  - d \left(\frac{\p_\tau}{i}\right)^2  I_{2\Delta-2d-2} 
	\right], \\
	\braket{O(\tau,x)O(0)}_{d,\Delta}^{(nc)[2d]} &=  \alpha^{(2)}_0 \,
		I_{2\Delta-3d} + ( \alpha^{(2)}_1 - 2  \alpha^{(2)}_0) \left(\frac{\p_\tau}{i}\right)^2  I_{2\Delta-3d-2} \nonumber \\
  &  \quad +  ( \alpha^{(2)}_2-  \alpha^{(2)}_1 + \alpha^{(2)}_0) \left(\frac{\p_\tau}{i}\right)^4  I_{2\Delta-3d-4}, \label{BBKG_Hankel2}
    \end{align}
where the superscripts in square brackets indicate the order.    
These relations can be straightforwardly obtained to arbitrarily high order.

Using these expressions on \eqref{BBKG_1stSOLGeneraldD} one finds the position space correlator at general $d$ and $\Delta$ to first order. Normalising the operators so that the leading order constant is normalised to 1, it reads
\begin{equation}\label{} \label{BBKG_1stSOLGeneraldD_positionspace}
\braket{O(\tau,x)O(0)}^{(nc)}_{d,\Delta} = \frac{1}{|x|^{2\Delta}}  \left[
	1 + \tilde{\lambda}_1 
	\left(x^2\!-\!(d-1) \tau ^2\right)\!
	\frac{|x|^{d-2}}{\beta^d}
	\right] + O \left(\frac{|x|}{\beta}\right)^{2d} ,
\end{equation}
with
\begin{equation}
    \tilde{\lambda}_1 = \left(\frac{4\pi}{d}\right)^d\frac{\sqrt{\pi } (-1)^{d+1} \Delta   \, \Gamma \left(-\frac{d}{2}-\frac{1}{2}\right) \sin (\pi  (d-\Delta ))}{2^{d+2} \Gamma \left(1-\frac{d}{2}\right) \tan \left(\frac{\pi  d}{2}\right) \sin (\pi  \Delta ) }.
\end{equation}
One can check that this expression matches both the geodesic approximation and the ambient correlator \eqref{BBambientPrediction} to first order in $\beta^{-d}$ for $d=4$. This result hence substantiates the universality of the geodesic approximation for the energy-momentum tensor contribution as predicted by the ambient space formalism.

Furthermore, using \eqref{BBKG_Hankel0}-\eqref{BBKG_Hankel2} on \eqref{eq:HoloBBMomentumSpaceGeneric} one is able to check for odd $2\Delta+d$ that the higher orders in $|x|/\beta$  of such position space correlators can be decomposed in terms of the ambient curvature invariants. For instance, for $d=4$ and $\Delta=\frac{3}{2}$ the correlator up to second order reads
\begin{align}\label{BB2ptd4D32}
    	\braket{O(\tau,x)O(0)}^{(nc)}_{d=4,\Delta  =\frac{3}{2}} &= \frac{1}{|x|^3} \left[
	1- \pi^4 \frac{|x|^2 \left(x^2-3 \tau ^2\right) }{80 \, \beta^4} \right. \\
 & \qquad \quad \;\; \left. - \pi^8 \frac{|x|^4 \left(479 \tau ^4-1162 \tau ^2 x^2+199 x^4\right)}{268800 \, \beta^8}+O\left(\frac{|x|^{12}}{\beta^{12}}\right) 
	\right], \nonumber
\end{align}
which fixes the coefficients in \eqref{BBambientPrediction} to
\begin{equation}
  c_0=  -\frac{53}{1575}, 
  \qquad c_1= -\frac{11}{1120},
  \qquad c_2= -\frac{11}{16800}\, .
\end{equation}
This brings further evidence that the ambient curvature invariants form a basis for the multi-energy-momentum tensor spectrum and it confirms the ambient prediction \eqref{BBambientPrediction}. 
Via the relations between ambient and thermal OPE coefficients  \eqref{OPEbeta4}-\eqref{OPEbeta8_3}, this also represents a non-trivial check of the expansion in terms of thermal conformal blocks \eqref{ThermalCorrOPETn} in a non-trivial thermal state, in particular beyond the large-$\Delta$ regime studied in \cite{RodriguezGomez2021,RodriguezGomez2021a} and to arbitrarily high order in $|x|/\beta$.

\subsubsection{Non-perturbative 2-point function} \label{SSSec_BBNonPert}

In this subsection we  discuss possible non-perturbative effects in $|x| / \beta \to 0$ entering the thermal holographic correlator on the planar black hole background. In momentum space such contributions can be studied along the lines of \cite{Dodelson:2022yvn,Dodelson:2023vrw}, at least perturbatively in the instanton number. We are however interested in the correlator in position space and for this purpose we resort to a numerical calculation, fully non-perturbative in the boundary temperature.

Since we are not working perturbatively in $|x| / \beta$, to compute the position space two-point function we must solve \eqref{eq:BBKGeq} on the Euclidean cigar geometry with period $\beta$. The boundary conditions are a delta-function source at $\tau = |\vec{x}| = 0$ and we demand regularity in the interior. With the Euclidean time circle $\tau$, the holographic radial direction $z$, and noting  a rotational symmetry in the spatial boundary directions $x^i$, this leaves a 3d PDE problem. Without loss of generality we set $z_H = 1$ so that $\beta = \pi$. Next, we make the following coordinate changes,
\be
 z = 1 - \rho^2, \qquad \tau = \frac{1}{2}\phi, \qquad |\vec{x}| = \frac{R}{1-R^2}. \label{newvars}
\ee
In these coordinates we have $\rho \in [0,1]$ where $\rho =0$ is the tip of the Euclidean cigar geometry and $\rho =1$ is the conformal boundary, $\phi = (0,2\pi]$ is the angle around the thermal circle, and $R\in [0,1)$ where $R=0$ is the origin of spatial coordinates on the boundary and $R=1$ is the compactification of spatial infinity. 

The principal numerical challenge is handling the delta function source at the origin on the boundary. We subtract a function from $\Phi$ with the correct singularity structure, i.e. we define a new field $\Psi$ via,
\be
\Phi = \Psi + \tilde{G}_{AdS}
\ee
where $\tilde{G}_{AdS}$ is an analytically known function containing the correct source behaviour. A candidate function is the vacuum AdS bulk-boundary propagator,
\be
\frac{z^\Delta}{(\tau^2 + r^2 + z^2)^\Delta} = \frac{(1-\rho^2)^\Delta}{\left(\frac{\phi^2}{4} + (1-\rho^2)^2 + \frac{R^2}{(1-R^2)^2}\right)^\Delta}\,,
\ee
which however is not periodic in $\phi$. To address this we make the replacement
\be
\phi^2 \to \frac{2}{3} (7-\cos(\phi)) \sin\left(\frac{\phi}{2}\right)^2.
\ee
The resulting function $\tilde{G}_{AdS}$ is then periodic $\phi \sim \phi + 2\pi$,  contains no additional singularities, and is regular in the interior. Hence to find the 2-point function we now need to solve,
\be
\left(\Box - \Delta(\Delta-d)\right) \Psi  = -\left(\Box - \Delta(\Delta-d)\right) \tilde{G}_{AdS}\,, \label{KGsourced}
\ee
where $\Psi$ obeys a Dirichlet zero boundary condition at the conformal boundary, and is also regular in the interior. 

We work with $\Delta = 5/2$, so that the near boundary behaviour of $\Psi$ is,
\be
\Psi = a(\tau,r) z^\frac{3}{2} + b(\tau,r) z^\frac{5}{2} + \ldots = a(\tau,r) (1 - \rho^2)^\frac{3}{2} + b(\tau,r) (1 - \rho^2)^\frac{5}{2}  + \ldots
\ee
To this order the $z$ expansion is equivalent to the Fefferman-Graham expansion. Note that 
\be
\Phi = a(\tau,r) z^\frac{3}{2} + \left(b(\tau,r) + \frac{1152 \sqrt{6}}{\left(15 + 24 r^2 - 16 \cos(2\tau) + \cos(4\tau)\right)^\frac{5}{2}}\right) z^\frac{5}{2} + \ldots \label{psiFG}
\ee
We define
\be
\Psi = (1 - \rho^2)^\frac{3}{2} H\,,
\ee
enforce $a=0$ through a Dirichlet boundary condition $H_{\rho = 1} = 0$, and read off $b$ from the solution as $b = \partial_\rho H\big|_{\rho =1}$. The two point function is then given by the data $b$, corrected by the subtracted function,
\be
\left<O(\tau,r)O(0,0)\right>_{4,5/2}^{(\beta)} = b(\tau,r) + \frac{1152 \sqrt{6}}{\left(15 + 24 r^2 - 16 \cos(2\tau) + \cos(4\tau)\right)^\frac{5}{2}}. \label{twopointG}
\ee
For the rest of the problem we enforce tip of the cigar regularity with $\partial_\rho H\big|_{\rho=0} = 0$,  origin regularity on the boundary with $\partial_R H\big|_{R=0} = 0$, and at spatial infinity on the boundary the response to the delta should vanish, so we also set $H\big|_{R=1} = 0$. 

The PDE is discretised using a grid of $N_\rho, N_\phi, N_R$ points in the $\rho,\phi,R$ directions respectively. We utilize Chebyschev collocation in $
\rho$ with second-order finite difference methods for $\phi$ and $R$. This discretisation of \eqref{KGsourced} give rise to a linear problem
\be
M H = S
\ee
where $M$ is a matrix of size $(N_\rho N_\phi N_R)^2$ and $S$ is a vector of size $N_\rho N_\phi N_R$. We then solve for $H$, read off $b = \partial_\rho H\big|_{\rho =1}$ and compute $\left<O(\tau,r)O(0,0)\right> $ using \eqref{twopointG}.

\begin{figure}[t!]
	\begin{center}
		\includegraphics[width=\textwidth]{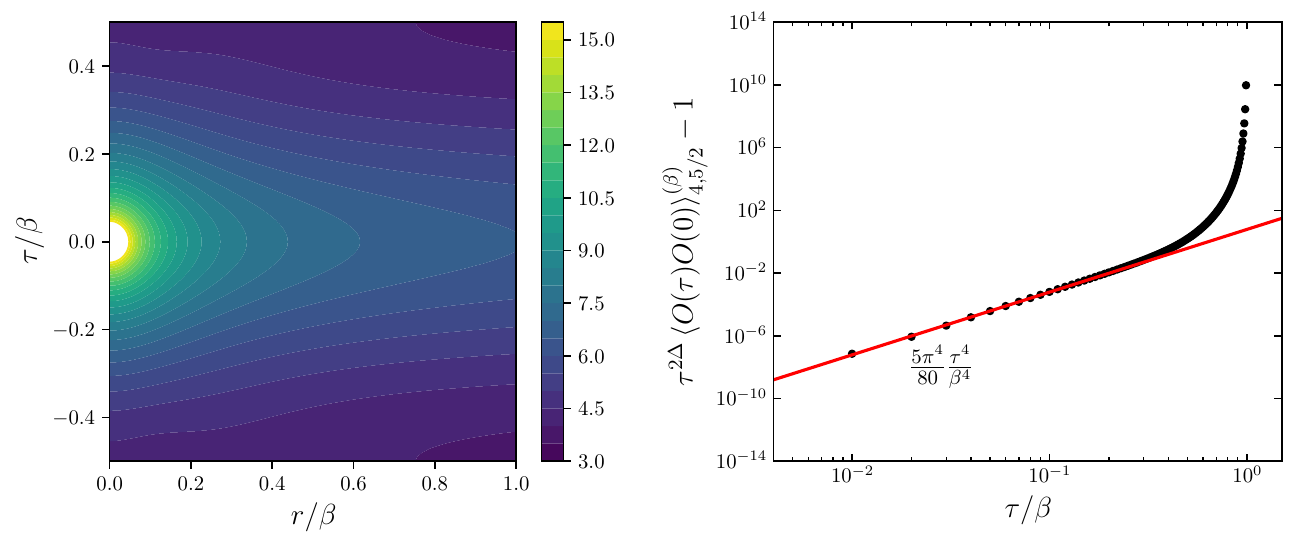}
		\caption{Non-perturbative thermal 2-point function for $d=4$, $\Delta = 5/2$ from holography. \textbf{Left:}. Contour plot of the 2-point function over the full range of the thermal circle. The colour scale is logarithmic, corresponding to $\log\left(\beta^{5}\left<O(\tau,r)O(0,0)\right>_{4,5/2}^{(\beta)}\right)$. \textbf{Right:} Showing a log-log plot to illustrate the leading behaviour at $x^i=0$ near $\tau = 0$ (black dots). The power-law behaviour is consistent with the analytically derived energy-momentum tensor contribution (red line).
  \label{fig:numerics_5over2}}
	\end{center}
\end{figure}

The results at $d=4$, $\Delta = 5/2$  are shown in figure \ref{fig:numerics_5over2}. In particular, the behaviour of the 2-point function in the limit $x\to0$ is consistent with the prediction of the ambient formalism \eqref{BBambientPrediction} and with  the perturbative holographic value  \eqref{BBKG_1stSOLGeneraldD_positionspace}.

\begin{figure}[t!]
	\begin{center}
		\includegraphics[width=\textwidth]{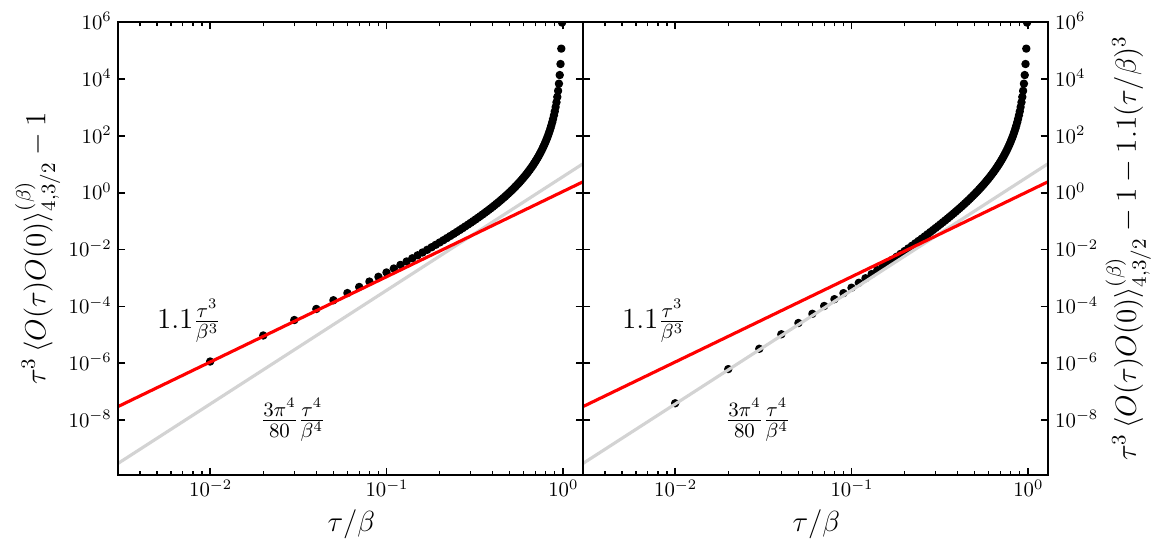}
    \caption{Non-perturbative thermal 2-point function for $d=4$, $\Delta = 3/2$ from holography. Showing a log-log plot to illustrate the leading behaviour at $x^i=0$ near $\tau = 0$ (black dots). \textbf{Left:} With the leading conformal behaviour subtracted, the remaining power-law at short distances is consistent with the leading term in the double-twist spectrum, $a_{0,0}^{(OO)}$ (red line). \textbf{Right:} Making a further subtraction to remove the leading double-twist contribution reveals the analytically derived energy-momentum tensor contribution (grey line).\label{fig:numerics_3over2}}
	\end{center}
\end{figure}

In Figure \ref{fig:numerics_3over2} we show the results for $d=4$, $\Delta = 3/2$. The behaviour of the first subleading term in $\tau\to0$ differs from that expected for the single energy-momentum tensor block, and it is compatible with the exchange of an operator of  dimension $\Delta=3$. This suggests the appearance of  the operator :$OO$: belonging to the so-called double-twist spectrum. These are operators of the schematic form :$O\Box^n\p_{i_1} \dots \p_{i_J}O$:, symmetric and traceless in the $J$ indices. They are primaries with scaling dimensions $\Delta_{p,J} = 2 \Delta + 2p + J$ and even spin $J$. Given their dimensions and tensorial properties, double-twist operators appear in the thermal OPE \eqref{eq:ThOPE} with blocks of the form,
\begin{equation} \label{eq:ThOPEDoubleTwists}
    \braket{O(\tau,x)O(0)}^\beta_{d,\Delta} \supset
    \frac{1}{\beta^{2\Delta}} \sum_{p=0} ^\infty \sum_{\substack{J=0 \\ J \text{even}}}^\infty 
    a_{p,J}^{(OO)} \, C_J^{(\nu)} (q)
    \left(\frac{|x|}{\beta}\right)^{2p+J}.
\end{equation} 
In the limit $x\to0$ the $n=0, J=0$ block precisely reproduces the scaling in $\tau$ displayed in Figure \ref{fig:numerics_3over2}, and our non-perturbative computation thus makes a prediction for the dynamical OPE coefficient
\begin{equation}
    a_{0,0}^{(OO)} \simeq 1.1 \,.
\end{equation}
We discuss the appearance of the double-twist spectrum and its non-perturbative nature  at length in section \ref{SSSec_Double_Twists}.

In Figure  \ref{fig:numerics_3over2} we also show the second-subleading behaviour which we recognise as the energy-momentum tensor block. Also in this case the value of its coefficient is compatible with the ambient prediction \eqref{BBKG_1stSOLGeneraldD_positionspace}, thus confirming the ambient prediction about the exactness of the energy-momentum tensor coefficient at the non-perturbative level. The numerical value of the coefficient also matches the  perturbative analytic correlator \eqref{BBKG_1stSOLGeneraldD_positionspace}, supporting the claim that multi-energy-momentum tensors do not receive non-perturbative corrections (as long as $\kappa$ is not integer).

\subsubsection{On the double-twist spectrum}
\label{SSSec_Double_Twists}

In section \ref{SSSec_BBNonPert} we saw that the exact thermal scalar 2-point function in position space contained terms consistent with double-twist operators in the thermal OPE, \eqref{eq:ThOPEDoubleTwists}. Such contributions were previously argued to arise in \cite{Iliesiu2018,Gobeil:2018fzy,Karlsson:2021duj,Fitzpatrick2019, Fitzpatrick2019a, Li:2019tpf}. However in perturbation theory in $\beta^{-1}$ double-twist contributions are not seen, for example \eqref{BBKG_1stSOLGeneraldD_positionspace}, \eqref{BB2ptd4D32} and the ambient space result, \eqref{BBambientPrediction}. In this section we discuss this shortcoming of the perturbative approach.

In Euclidean signature there is a claim  that double-twist contributions cannot appear as perturbative terms in an expansion of a momentum space 2-point function of the form \eqref{BB_MomentumSpaceGeneric2pf} \cite{Manenti:2019wxs}. That discussion however is based on Fourier transforming the thermal conformal blocks in \eqref{eq:ThOPE} to momentum space, and this is subtle on at least two grounds. First, the Fourier transform is performed over the full thermal cylinder $S^1 \times \R^{d-1}$, while the thermal OPE is convergent only for $|x|<\beta$. Second, OPEs do not capture contact terms and as the Fourier transform involves an integration over all space, the momentum space behaviour of the correlators even at short-distances may not be captured by the Fourier transform of the position space OPE, see \cite{Bzowski:2014qja}. One should thus view with caution the result in  \cite{Manenti:2019wxs}.
For Lorentzian thermal CFTs it has been argued in \cite{Dodelson:2023vrw} that the double twist spectrum arises by Fourier transforming non-perturbative contributions present in the momentum space 2-point function \cite{Festuccia:2005pi}.

To gain some intuition, consider the simple case of thermal AdS$_{d+1}$ in the bulk. The scalar 2-point function takes the form of a sum over images \eqref{eq:sum_of_images} of the correlator computed on Euclidean AdS (which consists solely of the identity block),  
\begin{equation} \label{eq:MFTcorr}
    \braket{O(\tau,x)O(0)}^{(\beta)}_{\Delta} =  \sum_{m \in \Z} \frac{1}{\left[ (\tau+ m \beta)^2 +x^2 \right]^{\Delta}}\,.
\end{equation}
Note that this sum over images is intrinsically non-perturbative in $|x| / \beta \to 0$, as $\beta$ is kept finite while the correlator on the non-compact bulk must be evaluated at parametrically large Euclidean time $\tau + m \beta$. In this case the sum over images can be carried out explicitly and gives rise to the double-twist spectrum.

Following this observation we now show how the double-twist spectrum arises by a sum over images of the non-compact correlator containing multi-energy-momentum tensor contributions as prescribed by \eqref{eq:sum_of_images}. Let us assume that the two insertions are separated along $\tau$ only. Following the discussion in section \ref{SSSec_BBPert}, the non-compact correlator contains only multi-energy-momentum tensor blocks, taking the form,\footnote{We comment later on how this discussion generalises when additional operators enter the OPE limit of the non-compact correlator.}
\begin{equation} \label{BB_NonCompCorrx0}
    \braket{O(\tau)O(0)}^{(nc)}_{d,\Delta} = \sum_{n=0}^\infty \, 
    \frac{a_{n}^{(T)}}{\beta^{2\Delta}} \left|\frac{\tau}{\beta} \right|^{nd-2\Delta},
\end{equation}
since in the limit $x\to 0$ the sum of the multi-energy-momentum tensor contributions of different spin at a given order $d$ reduces to a power of $\tau$ times a collective constant $a_{n}^{(T)}$. As an OPE this expression is valid in some interval $0 < |\tau| < \tau_\star$. 

In order to perform the image sum of \eqref{BB_NonCompCorrx0} it is convenient to analytically continue to the complex $\tau$ plane. Because of the absolute value in \eqref{BB_NonCompCorrx0}, we first focus on the case $\tau > 0$ where,
\begin{equation} \label{BB_NonCompCorrx0Positive}
    \braket{O(\tau)O(0)}^{(+)}_{d,\Delta} = \sum_{n=0}^\infty \, 
    \frac{a_{n}^{(T)}}{\beta^{2\Delta}} \left(\frac{\tau}{\beta} \right)^{nd-2\Delta}.
\end{equation}
We then subsequently continue to $\tau\in \mathbb{C}$ so that \eqref{BB_NonCompCorrx0Positive} is valid in an annulus $0 < |\tau| < \tau_\star$, with $\tau_\star$ corresponding to the smallest radius at which there will be singularities in the complex $\tau$ plane. We then attempt to analytically continue beyond $\tau_\star$, to a function which we denote $G_+(\tau)$. We assume that there are no singularities of $G_+(\tau)$ lying on the positive real axis, that is, the only singularity of the non-compact correlator is the one at coincident points.

To extend the range of validity of the sum \eqref{BB_NonCompCorrx0Positive} we take $G_+$ to be composed of singular and non-singular parts,
\begin{equation} \label{BB_NonCompCorrx0_Split}
    G_+(\tau) = \sum_{\ell} 
    W_\ell(\tau) + 
    \sum_{n=0}^\infty \, 
    \frac{\tilde{a}_{n}^{(T)}}{\beta^{2\Delta}} \left(\frac{\tau}{\beta} \right)^{nd-2\Delta},
\end{equation}
where the first sum includes all poles and branch points whose positions are governed by the parameters $y_\ell$,
\begin{equation}
    W_\ell(\tau) = \frac{1}{\tau^{2\Delta}} \frac{A_{(\ell)}}{
    \left(\left(\tau/\beta\right)^d - y_\ell\right)^{\mu_{(\ell)}}
    }\,,
\end{equation}
with non-negative real $\mu_{(\ell)}$.
The second sum in \eqref{BB_NonCompCorrx0_Split} has an infinite radius of convergence.

With $G_+(\tau)$ known, and the $\tau \to -\tau^\ast$ symmetry of the correlator, the sum over images \eqref{eq:sum_of_images} is given by
\bea 
\braket{O(\tau)O(0)}^{(\beta)}_{d,\Delta} &=& \sum_{m\in \mathbb{Z}}\braket{O(\tau+m\beta)O(0)}^{(nc)}_{d,\Delta},\\
&=& G_+(\tau) + \sum_{m =1}^\infty \left[G_+(\tau + m\beta) + G_+(-\tau + m\beta)\right].\label{BBKG-CompactF2SumImages_GPLUS}
\eea
In appendix \ref{AppendixDoubleTwists} we give a detailed account of how to perform this sum over images. The resulting thermal correlator arising from the non-compact correlator \eqref{BB_NonCompCorrx0_Split} reads for real $\tau$ in $0 < |\tau| < \tau_\star$,
\begin{equation} \label{BBKG_FullPosSpaceCorrThOPE}
    \braket{O(\tau)O(0)}^{(\beta)}_{d,\Delta} = \sum_{n=0}^\infty \, 
    \frac{a_{n}^{(T)}}{\beta^{2\Delta}} \left|\frac{\tau}{\beta} \right|^{nd-2\Delta} 
     + \frac{1}{\beta^{2\Delta}} \sum_{p=0}^\infty \, \left[ a_{\text{reg},\,p}^{(OO)} + \sum_\ell a^{(OO)}_{(\ell)\, p} \right] \,\frac{\tau^{2p}}{\beta^{2p}} \, ,  
\end{equation}
where we defined the coefficients
\begin{equation}
\label{BBKG_FullPosSpaceCorrThOPE_Coeff}
    a_{\text{reg},\, p}^{(OO)} = 
    2 
    \sum_{n=0}^\infty \frac{\Gamma(2p+2\Delta-nd)}{(2p)! \, \Gamma(2\Delta-nd)} \, \zeta(2p+2\Delta-nd) \, \tilde{a}_n^{(T)}\,,
\end{equation}
\begin{align}\label{DoubleTwists_aOOlp}
     &a^{(OO)}_{(\ell)\, p} =  \\
     &\frac{2 A_{(\ell)}}{(2p)!}
     \sum_{j=0}^\infty  \binom{\mu_{(\ell)}+j-1}{j} \Bigg[ 
     (y_\ell )^j \, \big( d(\mu_{(\ell)} +j ) + 2\Delta \big)_{2p} \, \zeta\big(2p +d(\mu_{(\ell)} +j ) + 2\Delta , n^*_{\ell} \big) \nonumber \\
     &  + (-1)^j (-y_{\ell})^{-\mu_{(\ell)}-j} \, (2 \Delta - dj)_{2p} \, \bigg( 
     \zeta \big(2p+2\Delta-dj\big) - \zeta\big(2p+2\Delta -dj, n^*_{\ell} \big) \bigg) \Bigg] \,, \nonumber
\end{align}
as well as $n_{\ell}^* = \ceil*{|y_\ell|^{1/d}/\beta}$, the least integer greater than or equal to $|y_\ell|^{1/d}/\beta$. Note that for non-integer $\kappa = \Delta - d/2$ there is no possible mixing between double-twist operators and the multi-energy-momentum tensors in the ambient expansion.  The first sum in \eqref{BBKG_FullPosSpaceCorrThOPE} contains the multi-energy-momentum tensor spectrum \eqref{BB_NonCompCorrx0}, left untouched by the sum over images. This result justifies our earlier claim that the non-compact 2-point function $\braket{O(\tau,x)O(0)}^{(nc)}_{d,\Delta}$ captures the multi-energy-momentum contributions. The absence of mixing also shows that the first-order exactness of the ambient geodesic term is robust in this case.
Furthermore, through the second sum this expression \eqref{BBKG_FullPosSpaceCorrThOPE} provides a prediction for the double-twist coefficients, taking as an input the multi-energy-momentum tensor coefficients $a_{n}^{(T)}$ and the singularities of the analytically continued non-compact correlator (in particular, their positions, orders and the factors $A_{(\ell)}$). Although this computation was carried out in the limit $x\to0$, these same techniques can be applied in the case of non-vanishing $x$, as well as for theories where space-like directions are compact.

Non-perturbative effects in momentum space $\sim e^{-\beta \omega}$  may yield additional regular and singular contributions besides the multi-energy-momentum tensor operators in the non-compact correlator \eqref{BB_NonCompCorrx0}. In that case this computation can be repeated without obstructions. The form of the double-twist coefficients changes accordingly, while the structure of \eqref{BBKG_FullPosSpaceCorrThOPE} is preserved. This suggests that under sum over images any operator entering the OPE limit of the non-compact correlator contributes to the double-twist coefficients in an analogous way to multi-energy-momentum tensors. We can conclude that the double-twist spectrum in the holographic Euclidean thermal 2-point function on the planar black hole arises from the sum over images of the non-compact position space correlator, and it may receive further contributions from possible non-perturbative pieces in momentum space, whose existence was not probed in our perturbative computation.

Let us consider how these results apply to the holographic thermal 2-point function on the planar black hole. Using the techniques of section \ref{SSSec_BBPert}, for $\Delta = 3/2$ and $d=4$ we computed the momentum space correlator \eqref{BB_MomentumSpaceGeneric2pf} up to order $O\left(k \beta\right)^{-360}$, from which we extracted the first 91 coefficients $a_{n}^{(T)}$ in \eqref{BB_NonCompCorrx0}. The asymptotic growth at large $n$ is captured by $a_{n}^{(T)} \approx (-1)^n 4^n$, indicating that the leading singularities in the complex $\tau$ plane are simple poles at $1+4 (\tau/\beta)^4=0$ giving an OPE radius of convergence $\tau_\star = \beta/\sqrt{2}$. Whilst this asymptotic growth is robustly identified, we have not determined the contributions of subleading singularities upon subtraction, so as to attain an infinite radius of convergence for the remainder as in equation \eqref{BB_NonCompCorrx0_Split}, and subsequently use \eqref{BBKG_FullPosSpaceCorrThOPE} to make predictions on the double-twist coefficients. The four singularities appearing at $|\tau| = \tau_\star$ can be related to singularities of holographic Lorentzian thermal 2-point functions, which originate in the existence of a bulk null geodesic connecting two boundaries via the singularity \cite{Fidkowski:2003nf, Festuccia:2005pi}. For more recent developments on such singularities see also \cite{Dodelson:2023vrw, Horowitz:2023ury, Dodelson:2023nnr}.

An interesting question is how to account for these double-twist contributions using ambient invariants. For CFTs in thermal states dual to thermal AdS and BTZ, the ambient space is a quotient of $(d+2)$-dimensional Minkowski space (being each ALAdS$_{d+1}$ slice simply a quotient of Euclidean AdS). One can check that the double-twist spectrum arises automatically from the sum over the distinct geodesics that connect the same two nullcone points, which is implicit in the prescription \eqref{eq:ProposalForm1}. In particular, the ambient correlator is equal to a sum of terms $\X_{12}^{-\Delta}$, each evaluated on one among the infinite distinct ambient geodesics that wrap the thermal circle, each characterised by a different winding.

However, as we detail in appendix \ref{AppendixLongGeodesics} there is a unique geodesic on the AdS planar black hole that connects two given boundary points, and hence there is only one ambient geodesic that connects a given pair of points on the nullcone. This entails that there are no periodic ambient geodesics to sum over, suggesting that double-twist contributions must be described by a novel class of invariants on the ambient space. We leave this interesting question to future work.

\subsection{The $d=2$ case and the BTZ black hole} \label{SSec_BTZ}
As a final example of the ambient formalism for thermal correlators, we consider the simplified case of $d=2$. We parametrise the thermal cylinder with coordinates $x^i = (\tau,\phi)$ with $0\leq \tau <\beta$ and consider for now a non-compact $\phi$. The states we are interested in are characterised by a energy-momentum tensor VEV of the form \eqref{BBthermalTij}, which for $d=2$ is
\begin{equation}
\braket{T_{ij}} dx^i dx^j = \frac{\pi}{4 G \, \beta^2} \left(d\tau^2 + d\phi^2\right).
\end{equation}
The appropriate ambient space can be written as a foliation of the Euclidean BTZ black hole \cite{Banados:1992wn, Banados:1992gq},
\begin{equation} \label{BTZ_Aspace}
    \tilde{g} = - ds^2 + s^2
    \left[ 
\frac{dr^2}{r^2-r_H^2} + \left(r^2-r_H^2\right) d\tau^2 +  r^2 d\phi^2
    \right] ,
\end{equation}
with $r_H = 2\pi / \beta$ and $r>r_H$. The radial $r$ coordinate is related to the usual ambient $\rho$ coordinate by
\begin{equation}
    r = \frac{1-2\rho^2}{\,2\,\sqrt{-2\rho\,}\,} \, r_H \,.
\end{equation}

Through the coordinate transformation
\begin{equation}
        X^0 = \frac{s \, r}{r_H} \cosh(r_H \phi), \qquad \quad X^1 = \frac{s \,r}{r_H} \sinh(r_H \phi)\,, \label{BTZMink1}
\end{equation}
\begin{equation}
        X^2 = s \, \sqrt{\frac{r^2}{r_H^2}-1} \, \cos(r_H \tau)\, \qquad \quad X^3 = s \, \sqrt{\frac{r^2}{r_H^2}-1} \, \sin(r_H \tau)\,,\label{BTZMink2}
\end{equation}
one can show the ambient metric \eqref{BTZ_Aspace} describes the geometry of $(d+2)$-dimensional Minkowski space in the causal future of the origin $X^M=0$, with line element $ds^2=\eta_{MN}dX^M dX^N$. As we mentioned in section \ref{SSec:RelToESpace}, any 4-dimensional ambient space is locally diffeomorphic to Minkowski space. This entails that all ambient curvature invariants are identically vanishing, and the only non-trivial scalar building block is the geodesic distance square $\X_{12}$. 

Let us now study scalar 2-point functions. Leveraging translational symmetries and turning to the ambient gauge $\X = (t,\rho, \tau,\phi)$ as in \eqref{AspGeneralFormTRHO}, we place the insertions at
\begin{equation} \label{BTZ_BCs}
    \X_1 = (t_1, 0, 0,0)\,, \qquad \quad 
    \X_2 = (t_2, 0, \tau, \phi)\,.
\end{equation} 
Ambient geodesics are simply straight lines on Minkowski. The boundary conditions \eqref{BTZ_BCs} fix the integration constants, yielding, in Minkowski coordinates \eqref{BTZMink1}-\eqref{BTZMink2},
\begin{equation} \label{BTZ_Geods1}
        X^0(\lambda) = \frac{1}{2} \Big[
    t_0 - t_0 \lambda + t_1 \lambda \cosh(r_H \phi)
    \Big]\,, \qquad \quad X^1(\lambda) = 
    \frac{1}{2} t_1 \lambda \sinh(r_H \phi)\,,
\end{equation}
\begin{equation}\label{BTZ_Geods2}
        X^2(\lambda) = \frac{1}{2} \Big[
    t_0 - t_0 \lambda + t_1 \lambda \cos(r_H \tau)
    \Big]\,, \qquad \quad X^3(\lambda) = 
    \frac{1}{2} t_1 \lambda \sin(r_H \tau)\,.
\end{equation}
Assembling the above results we obtain the invariant
\begin{equation} \label{BTZ_Xij}
    \X_{12} = \frac{t_0 t_1}{2}\, \Big[ 
    \cosh (r_H \phi) - \cos ( r_H \tau)
    \Big].
\end{equation}
Contrary to thermal AdS, there is only one geodesic connecting any pair of insertion points on the thermal cylinder for non-compact $\phi$, regardless of the periodicity in $\tau$. This is analogous to what happens with the higher dimensional black brane as discussed in section \ref{SSSec_Double_Twists} and appendix \ref{AppendixLongGeodesics}. 
The resulting ambient 2-point function is therefore 
\begin{equation} \label{BTZ_Acorr}
    \braket{ O(\tau,\phi)O(0)}^{(\beta)}_{d=2,\Delta}	
    = \frac{1}{\beta^{2\Delta}} 
    \frac{C_\Delta}{\big[ 
    \cosh \frac{2 \pi \phi}{\beta} - \cos \frac{2 \pi \tau}{\beta}
    \big]^{\Delta}} \,.
\end{equation}
Expanding this correlator in the OPE limit, only negative even powers of $\beta$ appear, describing the multi-energy-momentum tensor spectrum. 

The BTZ black hole geometry is however periodic in $\phi$ with period $2\pi$, and on the corresponding ambient space one has an infinite number of geodesics. Their form is the same as in equations \eqref{BTZ_Geods1}-\eqref{BTZ_Geods2} with $\phi \to \phi+ 2\pi m$, where $m \in \Z$ parametrises the winding around the $\phi$ circle. There is an invariant analogue to \eqref{BTZ_Xij} for each such geodesic, yielding a correlator of the form,
\begin{equation} \label{BTZ_Acorr_Periodic}
    \braket{ O(\tau,\phi)O(0)}^{(\beta)}_{d=2,\Delta}	
    = \frac{1}{\beta^{2\Delta}} 
    \sum_{m\in \mathbb{Z}}
    \frac{C_\Delta}{\big[ 
    \cosh \frac{2 \pi (\phi+2 \pi m)}{\beta} - \cos \frac{2 \pi \tau}{\beta}
    \big]^{\Delta}} \,.
\end{equation}
The expression \eqref{BTZ_Acorr_Periodic} matches the corresponding holographic result \cite{Keski-Vakkuri:1998gmz,Kraus:2002iv, Skenderis:2008dg}.

\section{CFTs on squashed spheres} \label{Sec:SqSphere}

Squashed spheres are a class of non-conformally flat and non-Einstein manifolds and represent an interesting case of study to make predictions using the ambient space formalism. Previous work on CFTs on squashed spheres includes \cite{Zoubos:2002cw,Zoubos2004,Hartnoll:2005yc,Bobev:2016sap,Bobev:2017asb,Bueno:2018yzo,Bueno:2020odt,Chester:2021gdw}. Fixing to $d=3$ for concreteness, the geometry reads 
\begin{equation}\label{eq:SSphereMetric}
	ds^2 = d\theta^2 +\sin^2\theta d\vp^2 + \frac{1}{1+\alpha}  \left( d\psi + \cos\theta d\vp\right)^2 ,
\end{equation}
where $0\leq \theta < \pi$, $0\leq \vp < 2 \pi$, $0\leq \psi < 4 \pi$ are the Euler angles. The real parameter $\alpha$ defines the squashing. For $\alpha=0$ we recover the round sphere, while in the limiting case  $\alpha\to \infty$ we obtain the cylinder $\R \times S^2$. \footnote{Performing this limit on the metric \eqref{eq:SSphereMetric} yields a degenerate geometry. We can attain a non-degenerate metric by first unwrapping the fibred $S^1$ so that $0\leq \psi < \infty$, and subsequently rescaling it as $\wt{\psi} = \frac{\psi}{\sqrt{1+\alpha}}$.}

As anticipated, these spaces are not Einstein but they are close to being Einstein in the sense that 
one can recast their Ricci tensor as
\begin{equation}\label{}
	R_{ij}(\theta) = \frac{R}{3} g_{ij}(\theta) + H_{ij}(\theta),
\end{equation}
where $H_{ij}$ is traceless, 
so that they have constant curvature $R=\frac{3+ 4 \alpha}{2(1+\alpha)}$. Their Cotton tensor is non-vanishing, hence they are not conformally flat. This means that regardless of the CFT state, their ambient space is not flat space.

For generic $\alpha$, the squashing breaks $SO(4)\simeq SU(2)_L\times SU(2)_R$ down to  $SU(2)_L \times U(1)_R$. Thus $d=3$ squashed spheres are endowed with only four out of the six isometries of round spheres. These isometries can be written as 
	\begin{align}
		K_1 &= -\sin\vp \partial_\theta + \frac{\cos\vp}{\sin\theta} \partial_\psi - 
		\cot \theta \cos\vp \partial_\vp  
		, \label{Squashed-Isometry1}\\
		K_2 &= \cos\vp \partial_\theta + \frac{\sin\vp}{\sin\theta} \partial_\psi - 
		\cot \theta \sin\vp \partial_\vp 
		,\\
	K_3 &= \partial_\vp , \label{Squashed-Transl1}\\
	K_4 &= \partial_\psi, \label{Squashed-Transl2}
	\end{align}
where $K_4$ generates the residual $U(1)_R$ symmetry. No additional conformal Killing vector is present for a generic squashing $\alpha$.

The Ward Identities associated to the vectors $K_1, \dots, K_4$ fix scalar 1-point functions of quasi-primary operators to be equal to constants, while operators with spin may have dependence on $\theta$. Defining the invariant 1-form
\begin{equation}
    \zeta = d\psi + \cos\theta d\vp, \qquad \quad \mathcal{L}_{K_i} \zeta = 0\,, 
\end{equation}
a generic spin-1 1-point function can be written as
\begin{equation}
    \braket{O_{i}(\theta)}_\alpha = u_1 \,\zeta_i \,,
\end{equation}
and a generic spin-2 1-point function can be written as
\begin{equation}
    \braket{O_{ij}(\theta)}_\alpha = 
    u_2 \,\zeta_i \, \zeta_j + u_2^{(\text{tr})} g_{(0)ij} \,,
\end{equation}
where the constants $u_2$ and $u_2^{(\text{tr})}$ are fixed by dynamics. Note that no antisymmetric part is allowed. $\braket{O_{ij}(\theta)}_\alpha$ has a non-trivial dependence on $\theta$, which entails that an infinite tower of descendants can be constructed acting with three-dimensional covariant derivatives on the squashed sphere.

The form of scalar 2-point functions of quasi-primary operators is partially fixed by the Ward Identities
\begin{equation}
  \left[  \mathcal{L}_{K_i}(\theta_1, \vp_1,\psi_1) +  \mathcal{L}_{K_i}(\theta_2, \vp_2,\psi_2) \right] \braket{O_1 O_2}_\alpha  = 0\,.
\end{equation}
Leveraging rotational symmetry along $\psi$ and $\vp$, we can move the first insertion to lie at $\psi_1 = \vp_1 = 0$ and the second insertion to be at $\psi_2 = \psi$, $\vp_2 = \vp$. Adopting the basis of cross-ratios from \cite{Zoubos2004}, scalar 2-point correlators must be of the form
\begin{equation}
    \braket{O_1(\theta_1,0,0) \, O_2(\theta_2,\psi,\vp)}_\alpha  = F(v_1, v_2),
\end{equation}
where
\begin{subequations}
	\begin{align}
		v_1 &=  \cos\frac{\theta_1}{2} \cos\frac{\theta_2}{2}
		\cos\frac{\psi + \vp}{2} + \sin\frac{\theta_1}{2} \sin\frac{\theta_2}{2}
		\cos\frac{\psi-\vp}{2} \,, \\
		v_2 &= \frac{1}{2} \left(
		1+ \cos \theta_1 \cos \theta_2+ \sin \theta_1 \sin \theta_2 \cos \vp
		\right)
		.
	\end{align}
\end{subequations}

\subsection{The ambient setup}

We intend to study CFTs on squashed spheres in states with a non-vanishing energy-momentum tensor VEV. To describe the multi-energy-momentum tensor contributions of their correlators using the ambient formalism we should identify suitable bulk solutions with a squashed 3-sphere as a boundary and a non-vanishing holographic energy-momentum tensor VEV. 
Four-dimensional AdS Taub-NUT and -bolt spaces allow one to study a wide class of such states. Their metric reads \cite{stephani_kramer_maccallum_hoenselaers_herlt_2003} 
\begin{equation}\label{MetricAdSTNUT}
	ds^2 = \frac{dr^2}{V(r)} 
	+ (r^2-n^2) (d\theta^2 +\sin^2\theta d\vp^2) + 4 n^2 V(r) \left(
	d\psi + \cos \theta d\vp\right)^2,
\end{equation}
where we defined
\begin{equation}\label{}
	V(r) = \frac{r^2 + n^2 -2 m r+\left(r^4 -6 n^2 r^2 -3 n^4\right)}{r^2 - n^2}\,.
\end{equation}
The NUT parameter is related to the squashing of the boundary by $n= (2\sqrt{\alpha +1})^{-1}$, and the boundary is reached for $r\to \infty$. The holographic energy-momentum tensor 1-point function in such geometries is parametrised by the mass parameter $m$ as 
\begin{equation}
    u_2 = - \frac{3}{8 \pi} \frac{m }{1+\alpha}\,, \qquad \quad u_2^{(\text{tr})} = \frac{m}{8 \pi} \,,
\end{equation}
and as such these geometries can be used to describe states where
\begin{equation}
    \frac{u_2}{\;u_2^{(\text{tr})}} = -\frac{3}{1+\alpha}\, .
\end{equation}

As an illustrative example, the CFT state that we consider is characterised by the energy-momentum tensor VEV associated to the self-dual AdS Taub-NUT geometry with no conical singularities. The ambient space we need is then \eqref{AspMetricAdSSlicing} with metric \eqref{MetricAdSTNUT} as $(d+1)-$dimensional hyperbolic slices, and with the choice of the mass parameter $m=  \frac{\alpha}{2\left(1+\alpha \right)^{3/2}}$. For later convenience, we write explicitly the energy-momentum tensor VEV, $\braket{T_{ij}}_\alpha = \frac{3}{16 \pi} g_{(3)ij}$, with
\begin{equation}
	\begin{split}
			g_{(3)ij} dx^i dx^j &= \frac{\alpha }{3 (\alpha +1)^{3/2}}
	\left[
	d\theta^2 -\frac{2 d\psi^2}{1+\alpha}  -\frac{4 \cos\theta}{1+\alpha} d\psi d\vp - \frac{(\alpha +3) \cos 2 \theta -\alpha +1}{2 (\alpha +1)} d\vp^2
	\right]  \\
	& = \frac{\alpha }{3}
	\left[
	d\theta^2 -2 d\psi^2 -4 \cos\theta \, d\psi d\vp -\frac{3 \cos 2 \theta +1}{2} d \vp^2
	\right] + O(\alpha^2).
	\end{split}
\end{equation}
In what follows we work perturbatively in small $\alpha$ for simplicity. 
To avoid cluttering in the expressions below we fix $\theta_1=0$, rename $\theta_2=\theta$ and define $\chi=(\vp + \psi) /2$. The two insertion points on the ambient space are thus $\X_1 = (s_1,r_1,0,0,0)$ and $\X_2 = (s_2,r_2,\theta,\psi,\vp)$, where the limit to the lightcone $s_i, r_i \to \infty$ with fixed $s_i / r_i = t_i = 1$ is understood.

\subsection{Geodesics}

Let us first solve the geodesic equations on this geometry between $\X_1$ and $\X_2$ so as to obtain the invariant $\X_{12}$. In this case it is convenient to compute the (divergent) geodesic length $L_{AdS}$ on a fixed hyperbolic slice and then use the relation \eqref{GeodApproxAdSAmbient} to find the finite ambient invariant $\X_{12}$. 

We hence consider the 4-dimensional self-dual AdS Taub-NUT metric  \eqref{MetricAdSTNUT} with $m=  \frac{\alpha}{2\left(1+\alpha \right)^{3/2}}$. We would like to study geodesics on this background with endpoints on the boundary $r\to \infty$ at the generic points $x_1 = (\theta_1,0,0)$  and $ x_2 = (\theta_2,\psi,\vp)$  corresponding to the values of the affine parameter $\lambda=0$ and $\lambda=1$ respectively. For simplicity we restrict to $\theta_1 = \theta_2 =0$.

The boundary isometries \eqref{Squashed-Isometry1}-\eqref{Squashed-Transl2} are also bulk isometries and one can use them to partially integrate the bulk geodesic equations. From the integrals of motion related to translational symmetries \eqref{Squashed-Transl1}-\eqref{Squashed-Transl2} along $\vp$ and $\psi$ one obtains the first-order equations
\begin{align} 
    \dot{\vp} &= \frac{A_{\vp }}{n^2-r^2} \, , \label{SquashedInt1}\\
    \dot{\psi} & = -\frac{A_{\psi } (n+r)}{4 n^2 (n-r) \left(-3 n^2+2 n r+r^2+1\right)}-\frac{A_{\vp }}{n^2-r^2}\,, \label{SquashedInt2}
\end{align}
where $A_{\psi }$ and $A_\vp$ are the constants of motion. Using equations \eqref{SquashedInt1}-\eqref{SquashedInt2} the 4-velocity constraint in the bulk $\dot{x}_\mu \dot{x}^\mu = L_{AdS}^2$ can be written as 
\begin{equation} \label{SquashedCons}
\begin{split} 
    & 4 n^2 (n+r(\lambda )) \dot{r}(\lambda )^2 + n A_{\psi }^2 + 4 \left(1-3 n^2\right) n^3 L_{\text{AdS}}^2 + \\
    &r(\lambda ) \left(A_{\psi }^2-4 n^2 L_{\text{AdS}}^2 r(\lambda ) (n+r(\lambda )) + 4 n^2 \left(5 n^2-1\right) L_{\text{AdS}}^2\right) = 0 \,.
\end{split}
\end{equation}
We use \eqref{SquashedCons} to eliminate $\dot{r}$ terms from the geodesic equation governing $\ddot{r}$, which then reads
\begin{equation} \label{SquashedRadial}
    (n+r(\lambda ))^2 \ddot{r}(\lambda )-L_{\text{AdS}}^2 \left(n-4 n^3+r(\lambda ) (n+r(\lambda ))^2\right)=0 \,.
\end{equation}

We first solve \eqref{SquashedRadial}. To regulate the divergence in the geodesic distance as one approaches the boundary we use boundary conditions $r(0)=r(1) = R$ with a radial regulator $R$ to be eventually set to infinity. We then plug the solution $r(\lambda)$ into the angular equations, which are solved subject to the Dirichlet boundary conditions at $x_1$ and $x_2$. This fully determines the trajectory. Finally, substituting $r(\lambda)$ and $A_\psi$ into \eqref{SquashedCons} allows one to find the value of $L_{\text{AdS}}$ in terms of the boundary points $x_1$ and $x_2$. Note that $A_\vp$ does not appear in \eqref{SquashedCons}, meaning that to obtain the geodesic distance $L_{\text{AdS}}$ it is sufficient to solve the reduced ODE 
\begin{equation} \label{SquashedChi}
   \dot{\chi}(\lambda )  +  \frac{A_{\psi } (n+r(\lambda ))}{8 n^2 (n-r(\lambda )) \left(-3 n^2+2 n r(\lambda )+r(\lambda )^2+1\right)} = 0\,,
\end{equation}
in terms of $\chi(\lambda) = \frac{\psi(\lambda) + \vp(\lambda)}{2}$ with boundary conditions $\chi(0) = 0$ and $\chi(1) = \chi$.

The explicit solution to \eqref{SquashedRadial} can be found in terms of inverse elliptic functions. However we are interested in an $\alpha \to 0$ expansion of the geodesic distance, which corresponds to an $n\to \frac{1}{2}$ expansion in the current parametrisation. We thus expand the unknown functions and integration constants as
\begin{equation}
    r(\lambda) = \sum_{k = 0}^\infty \left(n-\frac{1}{2}\right)^k r_k(\lambda)\,, \qquad \quad
    \chi(\lambda) = \sum_{k = 0}^\infty \left(n-\frac{1}{2}\right)^k \chi_k(\lambda)\,,
\end{equation}
\begin{equation}
    A_{\psi} = \sum_{k = 0}^\infty \left(n-\frac{1}{2}\right)^k A^{(k)}_{\psi}\,, \qquad \qquad 
    L_{\text{AdS}} = \sum_{k = 0}^\infty \left(n-\frac{1}{2}\right)^k L_{\text{AdS}}^{(k)}\,.
\end{equation}
At leading order $k=0$ the bulk is simply global Euclidean AdS$_4$ and the boundary is a round sphere. Following this integration scheme, one finds
\begin{align}
    r_0(\lambda) & = \frac{(8-8 \cos \chi )^{-\lambda } \left[64^{\lambda } R^{2\lambda +1} (1- \cos \chi )^{2 \lambda }+8 R^{3-2 \lambda } (1-\cos \chi )\right]}{1+8 R^2 (1-\cos \chi)}\,, \\
    \chi_0(\lambda) &= -\frac{\chi  \arctan \left[\frac{8 \sin \chi  (\cos \chi -1) \left(R^{4 \lambda } (8-8 \cos \chi )^{2 \lambda }-1\right)}{R^{4 \lambda -2} (8-8 \cos \chi )^{2 \lambda } \left(8 R^2 (\cos \chi -1) \cos \chi -1\right)-8 (\cos \chi -1) \left(\left(8 R^2-1\right) \cos \chi -8 R^2\right)}\right]}{\arctan \left[\frac{\sin \chi  \left(-32 R^4 (\cos 2 \chi -4 \cos \chi )-96 R^4+1\right)}{\cos \chi  \left(32 R^4 (-4 \cos \chi +\cos 2 \chi +1)+\left(1-8 R^2\right)^2\right)+16 R^2}\right]}\,, \\
    L_{\text{AdS}}^{(0)} &= \log \left[8 R^2 (1-\cos \chi )\right] \,, \\ 
     A_\psi^{(0)} &= 
    -\frac {\sqrt {32 R^4 \cos 2 \chi - 16 R^2 \cos \chi - 32 R^4 + 
     16 R^2 + 1}} {4 \left (8 R^2 \cos \chi - 8 R^2 - 1 \right) } \, \times \, \\
     & \!\!\!\!\!\!\!\!\!\!\!\!
     \text{arctanh}\left[\frac {\left (-8 R^2 \cos \chi + 8 R^2 - 
          1 \right) \sqrt {32 R^4 \cos 2 \chi - 16 R^2 \cos \chi - 
         32 R^4 + 16 R^2 + 1}} {64 R^4 (\cos \chi - 1) \cos \chi + 
       1} \right] ^{-1} \nonumber     
    .
\end{align}
The solution at first order is lengthy and we avoid displaying it here. The first order geodesic distance at leading order in $R\to\infty$ takes however a particularly compact form,
\begin{equation}
     L_{\text{AdS}}^{(1)} =  \left[4 (\pi -\chi ) \sin ^3\frac{\chi }{2}+\cos \frac{\chi }{2}+3 \cos \frac{3 \chi }{2}\right] \sec ^3\frac{\chi }{2}\,.
\end{equation}
Knowing $L_{\text{AdS}}^{(0)}$ and $L_{\text{AdS}}^{(1)}$, one can compute the invariant $\X_{12}$  to first order in $\alpha$ through \eqref{GeodApproxAdSAmbient} setting $R\to\infty$. It reads
\begin{equation} \label{SquashedX12}
    \X_{12} =  8(1-v_1) \left[ 
1 + \alpha \left( \frac{1-3 v_1}{1+v_1} + (\arccos v_1-\pi) \left( \frac{1-v_1}{1+v_1}\right)^{\frac{3}{2}} \right) + O(\alpha)^2
    \right] .
\end{equation}
In the limit $\theta_1 \to \theta_2$ we are considering here the cross ratios reduce to $v_1 = \cos \chi$ and $v_2 =1$. The fact that $v_2$ trivialises means that only $v_1$ can appear in invariants contributing to correlators such as \eqref{SquashedX12} in this limit.

\subsection{Invariants and ambient 2-point functions}

Using the techniques developed in appendix \ref{AppendixGeodPTranspT} one can straightforwardly compute the parallel transported vector $\hat{T}_1$ from $\X_1$ to $\X_2$. In particular, to order $O(\alpha^0)$ the $(d+2)-$dimensional background  geometry is simply Minkowski space. Thus, to find $\hat{T}_1$ to this order in $\alpha$ it is sufficient to take the Euler vector in Minkowski $X^M \p_M$ evaluated at the point $\X_1$, and make a transformation $X^M \to \X^M = (s,r,\theta,\psi, \varphi)$ to the ambient coordinates.  This change of coordinates reads
\begin{subequations}\label{}
	\begin{align}
		X_0&=t \left(1-\frac{\rho }{2}\right)  , \\
	X_1&=t \left(1+\frac{\rho }{2}\right) \sin \left(\frac{\theta }{2}\right) \cos \left(\frac{\phi -\psi }{2}\right), \\
	X_2&=t \left(1+\frac{\rho }{2}\right) \sin \left(\frac{\theta }{2}\right) \sin \left(\frac{\phi -\psi }{2}\right), \\
	X_3&=t \left(1+\frac{\rho }{2}\right)  \cos \left(\frac{\theta }{2}\right) \cos \left(\frac{\psi +\phi }{2}\right), \\
	X_4&=t \left(1+\frac{\rho }{2}\right)  \cos \left(\frac{\theta }{2}\right) \sin \left(\frac{\psi +\phi }{2}\right),
	\end{align}
\end{subequations}
where the ambient $t$ and $\rho$ are related to the AdS Taub-NUT radial coordinate (at zero-th order in $\alpha$) by
\begin{align}
    t &= \frac{s}{4 \left(\sqrt{r^2-\frac{1}{4}}+r\right)} \,
    ,\\
    \rho & = -8 \left(\sqrt{r^2-\frac{1}{4}}+r\right)^2 .
\end{align}
The resulting transported $\hat{T}_1$ is consequently
\begin{align}
		\hat{T}_1 &= \frac{ \cos \frac{\theta }{2} \cos \chi +1}{2 r} \p_s + \frac{\cos \frac{\theta }{2} \cos \chi-1}{64 r} \p_r \\ 
&\quad - 2 \sin \frac{\theta }{2} \cos \chi \p_\theta 
		- \sec \frac{\theta }{2} \sin \chi\p_\psi -\sec \frac{\theta }{2} \sin \chi\p_\vp + O(\alpha). \nonumber
\end{align}
From equations  \eqref{ARiemannCompg0ConfFlat}, \eqref{ARiemannCompg0ConfFlatE3} and \eqref{ARiemannCompd=3} the non-vanishing components of the ambient Riemann read
\begin{align} 
	\wt{R}_{rirj}	&= -\frac{3 t^2}{2r^5} g_{(3)ij} + O(\alpha)^2
	, \nonumber\\
	\wt{R}_{rijk}	&= \frac{t^2}{r^3} \left(
	\nabla_k g_{(3)ij} - \nabla_j g_{(3)ik}\right) + O(\alpha)^2
	,\label{SqS3-ARiemann}\\
	\wt{R}_{ijkl}	&= \frac{3 t^2}{2r}
	\left[
	g_{(0)ik} g_{(3)jl} + g_{(0)jl} g_{(3)ik} - (l \leftrightarrow k)
	\right] + O(\alpha)^2  \nonumber
	. 
\end{align}

These ingredients can be assembled to form the ambient curvature invariants that enter scalar correlators. As we showed in section \ref{SSec:BuildingBlocks} the ambient formalism predicts that the single-energy-momentum tensor contributions to scalar correlators are fully fixed by the leading  geodesic distance $(\X_{12})^{-\Delta}$.
Since $\braket{T_{ij}}_\alpha=O(\alpha)$,  the leading curvature invariants are of order $O(\alpha)^2$. As expected due to the infinite tower of non-trivial descendants of :$T^2$:, one can construct an infinite number of ambient curvature invariants at this order. However, in a short distance expansion the dominant contributions can be identified as the three invariants accounting for the three independent $\sim \braket{:\!T^2\!:}$ contributions, while the others include a higher and higher number of covariant derivatives and are thus subleading. Focusing on the dominant ones, a suitable basis is provided by the three curvature invariants
\begin{align}
	(\nabla \wt{\text{R}}\text{iem})^2  &= \frac{42 \alpha^2}{t^6} + O(\alpha)^3, \\
	\wt{\mathcal{R}}^{(1)}_{AC}\, \wt{\mathcal{R}}^{(1)AC} &=
	18 \alpha ^2 \sin ^2\!\frac{\theta }{2} \left[ 3 -\cos \theta- 2 \cos ^2\!\frac{\theta }{2} \cos 2 \chi \right]^{2} + O(\alpha)^3, \\
	\wt{\mathcal{R}}^{(0)}_{AC}\, \wt{\mathcal{R}}^{(2)AC} &=
	\frac{3}{2}  \alpha ^2 \! \left[-3 + 5 \cos \theta - 2 \cos ^2\!\frac{\theta }{2} \cos 2 \chi  \right] \! \left[ 3 - \cos \theta - 2 \cos ^2\!\frac{\theta }{2} \cos 2 \chi \right]^{2} \!\! + O(\alpha)^3,
\end{align}
where the tensors $\mathcal{R}^{(r)}$ are defined in \eqref{mathcalRweight0inv}. 
Using the expressions for the ambient Riemann  components in \eqref{SqS3-ARiemann} one can explicitly check that these invariants do not contain derivatives of the energy-momentum tensor VEV, ensuring they describe the independent  $:T^2:$ blocks. 

Note that weight-0 invariants of the form \eqref{mathcalRweight0inv} constructed as chains of tensors $\mathcal{R}^{(r)}$ are not sufficient to account for all the three :$T^2$: blocks, as opposed to the finite temperature example in section \ref{sec:thermal} where they can be used as a basis for
\emph{any} multi-energy-momentum tensor contribution :$T^n$: as we showed. This difference resides in the fact that in the present case the dimension of the CFT background is odd. As discussed in the end of section \ref{SSec:BuildingBlocks} this causes a number of ambient invariants to be either divergent or vanishing in the limit to the nullcone, as it is the case for instance for $\wt{\mathcal{R}}^{(0)}_{AC} \, \wt{\mathcal{R}}^{(0)AC}$ here.

Assembling the above ingredients following the prescription in section \ref{SSec:ProposalA2ptf} we arrive at our  main result for this example, the scalar 2-point function on this background and state,
\comment{\begin{equation}\label{eq:SSphere2ptf}
	\braket{O(\X_1) O(\X_2)}_\alpha
	= 
	\frac{C_\Delta}{(\X_{12})^{\Delta}}\!\left[1+
 c_1 (\nabla \wt{\text{R}}\text{iem})^2
+c_2 \wt{\mathcal{R}}^{(1)}_{AC}\wt{\mathcal{R}}^{(1)AC}
+ c_3 \wt{\mathcal{R}}^{(0)}_{AC} \wt{\mathcal{R}}^{(2)AC} +\dots
 \right] + O(\alpha^3),
\end{equation}}
\begin{align}\label{eq:SSphere2ptf}
	&\braket{O(X_1) O(X_2)}_\alpha
	= 
	\frac{C_\Delta}{(\X_{12})^{\Delta}}\Bigg[1+ \frac{3}{2}\alpha^2 \bigg[
 28 c_1 -  \Big(2 \cos ^2 \frac{\theta }{2} \cos 2 \chi +\cos \theta -3\Big)^2 \times \nonumber
 \\
 & \bigg( 6 c_2 (\cos \theta -1) + c_3 \Big(2 \cos ^2\frac{\theta }{2} \cos 2 \chi -5 \cos \theta +3 \Big) \bigg) +\dots
 \bigg] + O(\alpha)^3 
 \Bigg] , 
\end{align}
where $c_1, c_2$ and $c_3$ are theory-dependent constants, while the dots denote subleading terms in the short distance limit. These contributions can be constructed in a similar way using invariants containing more ambient covariant derivatives, while multi-energy-momentum-energy tensor blocks can be accounted for with invariants of higher order in the ambient Riemann.

Given the form \eqref{SquashedX12} of the invariant $\X_{12}$, in the case of $\theta=0$ we are able to explicitly write the form of the scalar 2-point function  to first order in $\alpha$,
\begin{align}
    \braket{O(\X_1) O(\X_2)}_\alpha &= 
	\frac{C_\Delta 8^{-\Delta}}{(1-v_1)^{\Delta}}
 \Bigg[ 
1 - \Delta  \left( \frac{1-3 v_1}{1+v_1} + (\arccos v_1-\pi) \left( \frac{1-v_1}{1+v_1}\right)^{\frac{3}{2}} \right) 
\alpha \, + \nonumber \\
& \quad\; + O(\alpha)^2
    \Bigg]. \label{eq:SSphere2ptf_geod}
\end{align}
A holographic computation for the 2-point function of the operator of dimension $\Delta=1$ was reported in \cite{Zoubos2004}. The computation relied on assumptions on the bulk propagator and in the limit $\theta_1 = \theta_2 = 0$, the result of \cite{Zoubos2004} reduces to the correlator on the round sphere 
\begin{equation}
   \braket{O(\X_1) O(\X_2)}^{\text{\cite{Zoubos2004}}}_\alpha = \frac{2}{1-v_1}\,, \label{Zoubos2pt}
\end{equation}
to all orders in $\alpha$, which is at odds with our results. The result \eqref{Zoubos2pt} also appears at odds with conformal perturbation theory, which suggests that the order $\alpha$ correction of the 2-point function should be given by an integral of a CFT 3-point function, and it is thus  expected to be non-zero. One should be able to explicitly check \eqref{eq:SSphere2ptf_geod} via such computation, and we leave this to future work.

\section{Relations with flat holography}
\label{sec_FlatHolography}

The ambient space is a $(d+2)$-dimensional Ricci-flat spacetime and its geometry encodes observables of a Euclidean CFT$_d$, as we have shown in the previous sections. Such setup has a manifest holographic flavour and in this section we  illustrate the connections of the ambient space construction with proposals of flat holography in $d+2=4$ dimensions.

Several proposals of flat holography on Minkowski spacetime has adopted the embedding space as a useful framework, in particular to address questions related to the matching of symmetries in the putative dual theories. In celestial holography  \cite{Strominger:2017zoo,Raclariu:2021zjz,Pasterski:2021rjz,Pasterski:2021raf,McLoughlin:2022ljp} the embedding space provides a convenient language to relate bulk scattering amplitudes to CFT correlators. The hyperbolic slicing of Minkowski space is also at the basis of the idea of flat holography as an uplifting AdS/CFT. First put on paper by de Boer and Solodukhin \cite{deBoer:2003vf,Solodukhin:2004gs}, more recently it has been leveraged in several different contexts including \cite{Cheung:2016iub,Ball:2019atb,Iacobacci:2022yjo}. In \cite{Salzer:2023jqv} an embedding space formalism on $\R^{4,2}$ was also developed to find the form of Carrollian 2- and 3-point functions up to spin 1.

The ambient space generalises the embedding space allowing for generic AL(A)dS slices and non-conformally flat celestial manifolds. We thus expect to be able to implement similar approaches with the ambient space as a way to extend the current understanding of flat holography beyond spacetimes perturbatively close to Minkowski. In such more general spacetimes, we expect non-trivial sources and VEVs to appear in the dual QFT description. Some instances have been recently discussed in \cite{deGioia:2022fcn,Gonzo:2022tjm,
Crawley:2023brz,
He:2023qha}.

The first question to address is which class of physical Ricci-flat spacetimes the ambient space can describe.
To help us answer this question, let us briefly present the Beig-Schmidt gauge \cite{Beig:1982bs,Beig}, a suitable set of coordinates to describe the neighbourhood of past, future and spatial infinity. This gauge has been used to analyse the well-posedness of the variational principle at spatial infinity, and to define the scattering problem on general asymptotically flat spacetimes by studying BMS charges and their antipodal matching \cite{deHaro:2000wj, Mann:2005yr,Mann:2006bd, Compere:2011ve,Virmani:2011gh,Compere:2011db,Troessaert:2017jcm,Capone:2022gme,Compere:2023qoa}.
The geometry of a 4-dimensional Ricci-flat spacetime in a neighbourhood of spatial infinity  can be written in Beig-Schmidt gauge as\footnote{Usually, metrics in Beig-Schmidt gauge are written with a non-trivial lapse function. One can however use a so-called log supertranslation to move to the form in normal coordinates that we discuss here.}
\begin{equation} \label{BS_gauge}
    \wt{g} = ds^2 + s^2 \left[ 
    g^+_{\alpha \beta}(x) + \frac{f_{\alpha \beta}(x)}{s} + \frac{\tilde{f}_{\alpha \beta}(x)}{s} \log s + O(s)^{-2}
    \right] dx^\alpha dx^\beta\, ,
\end{equation}
where $s$ describes the geodesic distance from spatial infinity, reached for $s\to \infty$. The Ricci-flatness condition can be solved order by order at large $s$, from which it follows that $g^+_{\alpha \beta}$ must be the metric of a three-dimensional ALdS spacetime. 

Importantly, if one imposes $s\p_s$ to be a dilational symmetry of this four-dimensional spacetime, one restricts to spacetimes where  $f_{\alpha \beta}$, $\tilde{f}_{\alpha \beta}$ as well as the higher order terms in the expansion vanish. What one finds in this case is simply the  ALdS slicing \eqref{AspMetricdSSlicing} covering part of the ambient space. A completely analogous Beig-Schmidt expansion can be made near future or past infinity, leading to higher order corrections in $1/s$ to the ambient ALAdS slicing \eqref{AspMetricAdSSlicing}.
Henceforth the ambient space represents a generalisation of Minkowski spacetime which maintains an analogue of the Euler vector, while a metric in Beig-Schmidt gauge allows one to describe more general Ricci-flat spacetimes where this dilational symmetry is absent.

As we mentioned in section \ref{SSec:RelToESpace}, all four-dimensional ambient spaces are locally flat. It is well known that the Goldstone mode of (Virasoro) superrotations is encoded in the $g_{(d)ij}$ term related to the holographic energy-momentum tensor appearing in the near-boundary expansion of the hyperbolic metric $g^+_{\alpha\beta}$ in \eqref{BS_gauge}   \cite{Compere:2016jwb,Ball:2019atb}. Thus four-dimensional ambient spaces describe superrotated Minkowski spacetimes. In dimensions higher than four, no analysis of the asymptotic charges of ambient or Beig-Schmidt geometries has been carried out, and it is not currently known what subset of Beig-Schmidt geometries is described by ambient space geometries. 

Returning to four-dimensional ambient spaces, in the  Beig-Schmidt expansion \eqref{BS_gauge} the supertranslation Goldstone mode is encoded in $f_{\alpha \beta}$. Enforcing the presence of the homothety $T=s\p_s$ requires $f_{\alpha \beta}$ to vanish and fully fixes the supertranslation mode. If we then consider asymptotically flat spacetimes which can be represented as ambient spaces, we expect to be able to describe the soft physics related to \Pcr transformations and superrotations, while this is less clearly so for supertranslations. Furthermore, \Pcr transformations together with Virasoro superrotations do not form a closed subalgebra of the extended BMS transformations. This entails that if one fixes supertranslations, the maximal consistent algebra of near-lightcone transformations is simply ISO($1,3$).
This hence seems to indicate that one has to give up the global homothety $s\p_s$ and resort to the more general geometries of the Beig-Schmidt class in order to describe the soft physics of gravity in flat spacetimes.

We note however that the  family of conformal structures at infinity that are captured  by ambient spaces is much richer than that of asymptotically flat spacetimes, since they allow for sections of null infinity which are not conformally flat. We studied an explicit example of this in section \ref{Sec:SqSphere}, where the sections of null infinity are non-Einstein manifolds. These are spacetimes whose analysis in gravity is limited in the literature (although exceptions exist \cite{Capone:2021ouo}), while they have not been studied in a holographic perspective at all.

\section{Outlook}
\label{sec:outlook}

The embedding space formalism is a mainstay of computations for CFTs in vacuum on conformally flat spaces. In this paper we have proposed the ambient space formalism for CFTs, which extends the embedding space to non-vacuum states and arbitrary curved manifolds. All the ideas and techniques usefully employed in embedding space may enjoy an analogous construction in ambient space. Even when all symmetries are broken, the ambient formalism encodes Weyl covariance and gives building blocks that appear universally in correlation functions. In this section we look at the prospects and remaining open questions.

In the ambient formalism, the natural Weyl invariant building blocks of correlation functions are geometric objects in the ambient space. In this work we constructed a class of these and -- through applications to thermal CFTs -- concluded that they only capture the contributions from multi-energy-momentum tensors in an OPE expansion. Thus an important open direction is to try to make the ambient approach complete by constructing invariants for other operator contributions. A family of contributions of this type we encountered in section \ref{SSec_BBHoloCorr} are the double-twist contributions which arise when summing over thermal images of the multi-energy-momentum tensors.

Another important generalisation is the treatment of states with VEVs of operators other than the energy-momentum tensor. The natural way to proceed is to introduce matter fields in the ambient space itself, and relax the $d+2$ dimensional Ricci flatness condition to Einstein's equations with a non-zero energy-momentum tensor for the matter fields. In this case the additional fields provide additional ways to build Weyl invariants and thus additional independent kinematic contributions in correlators.

One of the main technical conveniences of the embedding formalism is constructing correlation functions of operators with spin. We expect the ambient space to be similarly useful and we provided a brief discussion in section \ref{Ssec:SpinningCorr}. The first step down this road is to construct worldlines of spinning particles connecting the two insertion points, {\it i.e.} gyroscopes falling through ambient space. Unlike embedding space, ambient space has non-trivial Riemann curvature and this may lead to interesting new physical contributions in correlation functions with spin, or provide natural geometric interpretations for known expressions. Another tool available to move in this direction are the ambient generalisations of weight- and spin-shifting operators discussed in section \ref{Ssec:SpinningCorr}. 

For thermal CFTs, we placed particular emphasis on computing the 2-point function of scalar operators. The invariants we employed (section \ref{SSec:AcorrBB}) can also be assembled to build scalar $n$-point functions (section \ref{Ssec:Higher-ptf}) and it would be interesting to carry this out explicitly in some examples. We also performed a holographic computation of the 2-point function order by order in small temperature $\beta^{-1}$ which raised several questions. Firstly, are there non-perturbative contributions in $\beta^{-1}$, and how do they affect the construction of an OPE in momentum space? Secondly, what is the analytic structure of the 2-point function for generic $d, \Delta$: in our computation we identified singularities at complex $\tau$ at a radius less than $\beta$. These singularities appear to match those appearing in an asymptotic frequency-space analysis in \cite{Festuccia:2005pi} which originate in the existence of a null singularity that connects the two points \cite{Fidkowski:2003nf} signalling the presence of the black hole singularity. This suggests that the radius of convergence of the thermal OPE be less than $\beta$. 

In this work we also constructed the exact (non-perturbative) thermal holographic 2-point function with numerical methods in section \ref{SSSec_BBNonPert}. As far as we are aware, this constitutes the first computation of double-twist OPE coefficients for the black hole state, confirming that they are non-zero. It would be interesting to explore these correlators more systematically, which may benefit from developing a numerical method more naturally adapted to the delta-function sources such as finite-element methods or defect-adapted coordinate systems analogous to those used for extended sources in \cite{Janik:2015oja}. We have also shown that the sum of images relates the double-twist OPE coefficients to those of the multi-energy-momentum tensor. The exact relation depends on the analytic structure of the 2-point function, and it would be interesting to investigate this further and see if it is possible to reproduce the value of the leading double-twist OPE coefficient obtained non-perturbatively. 

For CFTs on squashed spheres our exploration of Weyl invariant contributions to correlation functions focussed on those associated to multi-energy-momentum  tensors, just as in the thermal case. However, the squashed sphere example is distinguished by the inhomogeneity of the space that the CFT lives on. Thus one should also seek to include ambient space building blocks that capture the contributions of descendants of the energy-momentum tensor. Aside from the ambient space formalism it would also be interesting to extend computations for squashed spheres using other techniques such as conformal perturbation theory in $\alpha$ in CFT, perturbation theory in $\alpha$ in holography using the AdS Taub-NUT metric, non-perturbative numerical computations, and to construct 1-point functions with generic spin in order to obtain OPE expansions of 2-point functions.

Finally, in section \ref{sec_FlatHolography} we discussed the relationship between the ambient space formalism and holography for asymptotically flat spacetimes. It may be fruitful to further develop this connection.

\section*{Acknowledgements}
It is a pleasure to thank Federico Capone, Matthew Dodelson, Kara Farnsworth, Robin Graham, Andrei Parnachev, Nicole Righi and Jakob Salzer for discussions. The work of EP is supported by the Royal Society Research Grants RGF/EA/181054 and RF/ERE/210267. KS and BW are supported in part by the Science and Technology Facilities Council (Consolidated Grant ``Exploring the Limits of the Standard Model and Beyond''). BW is supported by a Royal Society University Research Fellowship.

\appendix

\section{The ambient near-nullcone isometries
} \label{AppendixBIsometries}

As mentioned in section \ref{SubsecWeylTrAndIsometries}, one can show that any conformal Killing vector of the CFT background can be lifted to a near-nullcone isometry on the ambient space. Consider a conformal transformation generated in $d$ dimensions  by $E^{(0)}_j (x)$ satisfying 
\begin{equation}\label{KillingConfEq}
\nabla_i E^{(0)}_j + \nabla_j E^{(0)}_i = 2  \psi \, g_{(0)ij}(x),
\end{equation} 
with conformal factor $\psi(x)=\frac{1}{d}\,\nabla_l E^l_{(0)}$. One can extend it to the ambient space, at least close enough to the nullcone, as an isometry $K$ with components
\be\label{KAmbientExtensions}
K(t,\rho,x) = 
-t \psi(x) \, \p_t + 2\rho \, \psi(x) \, \p_\rho  + E^i(\rho,x)\,  \p_i.
\ee
Here we denote
\begin{equation}\label{KEjXRHO1}
E^j(\rho,x) = E^j_{(0)}(x) +  (\partial_i\psi)
\int_{0}^{\rho} d\rho' g^{ij} (x,\rho'),
\end{equation}
where the integral of the inverse metric expansion yields for the first few orders
\begin{equation}\label{KEjXRHO2}
E^j(\rho,x) = E^j_{(0)}(x) +  (\partial_i\psi)
\left[ g_{(0)}^{ij} \, \rho - P^{ij} \rho^2  + o\left(\rho^2\right)\right].
\end{equation}
In the context of the ALAdS realization it is well known that conformal symmetries are mapped to asymptotic symmetries in the bulk and a statement analogous to \eqref{KAmbientExtensions} can be found for instance in \cite{Papadimitriou2005}.

Consider for example the case of a flat ambient metric \eqref{AspEspaceMetric}. For such a flat CFT background the most general conformal transformation is generated by elements of $SO(1,d+1)$ of the form
\begin{equation}\label{KflatBdryConfTransf}
E_{(0)}=\left[a^i + \omega^i_{\,\;j} x^j + \lambda x^i + b^i x^2 -2 b^k x_k x^i \right] \partial_i,
\end{equation}
with conformal factor $\psi(x) = \lambda - 2 b\cdot x$. The corresponding ambient isometries read
\be \label{AspIsomEsp}
    K(t,\rho,x) = 
 -t \left(\lambda - 2 b\cdot x\right) \p_t + 2 \rho \left(\lambda - 2 b\cdot x\right) \p_\rho + \left( E^i_{(0)} - 2 \rho b^i\right) \p_i
\, .
\ee
These are Lorentz transformations on Minkowski as they appear in the ambient coordinates.

Besides those inducing conformal transformations on the $d$ dimensional background according to \eqref{KAmbientExtensions}, there is an additional class of ambient isometries that are the analogue of translations on the embedding space. In particular, when solving the ambient Killing equations in full generality, \begin{equation}\label{KillingEqAmbient}
\widetilde{\nabla}_M K_N + \widetilde{\nabla}_N K_M =0,
\end{equation}
another class of near-nullcone isometries turns out to be present if the CFT background has a constant scalar curvature $\nabla_i R=0$. These symmetries are parametrised by a scalar function $B(x)$ that must satisfy the equations
\begin{equation}\label{KsecondOrderEqforB}
\left[\nabla_i \partial_j - \frac{1}{d} g_{(0)ij} \, \Box 
+ \frac{1}{d-2}\left(R_{ij} - \frac{R}{d} g_{(0)ij}\right) \right] B(x) =0.
\end{equation}
As usual each integration constant appearing when solving \eqref{KsecondOrderEqforB} corresponds to an additional ambient isometry. In terms of the function $B(x)$ the components of their generators $K^M(t,\rho,x)$ read
\begin{subequations} \label{eq:KisometryB}
	\begin{align}
K^0(x) &= -\frac{1}{d} \left[\Box + \frac{R}{2(d-1)}\right]B,\\
K^\rho (t,\rho,x) &= \frac{1}{t d }\left[
\rho \, \Box + \frac{R}{2(d-1)}\,\rho +d \right] B,  \\
K^i(t,\rho,x) &= \frac{g^{ij}(x,\rho)}{t}
\left[\delta^m_j + \rho P^m_{\;\,j}\right] \partial_m B.
\end{align}
\end{subequations}

For a conformally flat $g_{(0)}$ one can check that they are simply the $d+2$ translations of the embedding space in disguise. 
More generally, the fact that the $d$-dimensional manifold must have constant curvature means that locally it is either a sphere, Euclidean space or a hyperboloid if we assume it to be a complete manifold. In all these cases the ambient space near the nullcone is locally Minkowski and hence there we expect translations to be isometries. In less trivial cases fewer such isometries may exist but nevertheless one can prove in full generality that they form a commutative subalgebra. In practice these additional symmetries are useful to find an adapted set of coordinates to rewrite the ambient space as Minkowski if $d+2$ such isometries exist.

We now wish to make a few comments about the utility of the ambient isometries \eqref{KAmbientExtensions} and \eqref{eq:KisometryB}. As mentioned, they help  find adapted coordinates to describe the ambient geometry. They also play a relevant role when solving the geodesic equations on an ambient space since they provide first integrals of motion that allow one to automatically reduce part of the geodesic equations to first order ODEs (see appendix \ref{AppendixGeodPTranspT} for more details). Finally, these ambient isometries may constrain further the form of  ambient correlators when considering more general states requiring additional ingredients other than the ambient metric and covariant derivatives of the curvature, and may enter themselves as ingredients for ambient building blocks.

\section{Ambient geodesics and parallel transport of $T$}

\label{AppendixGeodPTranspT}

In this appendix we provide more details and results concerning the solution of the ambient geodesic equations and the parallel transport of the homothety $T$ along such geodesics between two arbitrary points $\X_0$ and $\X_1$ on the ambient nullcone, that is $\X_0 = (t_0,0,x^i_0)$ and $\X_1 = (t_1,0,x^i_1)$.

Let us first focus on the geodesic problem. We indicate the geodesic  trajectories with $\X^M(\lambda)$, where $0\leq \lambda\leq1$. The boundary conditions are $\X(0) = \X_0$ and $\X(1) = \X_1$. In this affine parametrisation the velocity is normalised according to $\dot{\X}{}^M \dot{\X}{}^N \tilde{g}_{MN} = C$. Here $C$ is a constant fixed by the boundary conditions and its sign is related to the causal nature of the ambient trajectory. Its norm is the square of the geodesic length between the two points, \begin{equation}\label{}
	\ell(\X_0, \X_1) = \int_{0}^1 d\lambda \, \sqrt{\left| \tilde{g}_{MN} \dot{\X}{}^M \dot{\X}{}^N   \right|} = \sqrt{|C|}.
\end{equation}

The ambient geodesic equations
\begin{equation}\label{app:geod}
\dot{\X}{}^A(\lambda) \, \wt{\nabla}_A \dot{\X}{}^M(\lambda) =0	\quad \Rightarrow \quad \ddot{\X}{}^M(\lambda) + \wt{\Gamma}^M_{AB}(\lambda) \dot{\X}{}^A(\lambda) \dot{\X}{}^B(\lambda) =0,
\end{equation}
lead to
\begin{align}
	\label{PTeq1}\ddot{t} - \frac{1}{2} t g'_{ij} \dot{x}^i \dot{x}^j &=0, \\ 
	\ddot{\rho} + \frac{2}{t} \dot{t}\dot{\rho} -
	\label{PTeq2}\left(g_{ij} - \rho g'_{ij}\right)\dot{x}^i \dot{x}^j &=0, \\
	\label{PTeq3}\ddot{x}^k + \frac{2}{t} \dot{x}^k \dot{t} + \Gamma^k_{ij} \dot{x}^i \dot{x}^j + g^{kl }  g'_{il} \dot{\rho} \dot{x}^i &=0,
\end{align}
where the Christoffel symbols in \eqref{PTeq3} are computed using $g_{ij}(x,\rho)$ at fixed $\rho$. Here the prime denotes a derivative in $\rho$, while the dot stands for a derivative in $\lambda$. The velocity normalization condition reads 
\begin{equation}\label{}
\label{PTeqC} 2 \rho \,\dot{t}^2 +	2 \, t \,\dot{t}\, \dot{\rho} + t^2\, g_{ij}(x,\rho) \dot{x}^i \dot{x}^j = C.
\end{equation}
These equations cannot be integrated in full generality. As customary some of them can be reduced to first order ODEs using the ambient isometries $K^{(i)}_M$  described in appendix \ref{AppendixBIsometries}, if any is present. They lead to integrals of motion whose value is fixed by the boundary conditions of the problem,
\begin{equation}\label{}
	Q_i = K^{(i)}_M(\lambda) \dot{\X}{}^M(\lambda)\,.
\end{equation}

The geodesic equations can however be partially solved on general grounds. From \eqref{PTeq1} and \eqref{PTeqC} we can extract $g_{ij} \dot{x}^i \dot{x}^j$ and  $g'_{ij} \dot{x}^i \dot{x}^j$ as functions of $\rho$ and $t$. Plugging them into \eqref{PTeq2} one finds
\begin{equation}\label{}
	\ddot{\rho} + 4 \frac{\dot{t}}{t} \dot{\rho} - \frac{C}{t^2} + 2 \rho \frac{\dot{t}^2}{t^2} + 2 \rho \frac{\ddot{t}}{t} =0 \,,
\end{equation}
which can  be easily integrated. Imposing the boundary conditions specified above, the solution reads
\begin{equation}\label{eq:rhoLambdaGeods}
	\rho(\lambda) = -C \, \frac{\lambda(1-\lambda)}{2 t(\lambda)^2}\,.
\end{equation}
Observe that the sign of $C$ determines the sign of $\rho$, and this in turn determines the region of the ambient space where the geodesic is moving through -- either the region with ALAdS foliation or the one with ALdS slices. With $C>0$ the geodesic explores the ALAdS foliation and this region the equation \eqref{eq:rhoLambdaGeods} can be rewritten in terms of the coordinates \eqref{AspCoordsAdSSlicing} as
\be \label{eq:sLambdaGeods}
s(\lambda) =  t \sqrt{-2 \rho}= \sqrt{C \lambda(1-\lambda)},
\ee
so that the trajectory along $s$ is completely specified once we fix $C$. This relation turns out to be sufficient to find the explicit expression of the ambient invariant $\X_{ij}$.

Let us now move to the parallel transport of the homothetic vector $T$ from $\X_0$ to $\X_1$, defined by the equations
\begin{equation}\label{app:parellel}
	\dot{\X}{}^M(\lambda) \, \wt{\nabla}_M T^{A}(\lambda) =0 \,,
\end{equation}
which lead to
\begin{align}
	\label{PTeq4}\p_\lambda T^0 - \frac{t}{2} g'_{ij}\dot{x}^i T^j &=0,\\
	\label{PTeq5}\p_\lambda T^\rho +\frac{\dot{t}}{t} T^\rho +\frac{\dot{\rho}}{t} T^0 + (-g_{ij} + \rho g'_{ij}) \dot{x}^i T^j  &=0,\\
	\label{PTeq6}\p_\lambda T^l + \frac{1}{t} \dot{x}^l T^0 + \frac{1}{t} \dot{t} T^l + \frac{1}{2} g^{lm} g'_{jm} \left(
	\dot{x}^j T^\rho +\dot{\rho} T^j
	\right)
	+ \Gamma^l_{ij} \dot{x}^i T^j &=0.
\end{align}
One of them can be automatically integrated in view of the fact that the norm of $T$ must stay constant along a geodesic, and in this case the constant is zero (as its norm at $\X_0$ vanishes). Therefore $T^M(\lambda) T_M(\lambda) = 0$ entails
\begin{equation}\label{}
	\label{PTeq7}2 \rho T^0(\lambda)^2 + 2 t T^0(\lambda) T^\rho(\lambda) + t^2 g_{ij}(x,\rho) T^i(\lambda)T^j(\lambda) =0.
\end{equation}
Equations \eqref{app:geod} and \eqref{app:parellel} imply that $\dot{\X}{}^M T^N \tilde{g}_{MN}$ stays constant along the geodesic. Explicitly,
\begin{equation}\label{PTeqQ}
	2 \rho \dot{t} T^0 + t \dot{t} T^\rho +t \dot{\rho} T^0 + t^2 g_{ij}(x,\rho) \dot{x}^i T^j = -\frac{C}{2}.
\end{equation}
where the boundary conditions and equation \eqref{eq:rhoLambdaGeods} fix the constant on the r.h.s.

Observe that from \eqref{PTeq4} and \eqref{PTeqQ} we can obtain $g'_{ij}\dot{x}^i T^j$ and $g_{ij}(x,\rho) \dot{x}^i T^j$ in terms of $T^0$ and $T^\rho$. If we plug them into \eqref{PTeq5}, the resulting equation can be integrated in terms of $ T^\rho + \frac{2 \rho}{t} T^0$ to yield
\begin{equation}\label{eq:TrhoPTransport}
	T^\rho = - \frac{2 \rho}{t} T^0 - \frac{C \lambda}{2 t^2} \,.
\end{equation}

As anticipated these general features of the solutions are enough to compute  the invariant $\X_{ij}$. Since the homothetic vector at $\X_1$ has components $T_{(1)}^M = (t_1,0,0)$, using the ambient metric at $\X_1$ one has
\begin{equation}\label{}
	\X_{01} = -2 \, \T_{(0)}^M T_{(1)}^N \, \tilde{g}^{(1)}_{MN} = - 2 \, t_1^2 \, \T^\rho_{(0)}.
\end{equation}
Hence, evaluating \eqref{eq:TrhoPTransport} at $\lambda=1$ we can conclude that $\X_{ij} = C = \ell(\X_i,\X_j)^2$ as claimed in the main text.

This entails that solving parallel transport is not required to construct the invariant $\X_{ij}$ (even though it generally is when constructing other ambient invariants). One has to simply solve geodesic equations and extract $C$ from the norm of the velocity $\dot{\X}{}^M \dot{\X}{}^N \tilde{g}_{MN} = C$ at any point $\lambda$.

\section{Relation between ambient and AdS geodesics} \label{AppAmbientAdSGeods}

Consider an ambient space of the form \eqref{AspMetricAdSSlicing}. We intend to relate the geodesic distance between two nullcone points $\X_0 = (t_0,0,x^i_0)$ and $\X_1 = (t_1,0,x^i_1)$ on the ambient space with the geodesic distance $L_{AdS}$ between those same points $x^i_0$ and $x^i_1$ at the boundary of a single Euclidean ALAdS slice. In particular we wish to show that 
\begin{equation} \label{GeodApproxAdSAmbient_app}
    \frac{1}{(\X_{01})^\Delta} = \left. \frac{r^{-2\Delta}}{(t_0 t_1)^\Delta} e^{-\Delta L_{AdS}}\right|_{r=0}.
\end{equation}
We assume $t_0, t_1 > 0$ meaning that we are interested in spacelike geodesics on the ambient space. The same result can be attained with completely analogous computations in the cases of null and timelike ambient geodesics. We choose a slightly different parametrization consisting in a rescaling of the one used in appendix \ref{AppendixGeodPTranspT},
\begin{equation}
    \tilde{g}_{AB}  \dot{\X}{}^A \dot{\X}{}^B =1.
\end{equation}
We take advantage of the parent description 
\begin{equation} \label{S_AParentAction}
   S_A =  \frac{1}{2} \int_\gamma d\lambda \left[ \frac{1}{e}  \tilde{g}_{AB}  \dot{\X}{}^A \dot{\X}{}^B + e \right],
\end{equation}
where one can put the einbein onshell compatible with the parametrization above by setting
\begin{equation}
    e = \sqrt{\tilde{g}_{AB}  \dot{\X}{}^A \dot{\X}{}^B} = 1.
\end{equation}
The ambient action is then equal to
\begin{equation}
    S_A =  \int_0^{L} d\lambda \sqrt{\tilde{g}_{AB}  \dot{\X}{}^A \dot{\X}{}^B} =  L,
\end{equation}
{\it i.e.} $S_A$ is simply equal to the ambient geodesic length $L$, with $L^2 = C = \X_{12}$.

As shown in appendix \ref{AppendixGeodPTranspT} the presence of the homothetic vector $T=s\p_s$ allows one to automatically integrate the geodesic equation along $s$. In the current parametrization this entails $s(\lambda) = \sqrt{\lambda (L-\lambda)}$. Regulating the integrals by taking the domain $\lambda \in (\varepsilon_0, L-\varepsilon_1)$ (eventually to be set to $\varepsilon_0 \sim \varepsilon_1 \sim \varepsilon \to 0$) we can set $s(\lambda)$ onshell in \eqref{S_AParentAction} (after setting $e=1$) and rewrite $S_A$ as,
\begin{equation}
    S_A = L+    
    \frac{L}{4} \log\frac{\sqrt{\varepsilon_0 \varepsilon_1}}{L} + \frac{1}{2} \int_{\varepsilon_0}^{L-\varepsilon_1} d\lambda \, \lambda(L-\lambda) g^+_{\mu\nu} \dot{x}^\mu \dot{x}^\nu,
\end{equation}
where as customary $g^+_{\mu\nu}$ is the metric on an ALAdS slice. By rewriting the integral using the new parametrization
\begin{equation}
    p(\lambda) = \frac{1}{L} \log \frac{\lambda}{L-\lambda},
\end{equation}
and using $S_A=L$ one obtains the constraint
\begin{equation} \label{eq:constr}
    \frac{L}{4} \log\frac{\sqrt{\varepsilon_0 \varepsilon_1}}{L} + \frac{1}{2} \int_{\frac{1}{L} \log \frac{\varepsilon_0}{L}}^{-\frac{1}{L} \log \frac{\varepsilon_1}{L}} dp \; g^+_{\mu\nu} \dot{x}^\mu \dot{x}^\nu =0 \,.
\end{equation}

Normalizing the velocity on the Euclidean ALAdS slice as
\begin{equation}
    g^+_{\mu\nu} \dot{x}^\mu \dot{x}^\nu = q^2,
\end{equation}
for some constant $q$, which we will determine shortly, the parent action for a trajectory on such $(d+1)$ dimensional slice reads
\begin{equation} \label{eq:LAdS1}
    L_{AdS} = S_{EAdS} = \frac{1}{2q} \int_{\frac{1}{L} \log \frac{\varepsilon_0}{L}}^{-\frac{1}{L} \log \frac{\varepsilon_1}{L}} dp \; g^+_{\mu\nu} \dot{x}^\mu \dot{x}^\nu -  \frac{q}{L} \log \frac{\sqrt{\varepsilon_0 \varepsilon_1}}{L}.
\end{equation}
One also has 
\begin{equation} \label{eq:LAdS2}
    L_{AdS} =  \int_{\frac{1}{L} \log \frac{\varepsilon_0}{L}}^{-\frac{1}{L} \log \frac{\varepsilon_1}{L}} dp \; \sqrt{g^+_{\mu\nu} \dot{x}^\mu \dot{x}^\nu} =  - 2 \frac{q}{L} \log \frac{\sqrt{\varepsilon_0 \varepsilon_1}}{L}.
\end{equation}
We can then use the constraint \eqref{eq:constr} from the ambient space to fix $q$. Indeed, using \eqref{eq:LAdS1} and \eqref{eq:LAdS2} one finds from \eqref{eq:constr} that $q = L/2$, entailing 
\begin{equation} \label{eq_LAdS}
    L_{AdS} = - \log \frac{\sqrt{\varepsilon_0 \varepsilon_1}}{L}.
\end{equation}
Finally, note that in the current parametrization of geodesics,
\begin{equation} \label{eq_rLambda}
r(\lambda)= \frac{s(\lambda)}{t(\lambda)} = \frac{1}{t(\lambda)}\sqrt{\lambda(L-\lambda)}.
\end{equation}
From the boundary conditions, close to $\lambda=0$ and $\lambda=L$ one has respectively
\begin{align}
    \lambda \to 0&: \qquad t(\lambda)= t_0 + O(\lambda)\,, \\
    \lambda \to L&: \qquad  t(\lambda)= t_1 + O(L-\lambda)\,.
\end{align}
Using equation \eqref{eq_rLambda}, they lead to the following radial positions of the regulation parameters on an ambient geodesic, 
\begin{equation}
    r(\varepsilon_0) = \frac{1}{t_0} \, \sqrt{\varepsilon_0 L} + O(\varepsilon), \qquad \quad r(L-\varepsilon_1) = \frac{1}{t_1} \, \sqrt{\varepsilon_1 L} + O(\varepsilon), 
\end{equation}
Thus we can write the geodesic length $L_{AdS}$ in equation \eqref{eq_LAdS} in terms of the Fefferman-Graham radial position of the two endpoints as
\begin{equation}
    L_{AdS} = - \left. \log \frac{t_0 t_1 r^2}{L^2} \right|_{r \to 0}
    \quad \Rightarrow \quad L^2 = t_0 t_1 r^2 e^{L_{AdS}}\Big |_{r \to 0},
\end{equation}
which given $L^2 = \X_{01}$, precisely reproduces the expected relation \eqref{GeodApproxAdSAmbient_app} between the geodesic approximation on ALAdS and ambient spaces.

\section{Details on the curvature invariants at finite temperature} \label{App:DetailsCurvatureInvFiniteT}

Let us now discuss the completeness of the basis for weighted curvature invariants as provided by the weight-0 scalars \eqref{eq:Generic0weightCurvInv} in the finite temperature case presented in section \ref{sec:thermal}. Compatibly with equations \eqref{ARiemannCompg0ConfFlat}-\eqref{ARiemannCompg0ConfFlatE3}, the only non-vanishing components of the ambient Riemann are
\begin{subequations}
\begin{align}
       \wt{R}_{\rho j k\rho} &= \frac{d}{4} \left( \frac{d}{2}-1\right) g_{(d)jk} \, \rho^{\frac{d}{2}-2} t^2, \\
        \wt{R}_{i j k l} &= \frac{d}{4} \,
        \left[ 
        \delta_{il} g_{(d)jk} + \delta_{jk} g_{(d)il}-
        \delta_{ik} g_{(d)jl}-
        \delta_{jl} g_{(d)ik}
        \right] \,
        \rho^{\frac{d}{2}-1} t^2.
\end{align}
\end{subequations}
The subleading orders in $\rho$ in equations \eqref{ARiemannCompg0ConfFlat}-\eqref{ARiemannCompg0ConfFlatE3} are proportional to $d$-dimensional covariant derivatives acting on $g_{(0)}$ and $g_{(d)}$. In this case $g_{(0)ij}$ is the flat metric and $g_{(d)}$ is a constant tensor and hence the expansion in $\rho$ of the ambient Riemann truncates at the leading order.

Because of the homogeneity in $t$ of the Riemann and of the fact that the geometry does not depend on the boundary directions $x^i$, the action of the ambient covariant derivative on the Riemann decomposes as a derivative along $\rho$ plus terms which are proportional to the Riemann. Schematically,
\begin{equation} \label{BBactionACov}
    \wt{\nabla}_M \Riem
    = \delta^\rho_M \, \p_\rho \Riem + \Riem,
\end{equation}
and the same holds for higher order derivatives. Focusing on $d=4$, this means that the only independent tensorial structures containing the energy-momentum tensor VEV $g_{(d)ij}$ are $\delta_{ij} g_{(d)jk}$ (and symmetrisations) and $g_{(d)ij}$ itself, and they can be extracted from $\Riem|_{\rho=0}$ and $\wt{\nabla}\Riem|_{\rho=0}$. Higher order derivatives of the Riemann simply yield different linear combinations of those two structures. This entails that $\mathcal{R}^{(0)}$ and $\mathcal{R}^{(1)}$ are the only independent objects that one needs to construct the weight-0 invariants \eqref{eq:Generic0weightCurvInv}.
As a consequence, any order two weight-0 invariant can be written as a linear combination of the scalars $e_0$, $e_1$,$e_2$ defined in \eqref{BB_2ndOrderCurvatureInvariants}.

At a given order $\beta^{-nd}$ the weight-0 invariants are to be constructed as a chain of $n$ $\mathcal{R}$'s,
\be 
\mathcal{R}^{(r_1)\; M_2}_{\;\;\; M_1} \,
\mathcal{R}^{(r_2)\; M_3}_{\;\;\; M_2}
\dots \mathcal{R}^{(r_n)\; M_1}_{\;\;\; M_n}
,
\ee
where for each of them one has two possible choices, $r_i=0,1$. This means that they provide at most $2n$ different invariants, modulo cyclic permutations. Our aim is to reproduce the whole set of multi-energy-momentum tensor blocks entering the thermal OPE at order $\beta^{-nd}$. Each of them is proportional to a Gegenbauer polynomial $C^{(1)}_J$, with even $J=0 \dots 2n$, hence at this order one needs $n+1$ independent invariants.

At order $n=2$ this entails that the three independent scalars $e_0$, $e_1$,$e_2$ in \eqref{BB_2ndOrderCurvatureInvariants} form a basis of ambient invariants. At a generic order $n$, based on the counting above the weight-0 curvature scalars are in principle able to form an over-complete basis. We have checked explicitly that they generate a basis of $n+1$ invariants in $d=4$ up to $n=6$, and one may check it to arbitrarily high order $n$. 

With similar arguments based on the action \eqref{BBactionACov} of ambient covariant derivatives, this discussion can be easily extended to any even $d\geq4$, where now the two independent objects are $\mathcal{R}^{(d/2-2)}$ and $\mathcal{R}^{(d/2-1)}$ and any weight-0 invariant is built as a chain of them.

\section{Perturbative thermal holographic correlator for general $d$ and $\Delta$}
\label{App:HoloCorrgeneraldD}

In this appendix we solve the inhomogeneous second order differential equation \eqref{BBKG_1stOeq} for the first order correction $b_1$. We will consider here the case of generic $d$ and $\Delta$ with non-integer $\kappa$. Using the leading order solution, the onshell source on the RHS of \eqref{BBKG_1stOeq} reads
\begin{equation}\label{}
S(r) = -\sqrt{\frac{2}{\pi }} r^{\frac{3 d}{2}-2} \cos\!\left(\frac{2\kappa-1}{2} \pi\right) \left[K_{\kappa}(r) \left(\Delta ^2+\left(\eta ^2+1\right) r^2\right)-d r K_{\kappa+1}(r)\right],
\end{equation}
while the Wronskian of the homogeneous solutions is 
\begin{equation}\label{}
	W(u_1,u_2) = \cos\!\left(\frac{2\kappa-1}{2} \pi\right) r^{d-1}.
\end{equation}
Using the method of variation of parameter, the first order solution is of the form
\begin{equation}\label{}
	b_1(r) = (A(r)+a_1) u_1 + (B(r)+a_2) u_2\,,
\end{equation}
where $a_1$ and $a_2$ are integration constants while 
\begin{subequations} \label{BBAnB}
	\begin{align}
		A(r) &= - \int_0^r dr' \, \frac{u_2(r') S(r')}{W(r')}  \\
     &= - d \,\mathcal{I}^{(1)} (d,\kappa,1) + \Delta^2 \,\mathcal{I}^{(1)} (d-1,\kappa,0) +(1+\eta^2)  \,\mathcal{I}^{(1)} (d+1,\kappa,0)\,, \nonumber \\
	B(r) &=  \int_0^r dr' \, \frac{u_1(r') S(r')}{W(r')} \\
 & =  d \, \mathcal{I}^{(2)} (d,\kappa,1) - \Delta^2 \,\mathcal{I}^{(2)} (d-1,\kappa,0) -(1+\eta^2) \,\mathcal{I}^{(2)} (d+1,\kappa,0)\,. \nonumber
	\end{align}
\end{subequations}
Here we defined the following class of integrals involving two Bessel functions,
\begin{equation}\label{}
	\mathcal{I}^{(\ell)} (\alpha,\kappa,\delta)  = \int_0^r dr' \; r^{\prime\alpha} \, I_{(-1)^{\ell+1}\left(\kappa\right)}(r') \, K_{\kappa +\delta}(r')\,.
\end{equation}
The explicit expressions for these integrals are
{\small \begin{subequations}
	\begin{align}
\mathcal{I}^{(1)}(\alpha,\kappa,0) &= 
\frac{r^{\alpha +1} }{2 \kappa (\alpha +1)} \, _2F_3\left(\frac{1}{2},\frac{\alpha +1 }{2};1-\kappa,\kappa+1,\frac{\alpha +3 }{2};r^2\right)
\\
&\!\!\!\!\!\!\!\!\!\!\!\!\!\!\!\!\!\!\!\!\!\!\!\!\!\!\!\!\!\!\!\!\!  +\frac{2^{-2 \kappa-1} \Gamma \left(-\kappa\right) r^{\alpha +2 \kappa+1} }{(\alpha +2 \kappa+1) \Gamma \left(\kappa+1\right)} \, _2F_3\left(\kappa+\frac{3}{2},\frac{\alpha }{2}+\kappa+1;\kappa+2,2 \kappa+2,\frac{\alpha }{2}+\kappa+2;r^2\right), \nonumber \\
\mathcal{I}^{(1)}(\alpha,\kappa,1) &= \frac{r^{\alpha }}{2 \alpha }+
\frac{r^{\alpha } }{2 \alpha } \, _2F_3\left(\frac{1}{2},\frac{\alpha }{2};-\kappa,\kappa+1,\frac{\alpha }{2}+1;r^2\right)
\\
&\!\!\!\!\!\!\!\!\!\!\!\!\!\!\!\!\!\!\!\!\!\!\!\!\!\!\!\!\!\!\!\!\! 
+\frac{\pi  2^{-2(\kappa+1)} \sec\!\left(\frac{2\kappa-1}{2} \pi\right)
r^{\alpha +2 \kappa+2}}{(\alpha +2 \kappa+2) \Gamma \!\left(\kappa+1\right) \Gamma \!\left(\kappa+2\right)} \, _2F_3\left(\kappa+\frac{3}{2},\frac{\alpha }{2}+\kappa+1;\kappa+2,2\kappa+2,\frac{\alpha }{2}+\kappa+2;r^2\right)
, \nonumber \\
\mathcal{I}^{(2)}(\alpha,\kappa,\delta) &= \frac{\pi  2^{-\delta -1} \sec \left(\left(\delta+\kappa-\frac{1}{2}\right)\pi\right) r^{\alpha -\delta }}{\Gamma \left(1-\kappa\right)} \Bigg[ 
-\frac{2^{2 \delta +2 \kappa} r^{-2 \kappa+1} }{(-\alpha +\delta +2 \kappa-1) \Gamma \left(1-\delta -\kappa\right)}  \,\times
\\
& \!\!\!\!\!\!\!\!\!\!\!\!\!\!\!\!\!\!\!\!\!\!\!\!\!\!\!\!\!\!\!\!  _3F_4\!\left(\!-\kappa+\frac{1-\delta }{2},-\kappa-\frac{\delta }{2},-\kappa +\frac{\alpha -\delta +1}{2};1-\kappa,-2 \kappa+1-\delta ,1-\delta -\kappa,\frac{3}{2}-\kappa+\frac{\alpha -\delta}{2};r^2\!\right)
\nonumber \\
&\!\!\!\!\!\!\!\!\!\!\!\!\!\!\!\!\!\!\!\!\!\!\!\!\!\!\!\!\!\!\!\!\! 
-\frac{r^{2 \delta +1} \, _3F_4\left(\frac{\delta +1}{2},\frac{\delta }{2}+1,\frac{\alpha + \delta +1}{2};1-\kappa,\frac{\alpha + \delta +3 }{2},\delta +1,\kappa+\delta +1;r^2\right)}{(\alpha +\delta +1) \, \Gamma \left(\kappa+\delta +1\right)}
\Bigg]
, \nonumber 
	\end{align}
\end{subequations}
}

We have fixed the source at the leading $O(\epsilon^0)$ order, hence the order corresponding to the source in the near-boundary Fefferman-Graham expansion of the function $b_1$ must vanish. Integrating over $r' \in (0,r)$ as in equation \eqref{BBAnB} this is automatically true for $a_1=0$. The remaining integration constant $a_2$ is fixed by imposing regularity in the bulk interior $r\to \infty$. By studying the large--$r$ behaviour of $b_1$ this fixes
\begin{equation}\label{}
	a_2 =\frac{\pi ^{3/2} \left(d \eta ^2-1\right) \cot \left(\frac{\pi  d}{2}\right) \Gamma \left(-\frac{d}{2}-\frac{1}{2}\right) \csc (\pi  \Delta ) \sin \left(\frac{1}{2} \pi  (d-2 \Delta )\right) \csc (\pi  (d-\Delta ))}{4 \Gamma \left(1-\frac{d}{2}\right) \Gamma (-\Delta ) \Gamma (\Delta -d)}\,.
\end{equation}
The resulting holographic correlator to first order in $\epsilon$ is in equation \eqref{BBKG_1stSOLGeneraldD}. These results were first presented in \cite{Parisini:2022wkb}.
A similar approach was taken in \cite{Bajc:2022wws}, whose results match \eqref{BBKG_1stSOLGeneraldD} and \eqref{BBKG_1stSOLGeneraldD_positionspace}.

\section{Computation of the double-twist coefficients from the multi-energy-momentum tensor spectrum}
\label{AppendixDoubleTwists}

In this appendix we provide details on how to perform the sum over images \eqref{BBKG-CompactF2SumImages_GPLUS} to obtain the thermal correlator \eqref{BBKG_FullPosSpaceCorrThOPE}. 

First we consider the analytic part of $G_+$ in \eqref{BB_NonCompCorrx0_Split}. Setting for the moment $\beta=1$, we thus have to evaluate the sum
\begin{equation}
    S_\gamma(\tau) = \sum_{m =1}^\infty 
    \left| m+\tau \right|^{\gamma},
\end{equation}
with $\gamma = - 2 \Delta + n d$. As it is, this sum converges for $\gamma<-1$, however we can analytically continue it using the Hurwitz $\zeta$ function, in terms of which it reads
\begin{equation}
    S_\gamma(\tau) =  \zeta(-\gamma,1+\tau)\,.
\end{equation}
This expression is finite on the whole complex $\gamma$-plane except for a simple pole at $\gamma = -1$. To avoid it, it is sufficient to pick $\Delta \neq \mathtt{n} \, \frac{d}{2}
 + \frac{1}{2}$ for all non-negative integer $\mathtt{n}$. 

Using the expansion of the Hurwitz $\zeta$ for $\gamma \neq -1$ and $|\tau|<1$, 
\begin{equation} \label{eq_Hurwitz1}
    \zeta(-\gamma,1-\tau) = \sum_{p=0}^\infty \frac{\Gamma(p-\gamma)}{p! \, \Gamma(-\gamma)} \, \zeta(p-\gamma) \, \tau^p \, ,
\end{equation}
the sum over images of the analytic part then reads,
\begin{equation}
    \sum_{m =1}^\infty \, \sum_{n=0}^\infty \, 
    \frac{\tilde{a}_{n}^{(T)}}{\beta^{2\Delta}} \left(\frac{\tau}{\beta} +m \right)^{nd-2\Delta} = 
     \frac{1}{\beta^{2\Delta}}
    \sum_{p=0}^\infty Q_{\text{reg},\,p}^{(OO)} \frac{\tau^{p}}{\beta^{p}} \,,
\end{equation}
where we define the coefficients
\begin{equation}
    Q_{\text{reg},\, p}^{(OO)} = 
    \sum_{n=0}^\infty (-1)^p \frac{\Gamma(p+2\Delta-nd)}{p! \, \Gamma(2\Delta-nd)} \, \zeta(p+2\Delta-nd) \, \tilde{a}_n^{(T)}\,,
\end{equation}

Turning to the singular part of $G_+$ in \eqref{BB_NonCompCorrx0_Split}, for the moment we consider terms with poles of arbitrary positive integer order $\mu_{(\ell)}$,
\begin{equation}
    W_\ell(\tau) = \frac{1}{|\tau|^{2\Delta}} \frac{A_{(\ell)}}{
    \left(\left|\tau/\beta\right|^d - y_\ell\right)^{\mu_{(\ell)}}
    }\,.
\end{equation}
The sum over images that we wish to evaluate is then
\begin{equation}
    \sum_{m=1}^\infty W_\ell(\tau + m) = \sum_{m=-\infty}^\infty\frac{1}{(\tau+m)^{2\Delta}} \frac{A_{(\ell)}}{
    \big( (\tau+m)^d - y_\ell\big)^{\mu_{(\ell)}}
    }\,,
\end{equation}
where to simplify the presentation we have set $\beta=1$ -- we will reinstate $\beta$ at the end of the analysis. 
Due to the singularities, to ensure convergence we split this sum based on whether $\tau+m$, for any $\tau$ inside the circle of radius $|y_\ell|^{1/d}$, is inside or outside the circle of radius $|y_\ell|^{1/d}$. If $n^*-1 < |y_\ell|^{1/d} < n^*$, where $n^*$ is an integer, then $\tau+m$, with $m \geq n^*$, is outside the circle, and otherwise inside.  We thus 
split the sum as
\begin{equation}
    \sum_{m=1}^\infty W_\ell(\tau + m) = 
     Z_0(\tau) + Z_+(\tau)
    \,,
\end{equation}
with
\begin{equation}
Z_{0} (\tau) = \sum_{m=1}^{n^*-1} W_\ell(\tau + m)
     \,,\qquad \quad
 Z_{+} (\tau) = \sum_{m= n^*}^{ \infty} 
     W_\ell(\tau + m)  \,.
\end{equation}

Starting with $Z_+$, the range of $m$ in the sum guarantees that we can expand each summand as
\begin{equation}
\frac{1}{
    \big[ (\tau+m)^d - y_\ell\big]^{\mu_{(\ell)}}
    } = \sum_{j=0}^\infty \binom{\mu_{(\ell)}+j-1}{j}  \, (y_\ell)^j \,
    (\tau+m)^{-d(\mu_{(\ell)}+j)} \, .
\end{equation}
We can then rewrite
\begin{align}
    Z_+(\tau) &=  A_{(\ell)} \sum_{m=n^*}^\infty \sum_{j=0}^\infty \binom{\mu_{(\ell)}+j-1}{j}
    \, (y_\ell)^j \, (\tau+m)^{-d(\mu_{(\ell)}+j)-2\Delta} \,, \\
   &= A_{(\ell)} \sum_{j=0}^\infty \binom{\mu_{(\ell)}+j-1}{j}
    \, (y_\ell)^j \, \zeta\Big(d(\mu_{(\ell)}+j)+2\Delta, \tau +n^*\Big) \,,
\end{align}
which converges for $\Delta \neq \mathtt{n} \frac{d}{2} +\mu_{(\ell)}\,$ with integer $\mathtt{n}$. Using the expansion in $\tau$ of the Hurwitz $\zeta$,
\begin{equation} \label{eq_Hurwitz2}
  \zeta(y, \tau +n^*) = \sum_{p=0}^\infty \frac{(-1)^p}{p!} \, (y)_p \,
  \zeta(p+y, n^*) \, \tau^p \,,
\end{equation}
where $(a)_b$ indicates the Pochhammer symbol, the sum over images in the region outside the singularity takes the form
\begin{align}
    Z_+  = A_{(\ell)} \sum_{p=0}^\infty & \Bigg[ \frac{(-1)^p}{p!} \, \sum_{j=0}^\infty 
    \binom{\mu_{(\ell)}+j-1}{j}
    \, (y_\ell)^j \, \big(d(\mu_{(\ell)}+j)+2\Delta \big)_{p}  \\
    &
    \quad \times \, \zeta\Big(p+d(\mu_{(\ell)}+j)+2\Delta, n^*\Big)
    \Bigg] \tau^{p} \,. \nonumber
\end{align}
The finite sum $Z_0$ can be expanded in powers of $\tau$ in an analogous way, with the only difference that given $ |\tau+m| < n^*$  one must expand
\begin{equation}
    \frac{1}{
    \big[ (\tau+m)^d - y_\ell\big]^{\mu_{(\ell)}}
    } =
    \sum_{j=0}^\infty (-1)^j 
    \binom{\mu_{(\ell)}+j-1}{j}
    \, (\tau+m)^{d j} (-y_\ell)^{-\mu_{(\ell)} -j} \, ,
\end{equation}
so as to ensure convergence. Reinstating the factors of $\beta$, the sum over images of each singular piece hence takes the form,
\begin{equation}
 \sum_{m=1}^\infty W_\ell(\tau+m \beta) 
=  \frac{1}{\beta^{2\Delta}} \sum_{p=0}^\infty \, Q^{(OO)}_{(\ell)\, p} \,\frac{\tau^{p}}{\beta^{p}} \, ,   
\end{equation}
where we defined the coefficients
\begin{align}
     &Q^{(OO)}_{(\ell)\, p} =  \frac{ (-1)^p A_{(\ell)}}{p!}
     \sum_{j=0}^\infty  \binom{\mu_{(\ell)}+j-1}{j} \Bigg[ 
     \\ 
     &  \quad  \,\, (y_\ell )^j \, \big( d(\mu_{(\ell)} +j ) + 2\Delta \big)_{p} \, \zeta\big(p +d(\mu_{(\ell)} +j ) + 2\Delta , n^* \big) \nonumber \\
     &  + (-1)^j (-y_{\ell})^{-\mu_{(\ell)}-j} \, (2 \Delta - dj)_{p} \, \bigg( 
     \zeta \big(p+2\Delta-dj\big) - \zeta\big(p+2\Delta -dj, n^* \big) \bigg) \Bigg] \,. \nonumber
\end{align}
This expression has been obtained assuming integer order $\mu_{(\ell)}$. However, one may check that the results hold for all real positive $\mu_{(\ell)}$, thus allowing for branch points in the complex $\tau$-plane.

Overall, summing these contributions according to \eqref{BBKG-CompactF2SumImages_GPLUS}, only even powers of $\tau$ survive and one finds the thermal correlator \eqref{BBKG_FullPosSpaceCorrThOPE}.

\section{Non-perturbative geodesics on the planar black hole} \label{AppendixLongGeodesics}

In section \ref{SSec_BBHoloCorr} we focused on 2-point functions with the insertion points close to each other, $|x|/\beta \ll 1$. From the ambient perspective, this meant that we considered short geodesics of length of the same order as $|x|/\beta \ll 1$. Such geodesics can be obtained as perturbations in $|x|^d/\beta^d$ of geodesics on Minkowski spacetime, as shown in section \ref{SSec_B_Geods}. This implies that we do not expect to account for the double-twist spectrum since, as discussed in section \ref{SSSec_BBNonPert}, the double-twist spectrum is related to non-perturbative effects in $|x|/\beta$ and one has to take into account global properties of the background (in this case the periodicity of the $\tau$ direction) to fully describe them.

As mentioned in section \ref{SSSec_BBNonPert} a possibility to describe the double-twist spectrum in terms of the ambient formalism is that  long geodesics exist on the ambient space \eqref{eq:AmbientSpaceBrane}. By long geodesics here we mean ambient geodesics that connect nullcone points that are close to each other, and whose length is long compared to $|x|/\beta$. More explicitly, the length of such geodesics would scale like $\beta$ instead of the short geodesic, which scale like $|x|$ (see equation \eqref{eq:BBXij}). For example, this would be the case for geodesics that wrap the thermal circle. If this class of ambient geodesics existed, the double-twist contributions in the ambient correlator would likely emerge from the sum over geodesics of the ambient curvature invariants, paralleling what happens in the holographic correlator \eqref{eq:sum_of_images} where they arise from the sum over images of the multi-energy-momentum tensor spectrum. This is however not the case and we do not find any such geodesic as we detail below.

One can study long geodesics on the planar black hole background by finding exact solutions to the geodesic equations \eqref{eq:BBGeodEqtaux1}-\eqref{eq:BBGeodEqz} with boundary conditions \eqref{eq:BBsmallTt01BC0}. As already mentioned in the main discussion, generic solutions to \eqref{eq:BBGeodEqz} are  in terms of inverse elliptic functions and thus not easily tractable. Nonetheless, if one restricts to trajectories moving only along the $\tau$ direction (i.e. setting $x(\lambda)=0$ as an initial condition, meaning that $A_2=0$ in \eqref{eq:BBGeodEqtaux1}), the equation for $z(\lambda)$ is explicitly solvable for any $A_1$. Defining the dimensionless parameter $A= \frac{\sqrt{2}}{\pi} \beta A_1$, one finds
\be
z(\lambda)= \frac{\sqrt{2}}{\pi} \sqrt{\frac{(1-\lambda ) \lambda }{2 A^2 (1-\lambda) \lambda +\sqrt{1+A^4} (1-2 (1-\lambda) \lambda )}} \; \beta \,,
\ee
where we set the endpoints to lie on the same slice of the ambient nullcone, $t_0=t_f=1$. Through \eqref{eq:BBGeodEqtaux1} this leads to
\be \label{BB_longGeodTau}
\tau(\lambda) = \frac{\tau_f}{2} + \frac{\beta}{2 \pi } \left( \arctan\left[ Y_- \left(\lambda-\frac{1}{2}\right)\right]
+  \mbox{arctanh} \left[ Y_+ \left(\lambda-\frac{1}{2}\right)\right]\right)\,,
\ee
where
\be
Y_\pm = \sqrt{2} \, \frac{1\pm(A^2-\sqrt{A^4+1})}{ A}.
\ee
In expression \eqref{BB_longGeodTau} we still have to impose the condition $\tau(1)=\tau_f$, which fixes the value of the integration constant $A$. Renaming $\tau_f \to \tau$ and considering trajectories with $A>0$, the relation that $A$ must satisfy is, 
\be \label{eq:BBExactGeodFinalBC}
\frac{\tau }{\beta} = \frac{1}{\pi}
\left[\arctan \left(\frac{Y_-}{2}\right)+\mbox{arctanh}\left(\frac{Y_+}{2}\right)\right].
\ee
This equation is transcendental and cannot be inverted analytically to obtain an expression for $A(\tau)$. Nonetheless, we can extract interesting information from this class of orbits.\footnote{Note that these geodesics were also found in \cite{RodriguezGomez2021}.}

\begin{figure}[t!]
	\begin{center}
		\includegraphics[width=0.5\textwidth]{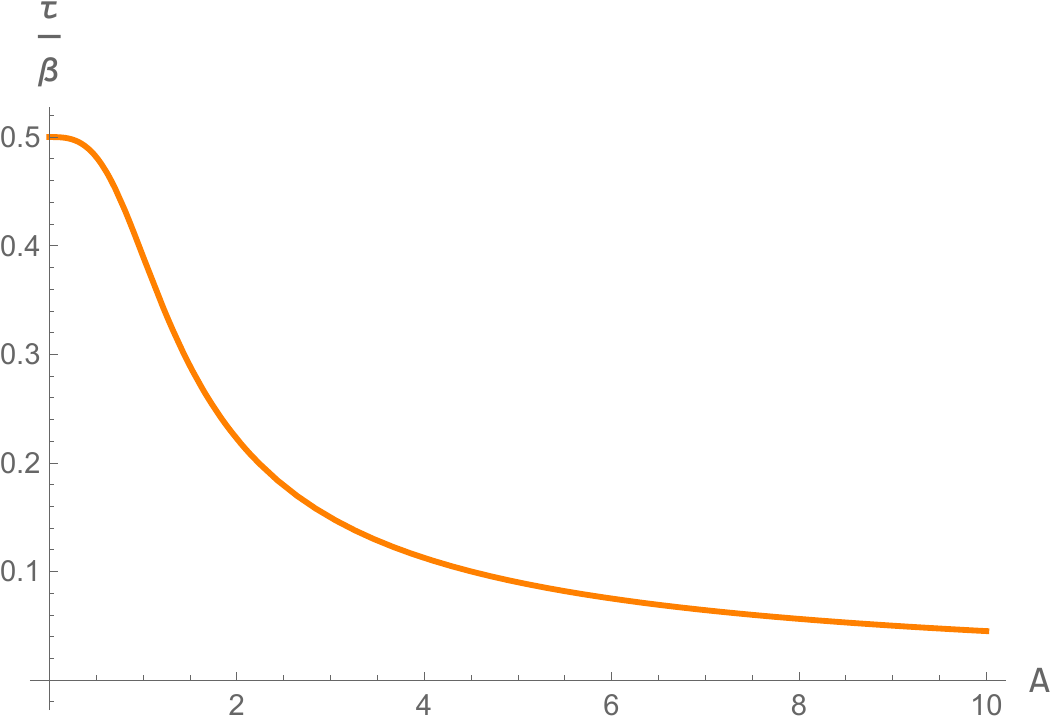}
		\caption{ This plot shows the behaviour of $\frac{\tau}{\beta}(A)$ for positive $A$ as fixed by the boundary condition \eqref{eq:BBExactGeodFinalBC} at $\lambda=1$.
			\label{fig:TauofA}}
	\end{center}
\end{figure}

For all $A>0$ these trajectories represent physical solutions. In particular $\tau(\lambda)$ is real and $0<z(\lambda)<z_H$ for any $0<\lambda<1$. We can interpret \eqref{eq:BBExactGeodFinalBC} as indicating which point $\tau$ on the thermal circle is reached by the geodesic as a function of the integration constant $A$. In Figure \ref{fig:TauofA} we plot $\tau(A)$. We see that the furthest $\tau$ one is able to reach is half of the circle, $\tau = \beta/2$, corresponding to $A=0$. An analogous behaviour is found if one repeats the analysis for $A<0$. One may also check (analytically) that $\tau(A)$ is a monotonically 
decreasing function of $A$.

The ambient geodesic distance square spanned by this class of trajectories is
\begin{equation}\label{}
	\X_{12}(\tau) = \frac{2}{\pi^2} \frac{\beta^2}{\sqrt{1+A(\tau)^4}}.
\end{equation}
One can test the small-$\tau$ behaviour of this geodesic distance to exclude the presence of long geodesics, which would correspond to $\X_{12} \sim \beta^2$  in $\tau/\beta\to0$, as opposed to short geodesics whose scaling is $\X_{12} \sim \tau^2$. By inverting the relation \eqref{eq:BBExactGeodFinalBC} in a series at small $\tau/\beta$ one can check that this class of exact ambient geodesics reduces to the perturbative geodesics of section \ref{SSec_B_Geods} for close insertions, meaning in particular that
\be
\X_{12}(\tau) = \tau^2 \left(1 + O(\tau^2)\right) \,.
\ee
Therefore, perturbatively close insertions always correspond to perturbatively short geodesics, contrary to what happens in thermal AdS.
To further confirm this picture we performed a numerical scan allowing for non-trivial momentum along the $x$ direction and still, no long geodesics were found. This suggests that the double twist spectrum arises in ambient correlators in a different way than a sum over long ambient geodesics, hinting at the existence of a class of ambient invariants not considered in this work.

\bibliographystyle{JHEP}
\bibliography{AEspace}

\end{document}